\documentclass[11pt]{My_preprint}

\title{Averaged equations for disperse two-phase flow with interfacial properties and their closures for dilute suspension of droplets}

\author[1,2]{Nicolas Fintzi}
\author[1]{Jean-Lou Pierson}
\affil[1]{IFP Energies Nouvelles, Rond-point de l'echangeur de Solaize, 69360 Solaize}
\affil[2]{Sorbonne Universit\'e, Institut Jean le Rond $\partial$'Alembert, 4 place Jussieu, 75252 PARIS CEDEX 05, France}

\begin{document}

\maketitle

\begin{abstract}
This article provides a derivation of the averaged equations governing the motion of dispersed two-phase flows with interfacial transport. 
We begin by revisiting the two-fluid formulation, as well as the distributional form of the interfacial transport equation which holds on the entire domain. 
Following this, a general Lagrangian model is introduced, which accounts for the effects of both internal and interfacial properties of the dispersed inclusions (bubbles, droplets, or particles) within a continuous phase.
This is achieved by derivation of conservation laws for particle surface and volume-integrated properties. 
By summing the internal and interfacial conservation laws, we derive a conservation equation for an arbitrary Lagrangian property associated with the inclusion. 
We then proceed by deriving the lesser-known conservation equations for the moments of the volume and surface distribution of an arbitrary Lagrangian property.  
Next, the averaged equations for the dispersed phase are derived through two distinct approaches: the particle-averaged (or Lagrangian-based) formalism, and the phase-averaged method. 
One important conclusion of this work is the demonstration of the relationship between the particle-averaged and phase-averaged equations. 
We show that the dispersed phase-averaged equations can be interpreted as a series expansion of the particle-averaged moment equations. 
We then present a "hybrid" set of equations, consisting of phase-averaged equations for the continuous fluid phase, complemented by an arbitrary number of moment conservation equations for the dispersed phase.
To further illustrate the methodology, we derive the mass, momentum, second moment of mass and first moment of momentum equations for droplets or bubbles suspended in a Newtonian fluid. 
In particular, we highlight the role of the second-order moment of mass equation and first-order moment of momentum equation, which link droplet deformation to the stresslet. 
We then derive closure laws in the dilute, viscous-dominated regime, with particular emphasis on the effects of surface tension gradients.
Additionally, we discuss several covariance closure terms that emerge in the averaged equations. 
Finally we demonstrate how the leading order deformation of the droplets can be obtained thanks to the second-order mass moment and first moment of momentum equation.
\end{abstract}

\section{Introduction}

Dispersed multiphase flows are encountered across a wide range of chemical engineering applications. 
These include gas-solid interactions in fluidized bed reactors, liquid-liquid flows in extractors, and gas-liquid bubbly flows in flotation processes. 
These systems exhibit a wide range of scales, from the size of individual inclusions (as small as a few micrometers) to the size of the reactor (often exceeding one meter), making fully resolved simulations computationally impractical. 
As a result, the current engineering practice relies on averaged equations of motion for both the dispersed and continuous phases. 
Therefore, developing a robust set of averaged equations that accurately captures the complexities of dispersed multiphase flows is essential. 
In this study, we aim to overcome the limitations of existing models by proposing a comprehensive set of averaged equations that can be applied to various types of dispersed multiphase flows, including interfacial transport processes.
Indeed, in many chemical engineering processes involving liquid or gas inclusions, mass transfer occurs between phases. 
This interphase mass transfer can generate gradients in surface tension, which in turn give rise to various phenomena such as Marangoni convection \citep{wegener2014fluid} or capillarity-induced motion of a droplet due to concentration gradient \citep{Subramanian_1985}.

The majority of averaged two-fluid models are based on the framework proposed by \citet{drew1983mathematical}, where both the dispersed and continuous phases are treated similarly (see, for example, \citet{hu2021cfd}). 
Despite its versatility-allowing applications beyond dispersed flows, such as stratified or slug flow, the physical interpretation of the closure terms remains difficult \citep{drew1983mathematical}, and the mathematical well-posedness of the entire system is still a matter of debate \citep{panicker2018hyperbolicity, lhuillier2013}.
An alternative approach treats the dispersed phase using a Lagrangian framework. 
This approach originates from the kinetic theory of gases, where the motion of particles is described by a single-particle distribution function governed by the Boltzmann equation \citep{chapman1990mathematical}. 
It has proven particularly successful in predicting the dynamics of dry granular materials \citep{rao2008introduction} and gas-solid flows \citep{simonin1996}. 
However, the fluid phase and the dispersed phase are handled through two distinct formalisms: phase averaging for the fluid and Lagrangian averaging for the dispersed phase.
In a pioneering study, \citet{buyevich1979flow} demonstrated that these two averaging methods could be connected by performing a Taylor expansion of the closure terms around the particle’s center of mass. 
This ``hybrid'' formalism was later revisited by \citet{lhuillier1992ensemble} and by \citet{zhang1994averaged, zhang1994ensemble}. The latter provided closures for the momentum transport equations in the inviscid flow limit. 
Subsequently, \citet{jackson1997locally} and \citet{zhang1997momentum} provided appropriate closures for the Stokes flow regime for dilute suspensions involving spherical solid particles and spherical droplets, respectively.

Many previous studies that used the ``hybrid'' formalism have focused on solid particles \citep{buyevich1979flow,jackson1997locally}. 
Notable exceptions include \citet{zhang1994ensemble}, who investigated spherical bubbles with time-varying radii, and \citet{zhang1997momentum}, who examined spherical droplets. 
Despite these advancements, the question remains: how can different inclusion shapes, internal fluid motion, and surface transport equations be incorporated into such Lagrangian models? 
To the best of our knowledge, important surface properties , such as variable or constant surface tension, the arbitrary shape of fluid particles, and the distribution of surfactants, have not yet been fully incorporated into current averaged hybrid models. 
While the Lagrangian formulation of the interfacial area equation is well established \citep{lhuillier2000bilan}, its application to define general Lagrangian quantities for a dispersed phase is still not fully explored.

Several authors have addressed the question of equivalence between Lagrangian (or particle-averaged) and phase-averaged equations in various contexts \citep{zhang1997momentum,lhuillier2000bilan,nott2011suspension}. 
In \citet[Appendix A]{zhang1997momentum}, the authors demonstrated that these two frameworks are equivalent for spherical inclusions when only considering the first-order moments. 
Later, \citet[Appendix A]{nott2011suspension} extended the proof of \citet{zhang1997momentum}, showing that the Lagrangian and phase-averaged equations are strictly equivalent for suspensions of solid spherical particles by considering an infinite number of higher-order terms. 
In addition, \citet{lhuillier2010multiphase} argued that the phase-averaged equations applied to the dispersed phase are, in fact, a Taylor series expansion of the particle-averaged moment equations.
Similarly, in the context of interfacial area balance equations, \citet{lhuillier2000bilan} reached a comparable conclusion when comparing the phase-averaged and particle-averaged area density equations for spherical particles. 
These studies focus on monodisperse suspensions of spherical particles and all arrive at the same conclusion: the particle-averaged equation is rigorously equivalent to the phase-averaged equation for the dispersed phase.

Following the above review, there is a need for a comprehensive hybrid model that encompasses all relevant physical aspects, including surface properties and arbitrarily shaped particles. Although this problem has been studied by several research groups including \citet{lhuillier1992ensemble} and \citet{zhang1997momentum}; our contribution lies in presenting a unified and simple theoretical framework for deriving the hybrid set of governing equations. Here the Lagrangian balance equations are derived without any simplifying assumptions, thus allowing for: (1) the incorporation of interfacial properties while retaining a Lagrangian framework, and (2) the representation of internal gradients within droplets through the use of distribution moments.
Additionnally we provide a general relationship between the phase-averaged and particle-averaged frameworks for inclusions of arbitrary shapes, including interfacial transport phenomena.
We rigorously demonstrate the relation between the Lagrangian-averaged and Eulerian-averaged formulations, thereby extending earlier results by \citet{nott2011suspension}.
A central feature of our approach is the emphasis on moment equations for the dispersed phase, as previously evidenced by \citet{lhuillier2009rheology} in the context of solid particles. We argue that, regardless of the specific problem under consideration - including those involving fluid particles of complex shape - the hybrid formulation offers a more physically grounded alternative to the traditional two-fluid model for describing dispersed flows.
To illustrate our methodology, we derive the mass, momentum, and two first-order moment equations for droplets or bubbles suspended in a Newtonian fluid. 
In particular, we present the first-order moment of momentum equation, which connects droplet deformation to the stresslet, extending the classical result of \citet{batchelor1970stress}.

Despite the extensive work on dispersed two-phase flows, only a few theoretical results exist that fully account for the whole closure problem. 
In the regime of Stokes flow and for very dilute suspensions, \citet{jackson1997locally} and \citet{zhang1997momentum} independently derived the momentum closure for spherical solid particles, with \citet{zhang1997momentum} extending the work to spherical drops. 
The inclusion of the symmetric part of the first moment of hydrodynamic forces, namely the stresslet, leads to Einstein's viscosity formula \citep{lhuillier1996contribution}. 
Even in this regime, the averaged fluid phase exhibits non-Newtonian behavior, induced by the relative motion of the particles due to the incorporation of the second moment of hydrodynamic forces \citep{lhuillier1996contribution,zhang1997momentum}. In this work we provide closure laws in the dilute, viscous-dominated regime, with special attention to the role of surface tension gradients and the closure of higher-order moment equations.

In \ref{sec:two-fluid}, we begin by presenting the two-fluid formulation and the distributional form of the interfacial transport equation. 
Next, in \ref{sec:Lagrangian}, we introduce a Lagrangian-based model that describes individual fluid particles with arbitrary shapes and surface properties. 
In \ref{sec:averaged_eq}, we highlight the link between the particle-averaged equations and the phase-averaged equation for any type of dispersed inclusions. 
Building on this demonstration, we introduce a hybrid set of conservation equations: one equation for the fluid phase and multiple equations for the dispersed phase, representing the conservation of moments of the distribution of the conserved quantity. 
We next derive the conservation equations for mass, momentum, and higher-order moments for both the dispersed and fluid phases, as presented in \ref{sec:averaged_surface}.
Subsequently, in \ref{sec:closure}, we focus on deriving the closure relations in the limit of dilute, viscosity-dominated flows.
Finally, we discuss the findings of this investigation in \ref{sec:conclusion}.

\section{The two-fluid formulation}
\label{sec:two-fluid}

In this section, we derive the conservation equations using a two-fluid formulation. While this approach has been employed in numerous studies, including those by \citet{kataoka1986local}, \citet{lhuillier2010multiphase}, \citet{ishii2010thermo}, \citet{morel2015mathematical}, and \citet{bothe2022sharp}, our method enables us to establish general results that encompass previous work.

\begin{figure}[h!]
    \centering
    \begin{tikzpicture}
        \foreach \x/\y/\ra/\r in {
        1/3/0.2/0.25,
        2.55/2.7/0.4/0.4,
        0.5/0.4/0.35/0.4,
        2/1/0.4/0.4,
        3/1.5/0.3/0.3,
        0.5/1.5/0.2/0.3,
        -0.5/2.5/0.35/0.5}{
            \draw[fill=gray!50](\x,\y) ellipse(\r cm and \ra cm);
        }
        \draw[dashed](3.2,1.7)circle(0.25);
\draw[thick,->](3.2,1.7)++(0.25,0)--++(1,0);
        \draw(2.55,2.7)node{$\Omega_d$};
        \draw(1.5,2)node{$\Omega_f$};
        \draw(1.45,1)node{$\Gamma$};
        \draw(2.5,0)node{$\Omega = \Gamma \cup \Omega_f \cup \Omega_d$};
\end{tikzpicture}
    \begin{tikzpicture}\draw[very thick](0:2)arc(0:90:2)node[above right]{$\Gamma_\alpha$};
        \draw[fill=gray!50](0:2)arc(0:90:2)arc(180:270:2);
        \draw[dashed](2,2)circle(2);
        \draw[->](1.42,1.42)--++(0.7,0.7)node[below right]{$\textbf{n}_d$};
        \draw[->](1.73,1)--++(-0.577,-0.333)node[below right]{$\textbf{n}_f$};
        \draw(2,3)node{$\Omega_f$};
        \draw(0.8,1.3)node{$\Omega_\alpha$};
    \end{tikzpicture}
    \caption{Topology of dispersed two-phase flows.}\label{fig:Scheme}
\end{figure}

We consider a system consisting of two phases, separated by a sharp interface $\Gamma(t,\FF)$,
where $t$ represents the current time and $\FF$ a flow realization over a set of possible dispersed multiphase flow realization.  
A realization can be represented by a list of variables such as particle positions, velocities and other factors that provide a complete and unique description of the flow \citep{zhang2021ensemble}. The phase subdomains are labeled as $\Omega_f(t,\FF)$ and $\Omega_d(t,\FF)$, corresponding to the continuous (fluid) phase ($f$) and the dispersed phase ($d$), respectively (see \ref{fig:Scheme}). 
The complete domain, $\Omega$, is composed of the union of $\Omega_f$, $\Omega_d$, and $\Gamma$. To track the position of phase $k$ and the interfaces, we introduce the phase indicator function $\chi_k$ defined as follows

\begin{align}
    \chi_k(\textbf{x},t,\FF) =  \left\{
      \begin{tabular}{cc}
        $1 \;\text{if} \;\textbf{x} \in \Omega_k(t,\FF)$\\
        $0 \;\text{if} \;\textbf{x} \notin \Omega_k(t,\FF)$
      \end{tabular}
      \right.
      \text{for $k = f,d$}.
      \label{eq:PIF}
\end{align}
Additionally, we define $\Omega_\alpha(t, \FF)$ as the region occupied by particle $\alpha$ at time $t$ within the configuration $\FF$ (\ref{fig:Scheme}). The collective union of $\Omega_\alpha(t, \FF)$ for $\alpha = 1, \ldots, N$, where $N$ denotes the number of particles present in the flow, corresponds to $\Omega_d(t, \FF)$. Similarly, we assume that $\Gamma(t, \FF)$ can be partitioned into $N$ subregions, labeled $\Gamma_\alpha(t, \FF)$, each representing the surface of particle $\alpha$.

\subsection{Topological equations}
Using the distribution formalism, one may show that $\chi_k(\textbf{x},t,\FF)$ obeys the following relations \citep{drew1983mathematical}. 
\begin{align}
    \pddt \chi_k
    + \textbf{u}_\Gamma^0 \cdot \grad \chi_k
    &= 0,
    \label{eq:dt_chi_k}\\
    \label{eq:grad_chi_k}
    \grad \chi_k
    &= - \delta_\Gamma \textbf{n}_k, 
\end{align}
where $\textbf{u}^0_\Gamma(\textbf{x},t,\FF)$ is the velocity of the interface and $\delta_\Gamma(\textbf{x},t,\FF)$ denotes the Dirac function localized on the interface, also called the interfacial indicator function \citep{drew1983mathematical,junqua2003}. 
More specifically the distribution $\delta_\Gamma$ is defined as \citep{appel2007}
\begin{equation}
<\delta_\Gamma,\varphi> =\int_{\Gamma} \varphi d\Gamma 
\label{eq:def_surf_distribution}
\end{equation}  
where $\varphi$ is a function with compact support. In this work, we use the subscript $_\Gamma$ to denote any quantity inherently defined on the interfaces, such as $\textbf{u}_\Gamma^0(\textbf{x}, t, \FF)$ and $\delta_\Gamma(\textbf{x}, t, \FF)$. Additionally, we employ the superscript $^0$ on a field to indicate that it is defined at the local, non-averaged level, meaning it is a function of the flow configuration $\FF$, the position vector $\textbf{x}$, and time $t$. For clarity and readability, we omit the arguments $\FF$, $\textbf{x}$, and $t$ for any local function denoted by $^0$ (as well as for $\delta_\Gamma$ and $\chi_k$), as their dependence on $\textbf{x}$, $t$, and $\FF$ is implied.

To derive a conservation equation for $\delta_\Gamma$, one can take the gradient of \ref{eq:dt_chi_k} and then compute the dot product of the resulting expression with $\textbf{n}_k$. This approach yields the following result \citep{marle1982macroscopic,drew1990,lhuillier2000bilan,junqua2003}

\begin{equation}
    \pddt \delta_\Gamma
    + \div [(\textbf{u}_\Gamma^0\cdot \textbf{n}) \textbf{n}\delta_\Gamma ]
    = (\textbf{u}_\Gamma^0 \cdot \textbf{n})(\div\textbf{n}) \delta_\Gamma ,
    \label{eq:dt_delta_I}
\end{equation} 
where we make use of the relation $\textbf{n}_k\cdot \pddt\textbf{n}_k= \frac{1}{2}\pddt(\textbf{n}_k\cdot \textbf{n}_k) = 0$. 
We did not specify the index of the normal $\textbf{n}$ in \ref{eq:dt_delta_I} to emphasize that this equation is independent of the index $k$. This is because $\textbf{n}$ appears twice in each term of this equation. 
Likewise, we will omit the index of the normal vector in subsequent discussions unless it is required.
Additionally, note that in \ref{eq:dt_delta_I} only the normal component of the surface velocity is present, and the term $\div \textbf{n}$ represents twice the mean curvature of the surface \citep{aris2012vectors}.  A second equation is obtained by taking the gradient of the distribution $\delta_\Gamma $ (\ref{ap:delta_I}), namely, 
\begin{equation}
    \grad\delta_\Gamma  
    =   \grad (\textbf{n} \delta_\Gamma) \cdot \textbf{n}.
    \label{eq:grad_delta_I}
\end{equation}
\ref{eq:grad_delta_I} has been derived straightforwardly, yet it differs from the expression recently obtained by \citet{orlando2023evolution}.  The source of this discrepancy is unclear to the current authors. 
It might be due to differences in their notation, with their operator gradient potentially representing a normal derivative. 
\ref{eq:dt_chi_k}, \ref{eq:grad_chi_k}, \ref{eq:dt_delta_I} and \ref{eq:grad_delta_I}  are commonly referred to as the topological equations of two-phase flows.
These equations describe the spatiotemporal evolution of the interfaces topology.

\subsection{Local conservation equations}
\label{sec:local_eq}
We now introduce the local conservation laws that govern the fluid inside bulk phases (inside $\Omega_d$ and $\Omega_f$) and on the interfaces (on $\Gamma$). 
Let $f_k^0$ denote a volumetric quantity of arbitrary tensorial order defined in $\Omega_k$.
Similarly, let $f_\Gamma^0$ represent an arbitrary surface property defined on $\Gamma$. 
Note that $f_\Gamma^0$ can be defined as a volumetric quantity integrated over the thickness of the interfaces \citet[Chapter 2]{ishii2010thermo}. 
Therefore, if $f_k^0$ has units $[x]$, then $f_I^0$ has the unit $[x.L]$, where $L$ is the unit of a length. 
The subscript $_{||}$ is used to denote the projection of a field onto the plane tangential to the surface $\Gamma$. Specifically, for an arbitrary quantity $\textbf{g}$ defined on $\Gamma$, its tangential projection is given by $\textbf{g}_{||} = \bm\delta_{||}\cdot \textbf{g}$ where $\bm\delta_{||} = (\bm\delta-\textbf{nn})$ is a tangential projection operator and $\bm\delta$ is the identity tensor. 
This definition also extends to the gradient operator, where $\gradI= \bm\delta _{||}\cdot \grad$ is referred to as the surface gradient operator. In addition we may define the surface divergence operator as $\gradI \cdot ()= (\bm\delta _{||}\cdot \grad)\cdot()$.

Following the strategy outlined in \citep{e2001mechanical,ishii2010thermo,bothe2022sharp}, the local conservation equations for $f_k^0$ and $f_\Gamma^0$ can be written as follows,
\begin{align}
    \label{eq:dt_f_k}
    \pddt f_k^0
    +\div \left(
        f_k^0\textbf{u}_k^0
        - \mathbf{\Phi}_k^0
        \right)
    &= 
    s_k^0
    & \text{ in } \Omega_k,&\\
    \pddt f_\Gamma^0 
    + f_\Gamma^0 (\textbf{u}_\Gamma^0 \cdot \textbf{n})(\div \textbf{n})
    +\divI
    (f_\Gamma^0 \textbf{u}_{\Gamma||}^0
        - \mathbf{\Phi}_{\Gamma||}^0 )
    &= 
    s_\Gamma^0
    - \Jump{
       f_k (\textbf{u}_\Gamma^0 - \textbf{u}_k^0)
       + \mathbf{\Phi}_k^0
    } 
    & \text{ on } \Gamma.&
    \label{eq:dt_f_I}
\end{align}
The tensors $\mathbf{\Phi}_k^0$ and $\mathbf{\Phi}_{\Gamma||}^0$ represent the non-convective fluxes corresponding to the quantities $f_k^0$ and $f_\Gamma^0$, respectively. Similarly, $s_k^0$ and $s_\Gamma^0$ are the source terms corresponding to the quantities $f_k^0$ and $f_\Gamma^0$, respectively. We assume that only the tangential component of the tensor $\mathbf{\Phi}_{\Gamma}^0$ plays a role in the surface balance equation \citep{bothe2022sharp}. 
Consequently, the flux terms of the form $\divI[(\ldots)_{||}]$ in \ref{eq:dt_f_I} represent two-dimensional  convective and non-convective fluxes arising from tangential motions or diffusive processes at the interface.
Conversely, the term involving the local curvature $(\div \textbf{n})$ is related to the interface expansion or contraction resulting from the normal velocity of the interface $\textbf{u}_\Gamma^0\cdot \textbf{n}$. For practical uses, note that the advecting term in \ref{eq:dt_f_I} can be written in the more compact form $\divI(f_\Gamma^0 \textbf{u}_\Gamma^0)$. 
Indeed, one can show that $f_\Gamma^0 (\textbf{u}_\Gamma^0 \cdot \textbf{n})(\div \textbf{n})
+\divI(f_\Gamma^0 \textbf{u}_{I||}^0) = \divI(f_\Gamma^0 \textbf{u}_\Gamma^0)$, by noticing that $\textbf{n}\cdot\gradI(\ldots) = 0$ and $\divI\textbf{n} = \div\textbf{n}$ \citep{nadim1996concise}.
In \ref{eq:dt_f_I} we have introduced the notation $\Jump{\ldots}$, where $\Jump{\ldots} = \sum_{k=1}^2 [\ldots] \cdot \textbf{n}_k$ signifies the jump across the interface. Thus, the last term on the right-hand side of \ref{eq:dt_f_I} represent a source of $f_\Gamma^0$ due to the discontinuity of bulk properties on either side of the interface. 
It is important to note that \ref{eq:dt_f_k} and \ref{eq:dt_f_I} are uniquely defined within the domains $\Omega_k$ and $\Gamma$, respectively.
Consequently, these equations are referred to as local conservation equations.

\subsection{The two-fluid formulation}
The phase indicator function $\chi_k$, and the Dirac delta function $\delta_\Gamma $, allow the extension of \ref{eq:dt_f_k} and \ref{eq:dt_f_I} to the entire flow domain $\Omega$. 
This extension is achieved by employing the methodology introduced by \citet{drew1983mathematical} and \citet{kataoka1986local} for the conserving laws inside the volume (\ref{eq:dt_f_k}).
The two-fluid formulation may be obtained by multiplying \ref{eq:dt_f_k} by $\chi_k$. 
Using \ref{eq:dt_chi_k} and \ref{eq:grad_chi_k} we obtain
\begin{equation}
    \pddt (\chi_k f_k^0)
    + \div (
        \chi_k f_k^0 \textbf{u}_k^0
        - \chi_k \mathbf{\Phi}_k^0 
        )
    = 
    \chi_k s_k^0
    + \delta_\Gamma \left[
        f_k^0
        \left(
            \textbf{u}_\Gamma^0
            - \textbf{u}_k^0
        \right)
        + \mathbf{\Phi}_k^0
    \right]
    \cdot \textbf{n}_k.
    \label{eq:dt_chi_k_f_k}
\end{equation}
This yields an equation defined over $\Omega$, to be solved for the quantity $\chi_k f_k^0$ instead of $f_k^0$. 
Moreover, in contrast to \ref{eq:dt_f_k} we observe the emergence of the interfacial term $ \delta_\Gamma [\ldots]$ on the right-hand side of \ref{eq:dt_chi_k_f_k}. 
To the best of the author's knowledge, the general form of the interfacial transport equation, expressed using the distribution formalism, has not yet been derived. However, some specific forms applicable to a given $f_\Gamma^0$ can be found in \citet{marle1982macroscopic} and  \citet{teigen2009}.
The derivation of this equation is straightforward but requires some algebra which is detailed in \ref{ap:interface_proof}. This yields\begin{equation}
    \pddt (\delta_\Gamma f_\Gamma^0)  
    + \div (
        \delta_\Gamma  f_\Gamma^0 \textbf{u}_\Gamma^0
        - \delta_\Gamma  \mathbf{\Phi}_{\Gamma||}^0 
        )
    = 
    \delta_\Gamma s_\Gamma^0
    - \delta_\Gamma \Jump{
    f_k^0 (\textbf{u}_\Gamma^0 - \textbf{u}_k^0)
    + \mathbf{\Phi}_k^0},
    \label{eq:dt_delta_I_f_I}
\end{equation}
which corresponds to the conservation equation of $\delta_\Gamma f_\Gamma^0$, which is defined over $\Omega$.
Although \ref{eq:dt_f_I} and \ref{eq:dt_delta_I_f_I} appear similar in many respects, they possess a fundamental distinction: \ref{eq:dt_f_I} is defined exclusively on $\Gamma$, it is a two-dimensional partial differential equation with a time-dependent metric, whereas \ref{eq:dt_delta_I_f_I} is a three-dimensional partial differential equation. 
This distinction holds significant importance, as the formulation of \ref{eq:dt_delta_I_f_I} proves to be more practical for the numerical modeling of quantities on deformable interfaces \citep{teigen2009}. The set of equations formed by \ref{eq:dt_chi_k_f_k} for $k = f,d$ and the surface transport equation or \textit{jump condition} \eqref{eq:dt_delta_I_f_I} is commonly known as the \textit{two-fluid} formulation of multiphase flows \citep{morel2015mathematical,tryggvason2011direct,drew1983mathematical,kataoka1986local}. 

According to \ref{eq:def_surf_distribution} the distribution  $\delta_\Gamma$ has the dimension of the inverse of a length, $[L^{-1}]$ \citep{appel2007}. 
Consequently, if the dimensionality of $f_k^0$ is assumed to be $[x]$, it follows that the dimensionality of $f_\Gamma^0$ must be $[x.L]$. Therefore , the product $\delta_\Gamma f_\Gamma^0$ carries the dimensionality $[x]$. Consequently, we can define a \textit{bulk} property $f^0$ as $f^0 = \sum_{k} \chi_k f_k^0 + \delta_\Gamma  f_\Gamma^0$ where $f^0$ represents any property of the flow of arbitrary tensorial order.
Then by summing \ref{eq:dt_chi_k_f_k} for $k=f,d$ and \ref{eq:dt_delta_I_f_I}, one obtains the \textit{bulk} formulation of multiphase flows, namely,\begin{equation}
   \pddt f^0
   + \div (
       f^0\textbf{u}^0
       -  \mathbf{\Phi}^0 
    )
   = s^0. 
   \label{eq:dt_f}
\end{equation}
where $f^0 \textbf{u}^0 = \sum_{k} \chi_k {f}_k^0\textbf{u}_k^0 + \delta_\Gamma  {f}_\Gamma^0\textbf{u}_\Gamma^0$, $\mathbf{\Phi}^0 = \sum_{k} \chi_k \mathbf{\Phi}_k^0 + \delta_\Gamma  \mathbf{\Phi}_\Gamma^0$ and $s^0 = \sum_{k} \chi_k s_k^0 + \delta_\Gamma  s_\Gamma^0$. 
In the literature, the \textit{single-fluid} formulation is characterized by summing only the quantities within the two phases, while the interfacial components are treated as source terms \citep{tryggvason2011direct,morel2015mathematical}.
Nevertheless, we want to point out here that our definition of $f^0$ yields a straightforward transport equation for $f^0$ thereby ensuring the consistency of the entire system of equations.

 \subsection{The averaged conservation equations}
\label{sec:avg_def}
In this study, we employ the ensemble average technique to establish the averaged conservation equations. 
This method is just one of several averaging approaches, including the volume average method \citep{jackson1997locally} and time averaging \citep{ishii2010thermo}. 
Although these techniques differ, they ultimately produce the same set of averaged equations, as evidenced by comparing volume averaging theory with ensemble averaging theory \citep{jackson1997locally,zhang1997momentum}. However, for slightly non-uniform suspensions, ensemble averaging is better suited for deriving the averaged equations because it does not require assumptions about the characteristic length scale of the spatial filter \citep{lhuillier1992ensemble}.
In the following we recall some properties of the ensemble average operator. 
Let, $P(\FF)$ be the probability density function that describes the probability of finding the flow in the configuration $\FF$. 
We note $d\PP = P(\FF) d\FF$ the probable number of flows located in the incremental region of the phase space $d\FF$ around the point $\FF$. 
It follows from this definition, that the ensemble average of an arbitrary local property $f^0(\textbf{x},t;\FF)$ defined on the whole space $\Omega$, is,
\begin{equation}
    f(\textbf{x},t)
    = \avg{f^0}(\textbf{x},t)
    =\int f^0(\textbf{x},t;\FF) d\mathscr{P}. 
    \label{eq:avg}
\end{equation}  
Note that we dropped the super script $^0$ on $f(\textbf{x},t)$ to indicate that this is an averaged quantity. 
The macroscopic variables are averaged over all $\FF$, and therefore depend only on $\textbf{x}$ and $t$.
Thus, we omit the arguments of the averaged fields, as this notation eliminates any potential ambiguity. 
The ensemble average operator is assumed to satisfy the following properties \citep{drew1983mathematical}
\begin{align}
    \avg{f^0+h^0} = f+h, 
    && \avg{\avg{f^0}h^0} = fh, \nonumber\\
    \avg{\pddt f^0} 
    = \pddt f,  
    &&\avg{\grad f^0}
    = \grad f. 
    \label{eq:avg_properties}
\end{align}
were $f^0$ and $h^0$ are two arbitrary Eulerian fields defined over $\Omega$. 
The first two relations are called the Reynolds' rules, the third one is the Leibniz' rule and the last one, is the Gauss' rule \citep{drew1983mathematical}.
Additionally, for any phase quantity defined in $\Omega_k$ we introduce the definition, 
\begin{equation}
    \phi_k f_k (\textbf{x},t) = \avg{\chi_k f_k^0},
    \label{eq:1_avg}
\end{equation}
where $\phi_k(\textbf{x},t) = \avg{\chi_k}$ is the probability of finding the phase $k$ at the location \textbf{x} and time $t$.
And $f_k$ is the average of the field $f_k^0$ conditioned on the presence of the phase $k$ in the configuration $\FF$ at $\textbf{x}$ and time $t$.
Equally, for interface quantities we have 
\begin{equation}
    \phi_\Gamma f_\Gamma (\textbf{x},t) = \avg{\delta_\Gamma f_\Gamma^0},
\end{equation}
with $\phi_\Gamma = \avg{\delta_\Gamma}$ the interface area per unit volume, also called the specific area at the point \textbf{x} at time $t$. 
Here, $f_\Gamma$ is the average of $f^0_\Gamma$ conditioned on the presence of an interface in the configuration $\FF$ at $\textbf{x}$ and time $t$. 
Additionally, we define the field of fluctuation of a given quantity around its mean as,
\begin{align}
    f'(\textbf{x},t,\FF) = f^0(\textbf{x},t,\FF) - f(\textbf{x},t).
    \label{eq:def_fluctu}
\end{align}
This relation applies to phase averaged quantities such that $f'_k = f^0_k - f_k$ and $f'_\Gamma = f^0_\Gamma - f_\Gamma$.

Applying the ensemble average on \ref{eq:dt_chi_k_f_k} and \ref{eq:dt_delta_I_f_I} and considering the properties from \ref{eq:avg_properties} to \ref{eq:def_fluctu}, yields the general form of the averaged equations of multiphase flows, namely,
\begin{align}
    \pddt (\phi_k f_k)
    +\div (\phi_k f_k \textbf{u}_k + \mathbf{\Phi}_k^\text{eq})
    &= 
    \phi_k s_k
    + \avg{\delta_\Gamma\left[
        \mathbf{\Phi}_k^0
        + f_k^0
        \left(
            \textbf{u}_\Gamma^0
            - \textbf{u}_k^0
        \right)
    \right]
    \cdot \textbf{n}_k},
    \label{eq:avg_dt_chi_f}\\
    \pddt (\phi_\Gamma f_\Gamma)
    +\div (\phi_\Gamma f_\Gamma \textbf{u}_\Gamma+ \mathbf{\Phi}_\Gamma^\text{eq})
    &= 
    \phi_\Gamma s_\Gamma
    - \avg{\delta_\Gamma 
    \Jump{
    \mathbf{\Phi}_k^0
    + f_k^0 (\textbf{u}_\Gamma^0 - \textbf{u}_k^0)
    } 
     },
    \label{eq:avg_dt_delta_f}
\end{align}
with, 
\begin{align}
    \mathbf{\Phi}_k^\text{eq}
    = \avg{\chi_k f_k' \textbf{u}_k'}
    - \phi_k \bm\Phi_k,
    &&
    \mathbf{\Phi}_\Gamma^\text{eq}
    = \avg{\delta_\Gamma f_\Gamma' \textbf{u}_\Gamma'}
    - \phi_\Gamma \bm\Phi_\Gamma. 
\end{align}
These equations are to be solved for the averaged field $f_f,f_d$, and $f_\Gamma$.
The main differences between these equations and their microscale counterparts (\ref{eq:dt_f_k} and \ref{eq:dt_f_I}) are:
(1) The unknowns are now averaged quantities,
(2) factors $\phi_k$ and $\phi_\Gamma$ are introduced in front of all the terms, and
(3) the additional terms $\avg{\chi_k f_k' \textbf{u}_k'}$ and $\avg{\delta_\Gamma f_\Gamma' \textbf{u}_\Gamma'}$ appear, representing the covariance between the conserved quantity ($f_k$ or $f_\Gamma$) and the local velocities.

It is important to highlight that the two-fluid model fails to adequately distinguish between the two phases, as evidenced by the \textit{symmetry} $k = f$ and $d$ in the aforementioned equations. This symmetry does not hold physically because the dispersed phase possesses a distinct topological nature compared to the continuous phase. 
In a dispersed two-phase flow system, the closure terms are typically expressed as functions of the Lagrangian properties of the particles \citep{jackson2000}. In contrast, the current system of equations yields continuously averaged quantities, which are not directly connected to the Lagrangian properties.
Therefore, in the subsequent section, we will introduce a kinetic-based model specifically devoted to the dispersed phase. 
As illustrated below, the equations governing the dispersed phase are more comprehensive as they bear a resemblance to the equations governing a single particle.

 \section{Lagrangian equations for the dispersed phase}
\label{sec:Lagrangian}

Lagrangian-based modeling of the dispersed phase has been explored extensively in numerous studies \citep{buyevich1979flow, lhuillier1992volume, simonin1996, zhang1994averaged, zhang1994ensemble, zhang1997momentum, jackson1997locally, zaepffel2011modelisation}. 
However, these studies predominantly focus on solid particles \citep{buyevich1979flow, lhuillier1992volume, simonin1996, zhang1994averaged, jackson1997locally} or non-deformable spherical fluid inclusions \citep{zhang1994ensemble, zaepffel2011modelisation}. 
In this work, we aim to address a more general scenario by considering fluid particles with arbitrary shapes and surface properties. Thus, we propose a Lagrangian-based model for each inclusion capable of describing the dispersed phase with arbitrary accuracy. 

 \subsection{Fundamental properties of the dispersed phase}
We define the mass $m_\alpha$, the position of center of mass $\mathbf{x}_\alpha$ and the momentum $\textbf{p}_\alpha$ as
\begin{align}
    m_\alpha(t,\FF)
    &= \intOF{ \rho_d  }, 
    \\
    \textbf{x}_\alpha(t,\FF)
    &= \frac{1}{m_\alpha }\intOF{ \rho_d \textbf{x} }     \label{eq:mass_pos0},
    \\ \textbf{p}_\alpha(t,\FF)
    &= \intOF{ \rho_d \textbf{u}_d^0 }.
\end{align}
Note that while the mass of the interface might be considered in the center of mass calculation, in practice, the mass within the volume is typically much greater than the mass on the interface. 
Therefore, we have opted to use the more straightforward definition considering only the mass within the particle volume.
Subsequently, we define the velocity of the particle center of mass as
\begin{equation}
\textbf{u}_\alpha(t,\FF) = \frac{d \textbf{x}_\alpha}{dt}.
\label{eq:u_alpha}
\end{equation}
We also define the particle's internal relative motions or the \textit{inner velocity}  as $\textbf{w}_d^0 = \textbf{u}_d^0 - \textbf{u}_\alpha$. 
Similarly we define $\textbf{w}_\Gamma^0$ as $\textbf{w}_\Gamma^0 = \textbf{u}_{\Gamma||}^0 - \textbf{u}_\alpha$.
Replacing $\textbf{x}_\alpha$ by its definition \eqref{eq:mass_pos0} in \ref{eq:u_alpha} we obtain
\begin{equation}
    \textbf{u}_\alpha(t,\FF) = \frac{1}{m_\alpha}
    \frac{d}{dt} 
    \left(
        \intOF{ \rho_d \textbf{x} }
    \right)
    - \frac{1}{m_\alpha^2} \frac{d}{dt} \left(\intOF{ \rho_d } \right)
    \intOF{ \rho_d \textbf{x} }.
\end{equation}
As will become clear in \ref{subsec:Lcl}, using the Reynolds transport theorem \eqref{eq:reynolds_transport} for both terms in parentheses and the definition of $\textbf{x}_\alpha(t,\FF)$ in the last term gives
\begin{align}
    \textbf{u}_\alpha(t,\FF) &=  \frac{1}{m_\alpha}\intOF{
        \pddt (\textbf{x}\rho_d ) + \div\left(\textbf{u}_d^0 \textbf{x} \rho_d\right) 
    } 
    + \frac{1}{m_\alpha}\intSF{ \textbf{x} \rho_d(\textbf{u}_\Gamma^0 - \textbf{u}_d^0) \cdot \textbf{n}_d } \nonumber\\
    & - \frac{1}{m_\alpha^2}\intOF{
        \pddt (\rho_d ) + \div\left(\textbf{u}_d^0 \rho_d\right) 
    } -  \frac{\textbf{x}_\alpha}{m_\alpha}    \intSF{ \rho_d(\textbf{u}_\Gamma^0   - \textbf{u}_d^0) \cdot \textbf{n}_d }
    \label{eq:u_alpha2}
\end{align}
Making use of the conservation of mass, the first term on the second line of \ref{eq:u_alpha2} vanishes. This yields,\begin{multline}
    \textbf{u}_\alpha(t,\FF) = 
    \frac{1}{m_\alpha}\intOF{ \left[
        \pddt (\textbf{x}\rho_d ) + \div\left(\textbf{u}_d^0 \textbf{x} \rho_d\right) 
    \right]} \\
    + \frac{1}{m_\alpha}\intSF{ \textbf{x} \rho_d(\textbf{u}_\Gamma^0   - \textbf{u}_d^0) \cdot \textbf{n}_d }
    -  \frac{\textbf{x}_\alpha}{m_\alpha}    \intSF{ \rho_d(\textbf{u}_\Gamma^0   - \textbf{u}_d^0) \cdot \textbf{n}_d }
\end{multline}
Then by considering the mass conservation for the first term and noticing that $\grad \textbf{x} = \bm\delta$ gives\begin{equation}
    \textbf{u}_\alpha(t,\FF) = \frac{1}{m_\alpha(t,\FF)} \left(
        \textbf{p}_\alpha(t,\FF)
        +  \intSF{\rho_d \textbf{r} (\textbf{u}_\Gamma^0 - \textbf{u}_d^0)\cdot \textbf{n}_d }
        \right),
        \label{eq:dt_y_alpha}
\end{equation}
where $\textbf{r}(\textbf{x},t,\FF) = \textbf{x} - \textbf{x}_\alpha(t,\FF)$ is the distance between any point inside $\Omega_\alpha$, to the center of mass of the particle $\alpha$.
In \ref{eq:dt_y_alpha}, it can be observed that the first component of the velocity represents the linear momentum divided by the mass of the particle. 
This corresponds to the mass-averaged velocity over the volume of the particle.
The second term in \ref{eq:dt_y_alpha} arises from the contribution of anisotropic mass transfer across the surface of the particle. 
This mass transfer leads to the motion of the particle's center of mass, thereby contributing to the total velocity.
To illustrate this concept, let us consider a fixed drop with no momentum lying over a very hot plate.
In this scenario, we assume that the plate is sufficiently hot to induce evaporation, specifically on the bottom portion of the drop.
Hence, under the effect of an anisotropic evaporation flux one may expect the second term to be non-negligible.
Consequently, the center of mass of the drop has a non-zero velocity in the opposite direction of the plate, even though the momentum is assumed to be zero.
Note that \ref{eq:dt_y_alpha} generalizes usual expression of the center of mass velocity whom neglect the second term.
In the following, for the sake of brevity we discard the dependency on $t$ and $\FF$ on the notations for all Lagrangian quantities denoted by the subscript $_\alpha$ and in particular $\Gamma_\alpha$ and $\Omega_\alpha$.
Nevertheless, the reader must understand that all Lagrangian quantities and integration domains subscribed by $_\alpha$ are time and configuration-dependent.

\subsection{Lagrangian conservation laws}
\label{subsec:Lcl}

We assign to a particle indexed $\alpha$, occupying the domain $\Omega_\alpha$ (see \ref{fig:Scheme}) an arbitrary Lagrangian property $\text q_\alpha$ defined by $\text q_\alpha  = \intO{ f_d^0}$. The Reynolds transport theorem for the (material) particle volume can be written as \citep{leal2007advanced},

\begin{equation}
\ddt  \intO{f_d^0}
    = \intO{\pddt f_d^0 }\\
    + \intS{ f_d^0 \textbf{u}_\Gamma^0\cdot \textbf{n}_d }.
    \label{eq:reynolds_transport0}
\end{equation}
Adding and subtracting $f_d^0 \textbf{u}_d^0\cdot \textbf{n}_d$ in the last integral of this equation and using the divergence theorem, yields
\begin{equation}
    \ddt  \intO{f_d^0}
    = \intO{\left[ \pddt f_d^0 + \div\left(f_d^0\textbf{u}_d^0\right) \right]}\\
    + \intS{ f_d^0 (\textbf{u}_\Gamma^0-\textbf{u}_d^0)\cdot \textbf{n}_d }.
    \label{eq:reynolds_transport}
\end{equation}
By substituting \ref{eq:dt_f_k} into \ref{eq:reynolds_transport} and using the divergence theorem we obtain the conservation law of the quantity $\text q_\alpha$ \begin{equation}
    \frac{d \text q_\alpha}{d t}
    = \intO{ s_d^0 }
    + \intS{ \left[
        f_d^0 (\textbf{u}_\Gamma^0-\textbf{u}_d^0) 
        + \mathbf{\Phi}_d^0 
        \right] \cdot \textbf{n}_d }.
    \label{eq:dt_q_alpha}
\end{equation}
The first term on the right-hand side of \ref{eq:dt_q_alpha} accounts for the total contribution of the source term $s_d^0$ to the particle $\alpha$,
while the second term is the surface integration of the phase exhange flux $f_d^0 (\textbf{u}_\Gamma^0-\textbf{u}_d^0)$ and the non-convective flux $\mathbf{\Phi}_d^0$.

Similarly, we define $\text q_{\Gamma\alpha} = \intS{ f_\Gamma^0}$ as an integrated surface property of the particle $\alpha$.
To differentiate time-varying surface integrals with respect to time, we make use of the surface Reynolds transport theorem, which reads \citep[Appendix B]{morel2015mathematical} \begin{equation}
    \ddt  \intS{f_\Gamma^0 }
    = \intS{ \left[
        \frac{D_\Gamma f_\Gamma^0}{Dt}
        +   f_\Gamma^0\gradI \cdot \textbf{u}_\Gamma^0
    \right]}.
    \label{eq:surface_derivative}
\end{equation}
where $\frac{D_\Gamma}{Dt}  = \pddt + \textbf{u}_\Gamma^0\cdot \gradI $ is the material derivative operator on the surface of the particle. 
Upon expanding the acceleration term we obtain the expression,
\begin{equation}
    \ddt  \intS{f_\Gamma^0 }
    = \intS{ \left[
        \pddt f_\Gamma^0
        +   \gradI \cdot (\textbf{u}_\Gamma^0f_\Gamma^0)
    \right]}.
    \label{eq:surface_derivative}
\end{equation}
Then inserting \ref{eq:dt_f_I} into \ref{eq:surface_derivative} we obtain
\begin{equation}
    \ddt  \text q_{\Gamma\alpha}
    = \intS{ 
        \gradI \cdot \boldsymbol{\Phi}_{\Gamma||}^0
    }
    +\intS{ 
        s_\Gamma^0
    }
    - \intS{
 \Jump{
        f_k^0 (\textbf{u}_\Gamma^0 - \textbf{u}_k^0)
        + \mathbf{\Phi}_k^0
    }
    }.
    \label{eq:dt_q_I_alpha}
\end{equation}
The surface divergence theorem applied to closed surfaces  reads \citep{nadim1996concise}
\begin{equation}
    \intS{\gradI \cdot \textbf{F}}
    = 
    \intS{ \textbf{F} \cdot  \textbf{n} (\div \textbf{n})},
    \label{eq:gauss_surface}
\end{equation} 
where $\textbf{F}$ is an arbitrary field.
This theorem demonstrates that any surface property parallel to the tangential plane of $\Gamma$, such as $\bm\Phi_{\Gamma||}$, fulfills the condition $\intS{\divI \bm\Phi_{\Gamma||}^0}
= 0$.
Therefore \ref{eq:dt_q_I_alpha} yields, \begin{equation}
    \ddt  \text q_{\Gamma\alpha}
    = \intS{ 
        s_\Gamma^0
    }
    - \intS{
 \Jump{
        f_k^0 (\textbf{u}_\Gamma^0 - \textbf{u}_k^0)
        + \mathbf{\Phi}_k^0
    }
    }.
    \label{eq:dt_q_I_alpha}
\end{equation}
In \ref{eq:dt_q_alpha}, the non-convective flux within the dispersed phase appears on the right hand side. While this is not problematic for fluid particles for which the internal stress is defined, the stress within rigid particles is not defined \citep{batchelor1970stress}.
Hence to remove the internal stress from \ref{eq:dt_q_alpha}, we treat the particle's volume and surface as a single entity and derive a conservation equation for $\text Q_\alpha = \text q_\alpha + \text q_{\Gamma\alpha}$. 
By summing \ref{eq:dt_q_alpha} and \ref{eq:dt_q_I_alpha} we directly obtain 
\begin{equation}
    \ddt  \text Q_\alpha
    = 
    \intO{ s_d^0 }
    + \intS{ s_\Gamma^0 }
    + \intS{ \left[
        f_f^0 (\textbf{u}_\Gamma^0-\textbf{u}_f^0) 
        + \mathbf{\Phi}_f^0 
        \right] \cdot \textbf{n}_d }. 
    \label{eq:dt_q_alpha_tot}
\end{equation}
This equation is the general form of the linear conservation law for the quantity $\text Q_\alpha$.
It applies to any particle immersed in a continuous phase following the conservation laws given by \ref{eq:dt_f_k} and \ref{eq:dt_f_I}.
We refer to this equation as the zeroth-order conservation equation of $f_d^0$, or alternatively, the linear conservation law of $f_d^0$, for the particle $\alpha$.
We would like to highlight that due to the consideration of closed surface, the non-convective flux $\mathbf{\Phi}_{\Gamma||}^0$, does not appear in \ref{eq:dt_q_alpha_tot}.
Consequently, for the conservation of linear momentum, the surface stresses do not contribute to the momentum balance of a particle.
As a result, surface tension or surface viscous stresses do not directly affect the linear momentum balance \footnote{However, Marangoni effect can induce motion through the imbalance of tangential stress \citep{young1959}. Indeed the resulting imbalance alters the stress within the continuous phase, generating a hydrodynamic force.}. 
This property has already been demonstrated by \citet{hesla1993note} who showed that the surface tension force does not contribute to the linear and angular momentum balance. 
Here, we have provided the general proof that the interfacial non-convective flux $\mathbf{\Phi}_{\Gamma||}^0$, which is present at the local scale according to \ref{eq:dt_f_I}, does not contribute to the zeroth-order conservation law of a particle with a closed surface.
It is important to note that this conclusion is valid only if the non-convective flux is parallel to the interface, which is what we assumed here.

\subsection{Higher order moment equations}

Because $f_d^0$ and $f_\Gamma^0$ are not always constant over the volumes and surfaces of the particles, it is interesting to introduce the first moment of the quantities $f_d^0$ and $f_\Gamma^0$. 
They are defined as $\textbf{q}_\alpha^{(1)}     = \intO{ \textbf{r} f_d^0 }$ and $\textbf{q}_{\Gamma\alpha}^{(1)}    = \intS{ \textbf{r} f_\Gamma^0 }$.  
In general, the first moments $\textbf{q}^{(1)}_{\alpha}$ and $\textbf{q}^{(1)}_{\Gamma\alpha}$ hold significant importance when considering particles with high internal gradients, i.e. when $\grad f_d^0$ or $\gradI f_\Gamma^0$ are non-negligible at the scale of one particle. 
Typically, when considering the motion of a fiber, one must consider the angular momentum balance \citep{guazzelli2011}, which corresponds to the antisymmetric part of the first moment of momentum.
It is possible to differentiate the moments $\textbf{q}^{(1)}_\alpha$ and $\textbf{q}^{(1)}_{\Gamma\alpha}$  with respect to time to obtain their conservation laws.
We use \ref{eq:reynolds_transport} to describe the evolution of $\textbf{q}^{(1)}_\alpha$ within time. This yields, 
\begin{equation}
    \frac{d}{dt} \textbf{q}^{(1)}_\alpha
      =  \intO{\left[
        \pddt(  f_d^0\textbf{r})
        + \div \left(  f_d^0 \textbf{r}\textbf{u}_d^0\right)
    \right]} 
    + \intS{  f_d^0 \textbf{r}  (\textbf{u}_\Gamma^0-\textbf{u}_d^0)\cdot \textbf{n}_d}.
\end{equation}
The first term on the right-hand side may be rewritten as
\begin{equation}
\intO{ \left[
        \pddt(\textbf{r}  f_d^0)+ \div \left( \textbf{u}_d^0 \textbf{r} f^0_d\right) 
    \right]}
    = \intO{\textbf{r}\left[
        \pddt f_d^0
        + \div \left(f_d^0 \textbf{u}_d^0\right)
    \right] }
    + \intO{ f_d^0 \left[
        \pddt \textbf{r}
        +(\textbf{u}_d^0 \cdot \grad) \textbf{r}
    \right]}
\end{equation}
Substituting \ref{eq:dt_f_k} in the first integral on the right-hand side, and considering the relation,
$  \pddt \textbf{r}
+ (\textbf{u}_d^0 \cdot \grad) \textbf{r}
= - \frac{d}{dt} \textbf{x}_\alpha  + \textbf{u}_d^0 
= \textbf{w}_d^0$,
for the second integral yields 
\begin{align}
    \frac{d}{dt} \textbf{q}^{(1)}_\alpha = \intO{\textbf{r} (s_d^0 +\div \bm\Phi_d^0)  }
    + \intO{ f_d^0\textbf{w}_d^0 }  + \intS{  f_d^0 \textbf{r}  (\textbf{u}_\Gamma^0-\textbf{u}_d^0)\cdot \textbf{n}_d}
\end{align}
Then using relation $\intO{\textbf{r}  \div \bm\Phi_d^0 }
= \intS{ \textbf{r} \bm\Phi_d^0 \cdot \textbf{n}_d }
- \intO{ \bm\Phi_d^0 }$ we get
\begin{align}
    \frac{d}{dt} \textbf{q}^{(1)}_\alpha
= \intO{\left( 
        \textbf{r} s_d^0  
        + f_d^0  \textbf{w}_d^0
        - \bm\Phi_d^0
    \right) }
    + \intS{\textbf{r} \left[
        \bm\Phi_d^0
        + f_d^0 (\textbf{u}_\Gamma^0-\textbf{u}_d^0)
    \right]\cdot \textbf{n}_d}  .
    \label{eq:dt_Q_alpha}
\end{align}
 \ref{eq:dt_Q_alpha} is the first order moment conservation equation for the particle $\alpha$. 
 In \ref{eq:dt_Q_alpha}, we recognize the first moment of the source term $s_d^0$, the first moment of the non convective flux term $\bm\Phi_d^0\cdot\textbf{n}_d$ and the first moment of phase exchange term, $f_d^0 (\textbf{u}_\Gamma^0-\textbf{u}_d^0)\cdot\textbf{n}_d$. 
 Additionally, two supplementary terms appear in \ref{eq:dt_Q_alpha}, namely: the volume integral of the non convective flux $\bm\Phi_d^0$, and a term related to the fluctuation of the internal velocity $f_d^0 \textbf{w}_d^0$.

Following the same procedure, and making use of \ref{eq:dt_f_I}, \ref{eq:surface_derivative} and \ref{eq:gauss_surface} , one can equally show that 
\begin{align}
    \ddt {\textbf{q}^{(1)}_{\Gamma\alpha}}
    &= \intS{ \left(
        \textbf{r}s_\Gamma^0
        + f_\Gamma^0 \textbf{w}_\Gamma^0
        - \mathbf{\Phi}_{||\Gamma}^0
    \right) }
    - \intS{\textbf{r} 
    \Jump{\mathbf{\Phi}_k^0
        + f_k^0 (\textbf{u}_\Gamma^0 - \textbf{u}_k^0)
    }
    },
    \label{eq:dt_Q_I_alpha}
\end{align}
Observations similar to those for \ref{eq:dt_Q_alpha} can also be made for the first moment of surface equation \eqref{eq:dt_Q_I_alpha}.
In particular, it is worth noting the presence of the surface non-convective flux $\mathbf{\Phi}_{\Gamma||}^0$ in \ref{eq:dt_Q_I_alpha}.

For similar reason than the linear conservation equations, we sum \ref{eq:dt_Q_alpha} and \ref{eq:dt_Q_I_alpha} to express the conservation equation of the total first moment $\textbf{Q}^{(1)}_\alpha = \textbf{q}^{(1)}_\alpha + \textbf{q}^{(1)}_{\Gamma\alpha}$, this yields 
\begin{multline}
    \ddt {\textbf{Q}^{(1)}_\alpha}
    = \intO{ \left(
        \textbf{r} s_d^0         
        + f_d^0  \textbf{w}_d^0 
        - \mathbf{\Phi}_d^0
    \right) }
    + \intS{ \left(
        \textbf{r}s_\Gamma^0
        + f_\Gamma^0 \textbf{w}_\Gamma^0
        - \mathbf{\Phi}_{\Gamma||}^0
    \right) } \\
    + \intS{ \textbf{r} \left[
        \mathbf{\Phi}_f^0
        + f_f^0 (\textbf{u}_\Gamma^0-\textbf{u}_f^0)
    \right]\cdot \textbf{n}_d  }. 
    \label{eq:dt_Q_alpha_tot}
\end{multline}
Likewise, conservation laws can be derived for the $n^{th}$ order moments of volume and surface, i.e. for
\begin{align}
    \textbf{q}^{(n)}_{\alpha}
    = \intO{
         \underbrace{\textbf{rr}\ldots \textbf{rr}}_{n\text{ times}}
        f_d^0 },
        && \text{and} &&
    \textbf{q}^{(n)}_{\Gamma\alpha}
    = \intS{
         \underbrace{\textbf{rr}\ldots \textbf{rr}}_{n\text{ times}}
    f_\Gamma^0 },
    \label{eq:Q_n_definition}
\end{align} 
respectively. The time derivatives of $\textbf{q}^{(n)}_{\alpha}$ and $\textbf{q}^{(n)}_{\Gamma\alpha}$ do not introduce any additional terms beyond those already present in equations \ref{eq:dt_Q_alpha} and \ref{eq:dt_Q_I_alpha}.
Instead, they only involve the $n^{th}$ order moments of the existing terms.
We provide the full derivation of $\ddt{ \textbf{q}^{(n)}_{\alpha}}$ in \ref{ap:Moments_equations}.
The higher-order moments characterize the distributions of the local quantities $f_d^0$ and $f_\Gamma^0$ within the regions $\Omega_\alpha$ and $\Gamma_\alpha$, respectively. 
To precisely reconstruct the fields $f_d^0$ and $f_\Gamma^0$ within $\Omega_\alpha$ and $\Gamma_\alpha$, an infinite number of moments would theoretically be required. 
Nevertheless, as pointed out by \citet[Appendix A]{zhang1997momentum}, when the degrees of freedom of the particles are limited (solid particles, spherical droplets in stokes flows\ldots), only a finite number of moments are necessary to reconstruct $f_d^0$ and $f_\Gamma^0$.

\subsection{Particle-averaged equations}

In the preceding subsections, we have described the dispersed phase using a Lagrangian framework. 
However, to ensure consistency with the Eulerian conservation equations that describe the continuous phase, it is necessary to extend the Lagrangian equations to an Eulerian description. 
The approach presented here follows the methodology pioneered by \citet{lhuillier1992ensemble}.
We introduce the function $\delta_\alpha$, which is defined as follows, 
\begin{align}
    \delta_\alpha(\textbf{x},\textbf{x}_\alpha(t,\FF)) 
    = \delta(\textbf{x}-\textbf{x}_\alpha(t,\FF)),
    \label{eq:delta_alpha}
\end{align}
where $\delta$ is the Dirac function.
Note that we explicitly note the arguments $(t,\FF)$ to highlight that the position of the particle $\alpha$ is a function of time and of the flow configuration $\FF$.
Taking the time derivative on $\delta_\alpha$ and 
applying the chain rule yields \citep{lhuillier1998,lhuillier2009rheology}\begin{equation}
    \pddt \delta_\alpha
    + \div (\textbf{u}_\alpha  \delta_\alpha)
    =0,
    \label{eq:dt_delta_alpha}
\end{equation}
where we used the identity, $\frac{\partial \delta_\alpha}{\partial \textbf{x}_\alpha}  = -\grad \delta_\alpha$ and the fact that $\textbf{u}_\alpha(t,\FF)$ is not a function of $\textbf{x}$. 
\ref{eq:dt_delta_alpha} does not apply in scenarios where topological changes occur, such as break-up or coalescence events. 
In these cases, a source term can be introduced on the right-hand side of \ref{eq:dt_delta_alpha}, similar to the approach used in population balance equations, to account for the birth or death of particles \citep{randolph2012theory}.
Consider a domain containing $N$ particles. We define the \textit{particle field} for a quantity $\text q_\alpha$ as the sum of $\delta_\alpha \text q_\alpha$ over all particles within the domain, expressed by $\displaystyle\sum_{\alpha=0}^N \delta_\alpha \text q_\alpha$. 
Note that the formula given by \ref{eq:dt_delta_alpha} remains valid for the sum of such fields, since the operations of differentiation and summation commute.
To obtain the averaged equations for the dispersed phase, we define the particle average of $\text q_\alpha$ as
\begin{equation}
     n_p \text q_p(\textbf{x},t) = \avg{\sum_\alpha\delta_\alpha \text q_\alpha},
     \label{eq:p_avg}
\end{equation}
where, $n_p(\textbf{x},t) = \avg{\sum_\alpha \delta_\alpha}$ is the probability density of finding a particle center of mass in the infinitesimal volume $d\textbf{x}$ around \textbf{x}, and $\text q_p(\textbf{x},t)$ is the average of $\text q_\alpha$ conditionally on the presence of a particle's center of mass at \textbf{x} and time $t$. 
To simplify the notations, we consider the shorthand \citep{lhuillier1998},
\begin{equation*}
    \sum_\alpha \delta_\alpha \to \delta_p, 
\end{equation*}
such that $\pavg{\text q_\alpha}=\avg{\sum_\alpha \delta_\alpha \text q_\alpha}=n_p\text q_p$.
Note that we used the subscript $_p$ on $\text q_p$ to denote that this represents a particle-averaged field, initially derived from Lagrangian quantities. 
Additionally, in light of \ref{eq:def_fluctu} we define the fluctuating part of a particle field $\text q_p$ as
\begin{equation}
    \text q_\alpha'(\textbf{x},t,\FF) = \text q_\alpha(\FF,t) - \text q_p(\textbf{x},t). 
    \label{eq:def_fluc_p}
\end{equation}

To derive the averaged equations for the particle phase, we begin by noting that multiplying a Lagrangian quantity $\text q_\alpha$ by $\delta_\alpha$ yields the corresponding Eulerian field  $\text q_\alpha(t,\FF)\delta_\alpha(\textbf{x},t,\FF)$, which is defined throughout the entire domain $\Omega$. 
Similarly, for any derivative of a Lagrangian quantity, such as $\ddt \text q_\alpha$, the corresponding Eulerian field is defined by multiplying $\ddt \text q_\alpha$ with $\delta_\alpha$.
Given that $\text q_\alpha(t,\FF)$ and $\textbf{u}_\alpha(t,\FF)$ do not depend on \textbf{x}, and by using \ref{eq:dt_delta_alpha}, we obtain
\begin{equation}
    \delta_\alpha \ddt \text q_\alpha
    = \pddt (\delta_\alpha \text q_\alpha)
    + \div (\delta_\alpha \text q_\alpha \textbf{u}_\alpha).
    \label{eq:dt_delta_alpha_q_alpha}
\end{equation}
Multiplying \ref{eq:dt_q_alpha_tot} and \ref{eq:dt_Q_alpha_tot} by $\delta_\alpha$, using \ref{eq:dt_delta_alpha_q_alpha} and applying the ensemble average (\ref{eq:avg}), yields
\begin{align}
    \pddt (n_p\text Q_p)
    + \div (n_p \text Q_p \textbf{u}_p + \pavg{\textbf{u}_\alpha' \text Q_\alpha'})
    = \pOavg{ s_d^0 }
    + \pSavg{ s_\Gamma^0 }\nonumber\\
    + \pSavg{ \left[\mathbf{\Phi}_f^0 + f_f^0 (\textbf{u}_\Gamma^0-\textbf{u}_f^0) \right] \cdot \textbf{n}_d },
    \label{eq:avg_dt_dq_alpha_tot}\\
    \pddt (n_p\textbf{Q}_p^{(1)})
    + \div \left(n_p \textbf{Q}_p^{(1)} \textbf{u}_p + \pavg{\textbf{u}_\alpha' \textbf{Q}_\alpha^{(1)'}}\right)
    =\pOavg{ \left(
        \textbf{r} s_d^0         
        + f_d^0  \textbf{w}_d^0 
        - \mathbf{\Phi}_d^0
    \right) }\nonumber\\
    + \pSavg{ \left(
        \textbf{r}s_\Gamma^0
        + f_\Gamma^0 \textbf{w}_\Gamma^0
        - \mathbf{\Phi}_{\Gamma||}^0
    \right) }
    + \pSavg{ \textbf{r} \left[
        \mathbf{\Phi}_f^0
        + f_f^0 (\textbf{u}_\Gamma^0-\textbf{u}_f^0)
    \right]\cdot \textbf{n}_d  }.
    \label{eq:avg_dt_dQ_alpha_tot}
\end{align}
The derivation of the higher moment particle-averaged equations is provided in \ref{ap:Moments_equations}.
The only fluxes appearing in \ref{eq:avg_dt_dq_alpha_tot} and \ref{eq:avg_dt_dQ_alpha_tot} are the fluctuation tensors $\pavg{\textbf{u}_\alpha' \text q_\alpha'}$ and $\pavg{\textbf{u}_\alpha' \textbf{Q}_\alpha'}$. 
Therefore, the non-convective fluxes $\bm\Phi_d^0$ and $\bm\Phi_\Gamma^0$ do not play the role of macroscopic fluxes, as it is the case in \ref{eq:avg_dt_chi_f} and \ref{eq:avg_dt_delta_f}. Instead, they act as source terms in the first moment and higher moment equations. 
This distinction is the main structural differences between the Kinetic-like model (\ref{eq:avg_dt_dq_alpha_tot} and \ref{eq:avg_dt_dQ_alpha_tot}) and the two-phase flow model (\ref{eq:avg_dt_chi_f} and \ref{eq:avg_dt_delta_f}). 
In this study, \ref{eq:avg_dt_chi_f} and \ref{eq:avg_dt_delta_f} are referred to as the phase-averaged equations, while \ref{eq:avg_dt_dq_alpha_tot} and \ref{eq:avg_dt_dQ_alpha_tot} are called the particle-averaged equations. 

 \section{Hybrid description of the two phases}
\label{sec:averaged_eq}

In this section, we discuss the link between  the phase averaged equations for the dispersed phase presented in \ref{sec:two-fluid} and the particle or Lagrangian averaged equations presented in \ref{sec:Lagrangian}.

\subsection{Link between particle-averaged and phase-averaged equations}
\label{sec:equivalence}
To model the dispersed phase, there are two distinct approaches. 
We can either use \ref{eq:avg_dt_chi_f} with $k=d$, or we can employ the particle-averaged equations \ref{eq:avg_dt_dq_alpha_tot}, \ref{eq:avg_dt_dQ_alpha_tot} and potentially the higher moments equations found in \ref{ap:Moments_equations}.
To better understand the physical significance of this choice and determine which formalism is more appropriate, it is essential to discuss the relationship that connects these two formalisms. 

It has been demonstrated in various studies \citep{buyevich1979flow,lhuillier1992ensemble,zhang1994averaged}, that phase-averaged quantities can be expressed as a Taylor series expansion of particle-averaged quantities. 
The aforementioned studies used the single-particle conditionally averaged approach to demonstrate this equivalence.  
In this work, we follow the "distributional" approach proposed by \citet{pahtz2025general}, as it clarifies the relationship between phase quantities and moment expansions prior to the application of any averaging formalism.
As demonstrated by \citet{pahtz2025general} (see also \ref{app:expansion} for a demonstration in the sense of distribution) we may show
\begin{equation}
    \chi_\alpha f^0_d 
    =\delta_\alpha\intO{f^0_d}
    - \div\left(\delta_\alpha\intO{\textbf{r} f^0_d}\right)
    + \frac{1}{2}\grad\grad :\left(\delta_\alpha\intO{\textbf{rr} f^0_d}\right)-
    \ldots 
    \label{eq:fd_asympt0}
\end{equation}
We recall that $\chi_\alpha = 1$ within the particle domain $\Omega_\alpha(\FF, t)$ and $0$ otherwise. 
Additionally, we have extended this approach to surface quantities in \ref{app:expansion} , resulting in
\begin{equation}
    \delta_{\Gamma\alpha}f_\Gamma^0  
=\delta_\alpha\intS{f^0_\Gamma}
- \div\left(\delta_\alpha\intS{\textbf{r} f^0_\Gamma}\right)
+ \frac{1}{2}\grad\grad :\left(\delta_\alpha\intS{\textbf{rr} f^0_\Gamma}\right)-
\ldots 
\label{eq:fG_asympt0}
\end{equation}
Likewise, we recall that $\delta_{\Gamma\alpha} = 1$ on the particle surface $\Gamma_\alpha(\FF, t)$ and $0$ otherwise.
We can identify the zeroth, first and second order moments of $f_d^0$ and $f_\Gamma^0$, into \ref{eq:fd_asympt0} and \ref{eq:fG_asympt0}, respectively. 
Then summing \ref{eq:fd_asympt0} and \ref{eq:fG_asympt0} we obtain,
\begin{equation}
    \chi_\alpha f^0_d  + \delta_{\Gamma\alpha} f_\Gamma^0  = \delta_\alpha\text Q_\alpha
    - \div  
    (\delta_\alpha\textbf{Q}_\alpha^{(1)})        
    + \frac{1}{2} \grad\grad : (\delta_\alpha\textbf{Q}_{\alpha}^{(2)})
    - \ldots  
    \label{eq:f_asympt0}
\end{equation}
It is important to observe that, even before any averaging procedure is applied, \ref{eq:f_asympt0} already demonstrate the link between the dispersed phase fields, even for a single particle. 
Notably, this relationship is valid in a distributional sense at the local level.

The dispersed phase indicator function $\chi_d$ can be expressed as a sum of phase indicator function, $\chi_d(\textbf{x},t,\FF) = \sum_\alpha\chi_\alpha(\textbf{x},t,\FF)$. 
Likewise, the interface indicator function $\delta_\Gamma$ can be written as $\delta_\Gamma =  \sum_\alpha  \delta_{\Gamma\alpha}$.
Thus, summing \ref{eq:fd_asympt0} and \ref{eq:fG_asympt0} over all particles in the domain and averaging over all configurations, gives the general relations that link continuous-averaged and particle-averaged fields, namely \citep{lhuillier1992ensemble,lhuillier1998,lhuillier2000bilan}, 
\begin{align}
    \avg{\chi_df_d^0} 
    &=  \pavg{\text q_\alpha}
        - \div  
        \pavg{\textbf{q}_\alpha^{(1)}}        
        + \frac{1}{2} \grad\grad : \pavg{\textbf{q}_{\alpha}^{(2)}}
        + \ldots  \label{eq:f_exp_chi} \\
    \avg{\delta_\Gamma  f_\Gamma ^0} 
    &=  \pavg{\text q_{\Gamma \alpha}}        
        - \div \pavg{\textbf{q}_{\Gamma\alpha}^{(1)}}
        + \frac{1}{2} \grad\grad : \pavg{\textbf{q}_{\Gamma\alpha}^{(2)}}
        - \ldots  
    \label{eq:f_exp_delta}
\end{align}
Summing \ref{eq:f_exp_chi} and \ref{eq:f_exp_delta} we obtain
\begin{equation}
    \avg{\chi_df_d^0+\delta_\Gamma  f_\Gamma ^0} = \pavg{\text Q_\alpha}
    - \div  
    \pavg{\textbf{Q}_\alpha^{(1)}}        
    + \frac{1}{2} \grad\grad : \pavg{\textbf{Q}_{\alpha}^{(2)}}
    - \ldots  \label{eq:f_exp}
\end{equation}
When considering an infinite number of terms in \ref{eq:f_exp} one might eventually obtain a converged approximation of $\avg{\chi_d f_d^0+\delta_\Gamma  f_\Gamma ^0}$. 
However, it is important to note that Taylor series have what is known as a \textit{radius of convergence} beyond which adding more terms does not necessarily improve the approximation \citep[Chapter 1]{appel2007}. 
In particular, for distances beyond a certain limit \textbf{r} the series might diverges depending on the behavior of the function $\avg{\chi_d f_d^0+\delta_\Gamma  f_\Gamma ^0}$ near the point $\textbf{x}$. 
For the purposes of this article, we will assume that the Taylor series has an infinite radius of convergence, although this assumption warrants further investigation.

To demonstrate the relation between the two formalisms, we follow a strategy similar to \citep{lhuillier2000bilan,lhuillier2009rheology}. 
Let $\mathcal{C}_d$ denote the phase-averaged equation of conservation (\ref{eq:avg_dt_chi_f} with $k=d$) and $\mathcal{C}_\Gamma $ the averaged surface transport equation (\ref{eq:avg_dt_delta_f}).
Specifically, they are defined as follows
\begin{align}
    \mathcal{C}_d
    &=
    - \pddt \avg{\chi_df_d^0}
    - \div \avg{\chi_d \mathbf{\Phi}_d^0 - \chi_df_d^0 \textbf{u}_d^0}
    + \avg{\chi_d s_d^0}
    + \avg{\delta_\Gamma \left[
        \mathbf{\Phi}_d^0
        + f_d^0
        \left(
            \textbf{u}_\Gamma ^0
            - \textbf{u}_d^0
        \right)
    \right]
    \cdot \textbf{n}_d},\\
    \mathcal{C}_\Gamma 
    &= 
    -\pddt \avg{\delta_\Gamma f_\Gamma ^0}
    -\div \avg{\delta_\Gamma  f_\Gamma ^0 \textbf{u}_\Gamma ^0-\delta_\Gamma  \mathbf{\Phi}_{\Gamma||}^0 }
    + \avg{\delta_\Gamma s_\Gamma ^0} 
    - \avg{\delta_\Gamma  \Jump{
     \mathbf{\Phi}_k^0+
    f_k^0 (\textbf{u}_\Gamma ^0 - \textbf{u}_k^0)
    } }. 
\end{align}
It should be noted from \ref{eq:avg_dt_chi_f} and \ref{eq:avg_dt_delta_f} that $\mathcal{C}_d\equiv 0$ and $\mathcal{C}_\Gamma  \equiv 0$.
By applying the Taylor expansion to each term of $\mathcal{C}_d+\mathcal{C}_\Gamma $ as described in \ref{eq:f_exp} yields
\begin{equation}
    \mathcal{C}_d 
    + \mathcal{C}_\Gamma  
    = \mathcal{M}^{(0)} - \div \mathcal{M}^{(1)} + \frac{1}{2} \grad\grad : \mathcal{M}^{(2)} \ldots = 0,
    \label{eq:scheme_equivalence}
\end{equation} 
where the expressions for $\mathcal{M}^{(0)}$ and $\mathcal{M}^{(1)}$ are given by 
\begin{align}
    &\mathcal{M}^{(0)}
    = 
    - \avg{\delta_p \ddt {\text Q_\alpha}}
+ \pOavg{ s_d^0 }
    + \pSavg{ s_\Gamma ^0 }
    + \pSavg{ 
    \left[\mathbf{\Phi}_f^0 
    + f_f^0 (\textbf{u}_\Gamma ^0-\textbf{u}_f^0) \right] \cdot \textbf{n}_d },\\
    &\mathcal{M}^{(1)} =
    -  \avg{\delta_p \ddt {\textbf{Q}_\alpha^{(1)}}}
+ \pOavg{ \left(
        \textbf{r} s_d^0         
        + f_d^0  \textbf{w}_d^0 
        - \mathbf{\Phi}_d^0
    \right) }
    + \pSavg{ \left(
        \textbf{r}s_\Gamma ^0
        + f_\Gamma ^0 \textbf{w}_\Gamma ^0
        - \mathbf{\Phi}_{\Gamma||}^0
    \right) } \nonumber\\
    &+ \pSavg{ \textbf{r} \left[
        \mathbf{\Phi}_f^0
        + f_f^0 (\textbf{u}_\Gamma ^0-\textbf{u}_f^0)
    \right]\cdot \textbf{n}_d  }.
\end{align}
Using \ref{eq:scheme_equivalence}, we reach one of the main conclusion of this study. 
We observe that $\mathcal{M}^{(0)}$ and $\mathcal{M}^{(1)}$ correspond to the zeroth and first-order moment equations, respectively. 
Additionally, as demonstrated in \ref{ap:Moments_equations} the coefficient $\mathcal{M}^{(n)}$ in \ref{eq:scheme_equivalence} represents the $n^{th}$ order moment in the particle-averaged conservation equation. 
From \ref{eq:scheme_equivalence} we conclude that combining \ref{eq:avg_dt_chi_f} for $k=d$ and \ref{eq:avg_dt_delta_f} effectively captures the particle moment equations through a Taylor expansion around the particle center of mass. 
Therefore, it is clear that by considering an arbitrary order of particle moments equations, one can achieve a highly accurate description of the dispersed phase.

The particle-averaged equations ($\mathcal{M}^{(0)}$\ldots $\mathcal{M}^{(n)}$) form a system with $n$ equations, one for each moment. 
In contrast, the dispersed phase-averaged equations ($\mathcal{C}_d$ and $\mathcal{C}_\Gamma$) consist of only two equations, which aggregate all the particle-averaged equations. 
This indicates that the particle-averaged formalism provides more information since it yields a separate equation for each moment, as opposed to the phase-averaged equations, which are limited to just two. 
This enhanced level of detail is achieved by considering the topology of the dispersed phase as demonstrated in the previous sections.

\subsection{Conservation equations}

Given that the aim of this work is not only to demonstrate the relationship between particle and phase-averaged formalisms but also to establish a comprehensive framework for analyzing dispersed two-phase flows, we now present \textit{the hybrid} set of conservation equations.
The system of equations governing a macroscopic quantity $f$ consists of one equation for the fluid phase, which ensures the conservation of $f_f$ and $n$ equations for the dispersed phase, representing the conservation of the quantities $\textbf{Q}_p^{(n)}$.  
In its most general form, the hybrid description of $f$ can be expressed as
\begin{align}
    \pddt (\phi_f f_f)
    +\div (\phi_f f_f \textbf{u}_f + \mathbf{\Phi}_f^\text{eff})
    &= 
    \phi_f s_f
    - \pSavg{\left[
        \mathbf{\Phi}_f^0
        + f_f^0
        \left(
            \textbf{u}_\Gamma^0
            - \textbf{u}_f^0
        \right)
    \right]
    \cdot \textbf{n}_d} ,
    \label{eq:avg_hybrid_dt_chi_f}\\
\pddt (n_p\text Q_p)
        + \div (n_p \text Q_p \textbf{u}_p + \pavg{\textbf{u}_\alpha' \text Q_\alpha})
        &= \pOavg{ s_d^0 }
        + \pSavg{ s_\Gamma^0 }\nonumber\\
        &+ \pSavg{ \left[\mathbf{\Phi}_f^0 + f_f^0 (\textbf{u}_\Gamma^0-\textbf{u}_f^0) \right] \cdot \textbf{n}_d },
        \label{eq:avg_hybrid_q}
        \\
        \pddt (n_p\textbf{Q}_p^{(1)})
        + \div (n_p  \textbf{u}_p  \textbf{Q}_p^{(1)}+ \pavg{\textbf{u}_\alpha' \textbf{Q}_\alpha^{(1)}})
        &=\pOavg{ \left(
            \textbf{r} s_d^0         
            + f_d^0  \textbf{w}_d^0 
            - \mathbf{\Phi}_d^0
        \right) }\nonumber\\
        + \pSavg{ \left(
            \textbf{r}s_\Gamma^0
            + f_\Gamma^0 \textbf{w}_\Gamma^0
            - \mathbf{\Phi}_{\Gamma||}^0
        \right) }
        &+ \pSavg{ \textbf{r} \left[
            \mathbf{\Phi}_f^0
            + f_f^0 (\textbf{u}_\Gamma^0-\textbf{u}_f^0)
        \right]\cdot \textbf{n}_d  },
        \label{eq:avg_hybrid_q_1}
        \\\nonumber
        \vdots
\end{align}
where the effective continuous phase non-convective flux term reads 
\begin{align}
    \mathbf{\Phi}_f^\text{eff}
    = \avg{\chi_f f_f' \textbf{u}_f'}
    - \avg{\chi_f \bm\Phi_f^0}
    - \pSavg{\textbf{r}\left[
        \mathbf{\Phi}_f^0
        + f_f^0
        \left(
            \textbf{u}_\Gamma^0
            - \textbf{u}_f^0
        \right)
    \right]
    \cdot \textbf{n}_d}
    + \div[\ldots].
    \label{eq:effective_stress}
\end{align}
The only difference in the conservation equation for the continuous phase (between \ref{eq:avg_dt_chi_f} and \ref{eq:avg_hybrid_dt_chi_f}) is the expansion of the exchange term $\avg{\delta_\Gamma \left[
    \mathbf{\Phi}_f^0
    + f_f^0
    \left(
        \textbf{u}_\Gamma ^0
        - \textbf{u}_f^0
    \right)
\right]
\cdot \textbf{n}_d}$
 into a Taylor series in a similar way to \ref{eq:f_exp_delta}. 
The presence of the terms $\div[\ldots]$ in the expression for  $\mathbf{\Phi}_f^\text{eff}$ suggests that higher-order moments of the interphase exchange term are involved.
Similarly, the ellipsis below \ref{eq:avg_hybrid_q_1} implies that an arbitrary number of dispersed phase moment equations can be introduced. In this format, it becomes evident that the exchange term on the right-hand side of \ref{eq:avg_hybrid_dt_chi_f} is identical to that on the right-hand side of \ref{eq:avg_hybrid_q}. 
Additionally, in the effective flux $\mathbf{\Phi}_f^\text{eff}$ the exchange term from  \ref{eq:avg_hybrid_q_1} appears, and this property continues for higher-order moments.  
Consequently, the zeroth order exchange term in the equation for $\text Q_\alpha^{(0)}$ plays the role of a source term for $f_f$, while the first and higher order exchange terms act as a source into the higher moment equations, and contribute to the effective non-convective fluxes for $f_f$.

The system of equations presented here offers a clear understanding of the roles of the dispersed phase non-convective flux terms, $\bm\Phi_d$ and $\bm\Phi_\Gamma$, in the particle phase conservation equation. 
As evidenced in \ref{eq:avg_hybrid_q}, $\bm{\Phi}_d$ and $\bm{\Phi}_\Gamma$  do not influence the lowest order particle-phase averaged conservation equation, i.e. the equation of $n_p \text Q_p$. 
However, \ref{eq:avg_dt_chi_f} (for $k = d$) and \ref{eq:avg_dt_delta_f}, show that the phase-averaged quantities $f_d$ and $f_\Gamma$, are affected by the non-convective fluxes at the particle surface and internally, since $\bm{\Phi}_d$ and $\bm{\Phi}_\Gamma$ appear in these equations.
This may seem contradictory at first, but it is important to note that $\bm{\Phi}_d^0$ and $\bm{\Phi}_\Gamma^0$ serve as source terms in the conservation equations for higher moments such as $\textbf{Q}^{(1)}_p$ \ldots $\textbf{Q}^{(n)}_p$, which are related to $f_d$ and $f_\Gamma$ through \ref{eq:f_exp}.
In summary, the non-convective fluxes $\bm{\Phi}_d$ and $\bm{\Phi}_\Gamma$  are not explicitly related to $\text Q_p$, regardless of particle nature or volume fraction. 
Instead, their influence on $\text Q_p$ is mediated through the closure terms in \ref{eq:avg_hybrid_q}, which may depend on the higher moments $\textbf{Q}^{(1)}_p$ \ldots $\textbf{Q}^{(n)}_p$ or other higher-order particle-related moments. These moments, in turn, explicitly depend on $\bm{\Phi}_d$ and $\bm{\Phi}_\Gamma$ as indicated by \ref{eq:avg_hybrid_q_1}.

The hybrid model consists of $n+1$ equations for $n+1$ unknowns, specifically $f_f$, $\textbf{Q}_p^{(1)}$\ldots$\textbf{Q}_p^{(n)}$ \footnote{$\phi_f$ and $n_p$ can also be treated as unknowns by adding corresponding equations, achieved by setting $f_f=1$ and $Q_\alpha^{(n)} =1$ in \ref{eq:avg_hybrid_dt_chi_f} and \ref{eq:avg_hybrid_q}.}.
The closure problem then involves deriving explicit expressions for all terms of \ref{eq:avg_hybrid_dt_chi_f} to \ref{eq:avg_hybrid_q_1}, of the form $\avg{\ldots}$ in terms of the problem's unknowns, i.e. 
$f_f$, $\text Q_\alpha$, $\text Q_\alpha^{(1)}$ \ldots $ \text Q_\alpha^{(n)}$.  
Once these expressions are determined, the equations can be solved.

\subsection{Discussion}
The last question we would to address is: in what circumstances is the inclusion of higher moments, such as $\textbf{Q}_p^{(n)}$, necessary within this system of equations?
There are three primary reasons why the moment $\textbf{Q}_p^{(n)}$ may be required:
\begin{enumerate}
\item First when $\textbf{Q}_p^{(n)}$ provides critical information sought in the analysis. 
For example, consider the case of fiber orientation in a flow, which corresponds to the second moment of the particle mass distribution. 
This information might be the central objective of the study, especially when the goal is to understand how fiber orientation evolves in industrial applications like composite materials \citep{advani1987use}.

\item Second $\textbf{Q}_p^{(n)}$is essential for accurately modeling closure terms. 
Again, fiber orientation serves as a relevant example. 
Indeed in dilute flows of axisymmetric fibers within the Stokes regime the force acting on the fiber (which represents the closure term in the momentum equation) is dependent on the fiber orientation \citep{kim2013microhydrodynamics}. \item Finally, in the continuous phase conservation equation \eqref{eq:avg_hybrid_dt_chi_f}, an infinite number of moments are involved. 
    Hence, the last reason why one may need to obtain such higher moments is because they may be required directly by \ref{eq:avg_hybrid_dt_chi_f}. 
\end{enumerate}
In the following, we focus on dispersed fluid-fluid suspensions, emphasizing the role of mass and momentum moments in their dynamics. \section{Averaged equations for dispersed fluid-fluid flows with surface tension}\label{sec:averaged_surface}

We consider a monodisperse fluid-fluid suspension of droplets (or bubbles) which are not necessarily spherical with volume $ v_p $, without mass transfer. 
Both the dispersed and continuous phases are treated as incompressible Newtonian fluids, characterized by constant viscosities $ \mu_k $ and densities $ \rho_k $. 
The surface tension at the interface between the two fluids is denoted by $ \gamma $ which is not necessarily constant. 
We only consider in the next section averaged mass and momentum conservation equations to describe the motions of the phases. 
\begin{table}
    \centering
    \begin{tabular}{|c|ccl|}\hline
    & Conservation law & mass & momentum \\ \hline
    Conserved quantity & $f_k^0$  & $\rho_k$ & $\rho_k \textbf{u}_k^0$ \\
    Source term & $s_k^0$  & $0$ & $\rho_k \textbf{g}$ \\
    Diffusive flux & $\Phi_k^0$ & 0 & $\bm\sigma_k^0 = -p_k^0 + \mu_k (\grad \textbf{u}_k^0 + ^\dagger\grad \textbf{u}_k^0)$ \\
    Surface diffusive flux & $\Phi_\Gamma^0$ & 0 & $\bm\sigma_\Gamma^0 = \gamma (\bm\delta - \textbf{nn})$ \\\hline
    \end{tabular}

    \caption{Definition of the physical quantities and local constitutive laws for phases $k$.}
    \label{tab:qte_Newtonian}
\end{table}
The conserved physical quantities  relevant to the problem are summarized in Table \ref{tab:qte_Newtonian}.

In the following, we consider the influence of Marangoni effects on the suspension dynamics. Surfactants and/or non-uniform temperature gradients typically cause variations in the surface tension coefficient at the droplet interfaces. 
One must solve a transport equation for either the temperature field or the surfactant concentration in order to determine the proper surface tension distribution on the droplet surface \citep{Subramanian_1985,leal2007advanced}. 
This can be done using the surface transport equations derived in \ref{sec:local_eq}. 
However, as these complexities fall outside the scope of this work, we assume that the surface tension coefficient is known in advance, having been determined from a separate problem solved a priori.

After presenting the higher-order mass and momentum moments required to describe the particle rotation and deformation we derive the averaged equations for mass, momentum and first moment of momentum for a fluid-fluid suspension.
The presentation is inspired by \citep{lhuillier2009rheology}, while extending it beyond the scope of solid particles.
Then we turn our attention to the decomposition of stress tensor in the momentum exchange terms. 
The section concludes with a discussion of the symmetry properties of the effective stress tensor, employing arguments analogous to those presented by \citet{lhuillier1996contribution}.

\subsection{Higher-order mass and momentum moments}

To describe the dispersed phase in addition to the quantities defined in \ref{sec:Lagrangian} we define the second-order moment of mass and the first-order moment of momentum as 
\begin{equation}
    \textbf{M}_\alpha 
    = \intO{ \rho_d \textbf{r} \textbf{r} }
    \;\;\;\text{and}\;\;\;
    \textbf{P}_\alpha 
    = \intO{ \rho_d \textbf{r} \textbf{u}_d^0 },
    \label{eq:first_moment_of_momentum_def}
\end{equation}
respectively. 
Note that the tensor $\textbf{M}_\alpha$ plays a role analogous to the inertia tensor $\textbf{I}_\alpha$ in solid mechanics. 
The two are related by the expression  $\textbf{I}_\alpha = (\bm\delta : \textbf{M}_\alpha)\bm\delta - \textbf{M}_\alpha$.
The second order tensor $\textbf{M}_\alpha$ represents the second moment of the mass distribution relative to the particle center of mass.
Similarly, the tensor $\textbf{P}_\alpha$ is the first moment of the momentum distribution within the particle volume. 
To offer a clearer physical interpretation of the moment of momentum tensor, we decompose $\textbf{P}_\alpha$ into three distinct components  
\begin{equation}
\textbf{P}_\alpha = \textbf{S}_\alpha + \textbf{T}_\alpha + P_\alpha \bm\delta,
\end{equation}
where $\textbf{S}_\alpha = \frac{1}{2}\left(\textbf{P}_\alpha + \textbf{P}_\alpha^\dagger - \frac{2}{3}(\bm\delta:\textbf{P}_\alpha)\bm\delta\right)$ is the symmetric traceless part, representing the  deformation or stretching of momentum,
$\textbf{T}_\alpha = \frac{1}{2}(\textbf{P}_\alpha - \textbf{P}_\alpha^\dagger)$ is the antisymmetric part, associated with the angular momentum,
and $P_\alpha = \frac{1}{3} \bm\delta : \textbf{P}_\alpha$ is a scalar that quantifies the momentum of expansion or compression of the droplet. 
The angular momentum of the particle can also be expressed as the pseudo-vector  
$\bm\mu_\alpha = \int_\Omega \rho_d\, \textbf{r} \times \textbf{u}_d^0 \, dV$,
and is related to $\textbf{T}_\alpha$ via $(\bm\mu_\alpha)_i = \epsilon_{ijk} (\textbf{T}_\alpha)_{jk}$ where $\epsilon_{ijk}$ is the Levi-Civita symbol.
To illustrate the physical significance of these tensors, we present in \ref{eq:scheme} three representative inner velocity fields, each with its corresponding moment of momentum tensor.
\begin{figure}[h!]
    \centering
    \hfill
    \begin{tikzpicture}[ultra thick,scale=0.6]
        \def\nRows{6}
        \def\nCols{6}
        \draw[dashed,thin] (0,0)circle(2.5);
        \draw[fill=gray!30] (0,0)ellipse(2.8 and 2.2);
        \foreach \x in {-\nRows,...,\nRows} {
            \foreach \y in {-\nCols,...,\nCols} {
                \pgfmathsetmacro\distance{veclen(\x*0.4, \y*0.4)};
                \pgfmathparse{\distance < 2.45 ? "blue" : "white"}
                \edef\colour{\pgfmathresult};
                \ifthenelse{\equal{\colour}{blue}}{                    
                    \draw[thin,->](\x*0.4,\y*0.4)--++(0.08*\x,-0.08*\y);
                }
            }
        }
        \node (txt) at (0,3){(\textit{Case 1})};
        \node (txt) at (0,-3){($|\textbf{S}_\alpha| > 0$)};
    \end{tikzpicture}
     \hfill
    \begin{tikzpicture}[ultra thick,scale=0.6]
        \def\nRows{6}
        \def\nCols{6}
        \draw[fill=gray!30] (0,0)circle(2.5);
        \foreach \x in {-\nRows,...,\nRows} {
            \foreach \y in {-\nCols,...,\nCols} {
                \pgfmathsetmacro\distance{veclen(\x*0.4, \y*0.4)};
                \pgfmathparse{\distance < 2.5 ? "blue" : "white"}
                \edef\colour{\pgfmathresult};
                \ifthenelse{\equal{\colour}{blue}}{                    
                    \draw[thin,->](\x*0.4,\y*0.4)--++(0.08*\y,-0.08*\x);
                }
            }
        }
        \node (txt) at (0,3){(\textit{Case 2})};
        \node (txt) at (0,-3){($|\textbf{T}_\alpha| > 0$)};
    \end{tikzpicture}
    \hfill
    \begin{tikzpicture}[ultra thick,scale=0.6]
        \def\nRows{6}
        \def\nCols{6}
        \draw[dashed,thin] (0,0)circle(2.5);
        \draw[fill=gray!30] (0,0)circle(2.2);
        \foreach \x in {-\nRows,...,\nRows} {
            \foreach \y in {-\nCols,...,\nCols} {
                \pgfmathsetmacro\distance{veclen(\x*0.4, \y*0.4)};
                \pgfmathparse{\distance < 2.3 ? "blue" : "white"}
                \edef\colour{\pgfmathresult};
                \ifthenelse{\equal{\colour}{blue}}{                    
                    \draw[thin,->](\x*0.4,\y*0.4)--++(-0.08*\x,-0.08*\y);
                }
            }
        }
        \node (txt) at (0,3){(\textit{Case 3})};
        \node (txt) at (0,-3){($P_\alpha < 0$)};
    \end{tikzpicture}
    \hfill
    \caption{Graphical representation of the inner kinematics   of an arbitrary particle under three scenarios. 
        The arrows represent the velocity field inside the particle, $\textbf{w}_d^0$, with the corresponding value of the moment of momentum tensor indicated below. 
        The operator $|\ldots|$ refers to the norm of the tensors. 
        According to the inner velocity field:
        (\textit{Case 1}) The particle experiences a mean deformation, resulting in non-zero stretching of momentum along the principal axis of deformation;
        (\textit{Case 2}) The particle is rotating, leading to a non-zero angular momentum vector in the direction of rotation;
        (\textit{Case 3}) The particle undergoes compression, resulting in a negative trace of the moment of momentum.
    }
    \label{eq:scheme}
\end{figure}
Injecting, $f_d^0 = \rho_d$ in the second-order moment equation (derived in \ref{ap:Moments_equations}) we obtain,
\begin{equation}
    \ddt {\textbf{M}_\alpha}=2(\textbf{S}_\alpha+P_\alpha\bm\delta),
    \label{eq:dt_M_alpha}
\end{equation}
which is the second-order moment of mass conservation equation assuming that the fluid within the drop is divergence free. 
From \ref{eq:dt_M_alpha} we deduce that the evolution of the distribution of mass of a particle is solely determined by the  of momentum $\textbf{S}_\alpha$ and $P_\alpha$. 
This indicates that angular momentum does not influence the evolution of the second moment of mass, a consequence of the symmetry of the tensor $\textbf{M}_\alpha$, which must be preserved after differentiation with respect to time.
However, this does not imply that the particle angular velocity ($\bm\omega_\alpha$) does not appear in this equation. 
For instance, in the case of rigid body motion where $\textbf{w}_d^0 = \bm\omega_\alpha \times \textbf{r}$, we obtain the relation  $2(\textbf{S}_\alpha+P_\alpha\bm\delta)_{ij} = \epsilon_{iab} (\bm\omega_\alpha)_a (\textbf{M}_\alpha)_{bj}+ 
\epsilon_{jab} (\bm\omega_\alpha)_a (\textbf{M}_\alpha)_{bi}  $. 

Now that we have described the kinematics   of the particle shape, let us proceed to derive an equation for the moment of momentum.
This equation is derived by injecting $\textbf{Q}_\alpha^{(1)} = \textbf{P}_\alpha$ in \ref{eq:dt_Q_alpha_tot}, it reads, 
\begin{equation}
    \ddt {\textbf{P}_\alpha}
    - \intO{ \rho_d  \textbf{w}_d^0 \textbf{w}_d^0 }
    = 
    - \intO{\bm{\sigma}_d^0}
    - \intS{ 
        \gamma (\bm\delta - \textbf{nn})
    }
    + \intS{ \textbf{r}\bm{\sigma}_f^0\cdot \textbf{n}}.
    \label{eq:dt_P_alpha}
\end{equation}
In the following sections the normal vector noted \textbf{n} always refer to $\textbf{n}_d$. 
The conservation equation of the angular momentum $\bm{\mu}_\alpha$ is obtained by taking the double contracted product of \ref{eq:dt_P_alpha} with $\bm\epsilon$, which directly gives
\begin{equation}
    \ddt\bm{\mu}_\alpha
    =  
\intS{ \textbf{r} \times \bm{\sigma}_f^0\cdot \textbf{n} }
    \label{eq:dt_mu_alpha}
\end{equation}
Note that every term on the right-hand side of \ref{eq:dt_P_alpha} vanished due to their symmetric nature apart from the skew-symmetric part of the hydrodynamic stress, which is the hydrodynamic torque applied on the particle $\alpha$.
In particular, the surface tension terms do not appear in the angular momentum balance since the tensor $\bm\delta-\textbf{nn}$ is symmetric, which is consistent with the findings of \citet{hesla1993note}. 
As a consequence, the surface tension does not affect the angular momentum regardless of the particle shape.

Taking the symmetric part of \ref{eq:dt_P_alpha}, and substrating the trace yields, 
\begin{align}
    &\ddt {\textbf{S}_\alpha}
    - \rho_d\intO{\left(\textbf{w}_d^0 \textbf{w}_d^0 -\frac{1}{3} (\textbf{w}_d^0 \cdot  \textbf{w}_d^0)\bm\delta\right)}
    = \nonumber \\
    &- 2\mu_d\intO{\textbf{e}_d^0}
    -  \intS{ \gamma
        \left( \frac{1}{3}\bm\delta - \textbf{nn} \right)
    }
    + \frac{1}{2}\intS{\left(\textbf{r}\bm\sigma_f^0+\bm\sigma_f^0\textbf{r}-\frac{2}{3}(\bm\sigma_f^0 \cdot \textbf{r})\bm\delta \right)\cdot \textbf{n}},
    \label{eq:dt_S_alpha}
\end{align}
where we have introduced the rate of strain tensor for phase \( k \), defined as \( \mathbf{e}_k^0 = \frac{1}{2} (\grad \mathbf{u}_k^0 + ^\dagger \grad \mathbf{u}_k^0) \). 
Here $^\dagger$ represents the transpose operator. 
On the left-hand side of \ref{eq:dt_S_alpha}, two inertial contributions can be identified: the time derivative of $\mathbf{S}_\alpha$, and the stress arising from the product of internal velocity field within the droplet. 
These inertial effects are counterbalanced by the terms appearing on the right-hand side of the equation, which include: the volume integral of the particle viscous stress; the surface tension moment; and the first moment of the hydrodynamic force.
While surface tension effects do not influence the linear or angular momentum equations directly, they do impact the moment of momentum $\textbf{P}_\alpha$, specifically its symmetric part $\textbf{S}_\alpha$.
Consequently, surface tension influences the hydrodynamic behavior of a particle exclusively through its effect on $\textbf{S}_\alpha$, which is related to the shape of a particle represented by $\textbf{M}_\alpha$, via \ref{eq:dt_M_alpha}.
Inserting \ref{eq:dt_S_alpha} in  \ref{eq:dt_M_alpha} yields,
\begin{align}    
    &\frac{1}{2}\frac{d^2 (\textbf{M}_\alpha-\frac{1}{3}(\textbf{M}:\bm\delta)\bm\delta)}{dt^2}
    =  \rho_d\intO{\left(\textbf{w}_d^0 \textbf{w}_d^0 -\frac{1}{3} (\textbf{w}_d^0 \cdot  \textbf{w}_d^0)\right)}
    - 2\mu_d\intO{\textbf{e}_d^0} \nonumber\\
    &- \intS{\gamma  
        \left( \frac{1}{3}\bm\delta - \textbf{nn} \right)
    }
    + \frac{1}{2}\intS{\left(\textbf{r}\bm\sigma_f^0+\bm\sigma_f^0\textbf{r}-\frac{2}{3}(\bm\sigma_f^0 \cdot \textbf{r})\bm\delta \right)\cdot \textbf{n}}.
    \label{eq:dt2_M_alpha}
\end{align}
\ref{eq:dt2_M_alpha} is a second-order differential equation governing the second-order mass moment that characterizes the droplet shape. 
In other words, it should be interpreted as an equation describing the droplet shape, represented by the traceless part of the tensor $\textbf{M}_\alpha$.
Accordingly, the right-hand side of \ref{eq:dt2_M_alpha} contains terms that promote deformation-such as the product of the internal velocity field and the traceless component of the first moment as well as terms that oppose deformation, including droplet internal viscous resistance and surface tension.
\ref{eq:dt2_M_alpha} is particularly useful to compute the unknown internal stress within solid particles, in terms of surface integral, i.e. the first moment of the hydrodynamic force.
This relation plays a key role in expressing the bulk stress of a suspension and ultimately leads to the effective viscosity, once a closed-form expression for the average first moment is obtained \citep{batchelor1970stress}. 
In the inertial regime, and for solid particles, the tensors $\textbf{M}_\alpha$ and the internal velocity field $\textbf{w}_d^0$ are fully prescribed by the particle kinematics. 
As a result, in \ref{eq:dt2_M_alpha}, $\textbf{M}_\alpha$ and $\textbf{w}_d^0$ appear not as unknowns but rather as known source terms.
For spherical particles specifically, the inertial correction to the first force moment has been derived by \citet{hwang1989modeling} and  \citet{lhuillier1996contribution}.

\subsection{Averaged momentum and mass conservation equations}

Let us now focus on the averaged momentum equation of the continuous phase. 
By applying \ref{eq:avg_dt_chi_f} with $f_k = \rho_f$ and $\rho_f\textbf{u}_f$ we obtain the mass and momentum equation for the continuous phase under the two fluid formulation, 
\begin{align}
    (\pddt + \textbf{u}_f\cdot \grad)\phi_f&=-\phi_f \div \textbf{u}_f
    \label{eq:mass_init}\\
    \phi_f \rho_f(\pddt + \textbf{u}_f  \cdot \grad) \textbf{u}_f
&= 
    \phi_f \rho_f \textbf{g}
    + \div [\phi_f \bm\sigma_f-  \avg{\chi_f\rho_f \textbf{u}_f'\textbf{u}_f'}]
    - \avg{\delta_\Gamma \bm\sigma_f^0\cdot \textbf{n}}.
    \label{eq:two_fluid_momentum_init}
\end{align} 
To obtain the hybrid formulation one can expand the final term on the right-hand side of \ref{eq:two_fluid_momentum_init} using the Taylor expansion provided in \ref{eq:f_exp}.  
However, before doing so, it is important to examine the term \( \phi_f \bm\sigma_f \), as it contains a non-closed contribution originating from the averaging of the strain rate tensor. Properly identifying and isolating this contribution is a necessary step prior to performing the Taylor expansion.

We begin by noting that $\phi_f\bm\sigma_f$ can be expressed in terms of the averaged fluid pressure $p_f$, and continuous phase averaged velocity ($\textbf{u}_f$), or bulk-averaged velocity ($\textbf{u} = \phi_f \textbf{u}_f + \phi_d \textbf{u}_d$).
Expressing $\bm\sigma_f$ as a function of $\textbf{u}_f$ instead of $\textbf{u}$ (or vice versa) leads to two distinct forms of the momentum equation, each associated with different formulations of the closure terms. 
We also review several alternative choices found in the literature (see also \citet{jackson2000, pahtz2025general} for an overview focused on solid particles), emphasizing that while all these formulations are mathematically equivalent, they give rise to different closure problems. 
We argue that one particular formulation is probably the most suitable, especially in light of the closure terms available in the literature for non-dilute flows.

First we introduce what we refer to as the \textit{mean Newtonian stress}, based either on the continuous phase averaged velocity $\textbf{u}_f$ or on the bulk velocity \textbf{u}, namely,
\begin{align}
    \bm\Sigma_f 
    &
    = -p_f \bm\delta + 2\mu_f \textbf{E}_f    
\label{eq:Sigma_f_average}
    \\
    \bm\Sigma &
    = -p_f\bm\delta + 2 \mu_f \textbf{E}
\label{eq:Sigma_average}
\end{align}
where $\textbf{E}_f = [\grad \textbf{u}_f + (\grad \textbf{u}_f)^{\dagger}]/2$ and $\textbf{E} =[\grad \textbf{u} + (\grad \textbf{u})^{\dagger}]/2$ represent the \textit{mean strain rate} tensor based either on $\textbf{u}_f$ or on \textbf{u}. 
Additionally the ensemble averaged stress $\phi_f \bm\sigma_f$ may be written as 
\begin{equation}
    \phi_f \bm\sigma_f = - \phi _f p_f \bm\delta + 2 \mu_ f \phi_f \textbf{e}_f
    \label{eq:sigma_average}
\end{equation}
where $\phi _f \textbf{e}_f = \avg{\chi_f (\grad \textbf{u}_f^0 + ^\dagger \grad \textbf{u}_f^0)}$. 
Inserting \ref{eq:Sigma_f_average} and \ref{eq:Sigma_average} in  \ref{eq:sigma_average} yields
\begin{align}
    \phi_f \bm\sigma_f 
    &=
\phi_f \bm\Sigma_f
    - \mu_f \avg{\delta_\Gamma( \textbf{u}_f'  \textbf{n} +  \textbf{n} \textbf{u}_f' )}
    \label{eq:stress_closure}\\
    \phi_f \bm\sigma_f 
    &=
\phi_f \bm\Sigma
    - \avg{2\mu_f \chi_d \textbf{e}_d^*}
    \label{eq:stress_closure1}
\end{align}
where we have used the relations 
\begin{equation}
    \avg{\chi_f \grad \textbf{u}_f^0}
    = 
    \phi_f \grad  \textbf{u}_f
    + \avg{\delta_\Gamma \textbf{n}_f \textbf{u}_f'}
    \label{eq:first_rel}
\end{equation}
and
\begin{equation}
\phi _f\textbf{e}_f
= 
    \phi_f \textbf{E}
    - \avg{\chi_d \textbf{e}_d^*},
    \label{eq:sec_rel}
\end{equation}
to derive \ref{eq:stress_closure,eq:stress_closure1}, respectively. 
In \ref{eq:stress_closure1} we have introduced $\textbf{e}_d^* = \textbf{e}_d^0 - \textbf{E} = \grad (\textbf{u}_d^0-\textbf{u})+^\dagger\grad (\textbf{u}_d^0-\textbf{u})$ which is the droplet internal shear rate relative to the `bulk' shear rate \textbf{E}.
It is worth noting that \ref{eq:first_rel} remains valid even when the interfacial rate of strain is nonzero. 
In contrast, \ref{eq:sec_rel} holds under the assumption that the bulk rate of strain satisfies $\textbf{E} = \phi_f \textbf{e}_f + \phi_d \textbf{e}_d$, \textit{i.e.} that $\avg{\delta_\Gamma \textbf{e}_\Gamma^0} = 0$. 
This condition is supposed to be met here because we did not consider interfacial viscosity \citep{nadim1996concise}.
The averaged momentum equation  employing the stress formulation \ref{eq:stress_closure} read, 
\begin{align}
\phi_f \rho_f(\pddt + \textbf{u}_f  \cdot \grad) \textbf{u}_f
&= \phi_f 
    \left(\div \bm{\Sigma}_f
    + \rho_f \textbf{g}\right)
    - \div 
    [\avg{\chi_f\rho_f \textbf{u}_f'\textbf{u}_f'}
    +\avg{\delta_\Gamma \mu_f( \textbf{u}_f'  \textbf{n} +  \textbf{n} \textbf{u}_f')}]
    - \avg{\delta_\Gamma \bm\sigma_f^{(1)}\cdot \textbf{n}},
    \label{eq:two_fluid_momentum}
\end{align}
where, $\bm\sigma_f^{(1)}=\bm\sigma_f^0 - \bm\Sigma_f$ which also reads \begin{equation}
\bm\sigma_f^{(1)} = -p_f'\bm\delta
+ \mu_f [
    \grad \textbf{u}_f'
    + ^\dagger \grad \textbf{u}_f']
    \label{eq:disturbance_stress1}
\end{equation} 
Likewise, using \ref{eq:stress_closure1} one may also derive another form of the momentum equation, namely,
\begin{equation}
    \phi_f \rho_f(\pddt + \textbf{u}_f  \cdot \grad) \textbf{u}_f
= \phi_f 
    \left(\div \bm{\Sigma}
    + \rho_f \textbf{g}\right)
    - \div 
    [\avg{\chi_f\rho_f \textbf{u}_f'\textbf{u}_f'} + \avg{2\mu_f \chi_d \textbf{e}_d^*}]
    - \avg{\delta_\Gamma \bm\sigma_f^{(2)}\cdot \textbf{n}},
    \label{eq:two_fluid_momentum2}
\end{equation} 
where $\bm\sigma_f^{(2)}=\bm\sigma_f^0 - \bm\Sigma$. 
It can be also expressed as, 
\begin{equation}
    \bm\sigma_f^{(2)} 
    =
    -p_f'\bm\delta
    + \mu_f [
        \grad \textbf{u}_f^*
        + ^\dagger \grad \textbf{u}_f^*
    ]
    \label{eq:disturbance_stress2}
\end{equation}
where $\textbf{u}_f^*= \textbf{u}_f^0 - \textbf{u}$.
In both cases (\ref{eq:two_fluid_momentum} and \ref{eq:two_fluid_momentum2}) one has to compute the local stress relative to the averaged continuous phase motion or bulk phase motion. 
Note the difference between the last two terms in formulations \ref{eq:two_fluid_momentum} and \ref{eq:two_fluid_momentum2}, which is given by $\phi_f \div (\bm\Sigma_f -\bm\Sigma)$.
Hence, the transition from one formulation to the other is straightforward and only requires the addition or subtraction of this term. 
However, it is important to recognize its physical significance.
This difference can be expressed explicitly as,
\begin{equation}
    \phi_f \div(\bm\Sigma_f - \bm\Sigma) = \mu_f \phi_f \div \left[\grad (\phi_d (\textbf{u}_f - \textbf{u}_d)) + ^\dagger\grad (\phi_d (\textbf{u}_f - \textbf{u}_d)) \right]
    \label{eq:diff_sigma1}
\end{equation}
where the decomposition 
\begin{equation}
\textbf{u} = \phi_f \textbf{u}_f + \phi_d \textbf{u}_d,
\label{eq:u_mean} 
\end{equation}
has been used in the derivation.
\ref{eq:diff_sigma1} reveals that the difference between the two formulations introduces a non-Newtonian stress term.
This non-Newtonian stress shares similarities with the second-order force moment closure \citep{jackson1997locally,zhang1997momentum}, and it is typically related to intrinsic convection in sedimentation processes \citep{lhuillier2022}.
Therefore, when addressing the closure problem, it is essential to treat each stress formulation carefully, as the inclusion or exclusion of this additional term can significantly impact the resulting model.

Although the stress decomposition and momentum formulations proposed herein are mathematically equivalent, each presents distinct advantages and limitations, which we examine in detail below.
First, we may observe that a simplification arises for \ref{eq:stress_closure1} in the case of solid particles, for which $\textbf{e}_d^0 = 0$.
As a result $\phi _f \bm \sigma _f = - \phi _f p_f \bm\delta + 2 \mu_ f \textbf{E}$ \citep{joseph1990ensemble,jackson2000}. Owing to this simplification, the formulation based on $\bm\Sigma$ and the associated closure relation \eqref{eq:stress_closure1} makes it a good candidate to be used in practice for solid particles. Second, employing \ref{eq:two_fluid_momentum} with \ref{eq:disturbance_stress1} may appear more convenient as the governing equations are directly formulated in terms of the fluid velocity field $\textbf{u}_f$, thereby circumventing the need to explicitly include an expression for the bulk velocity ($\textbf{u}$). Third, there exists a conceptual reason for preferring the bulk stress $\bm\Sigma$ when deriving closure relations for non-dilute suspensions.  Specifically, because $\bm\sigma_f^{(2)}$ depends on the disturbance velocity $\textbf{u}_f^* = \textbf{u}_f^0 - \textbf{u}$, the averaged velocity $\textbf{u}$ naturally serves as the "far-field" or "undisturbed" velocity boundary condition in the conditionally averaged Navier-Stokes equations \citep{hinch1977averaged,fintzi2025}. Notably, most (if not all) existing theoretical solutions addressing the disturbance field generated by a particle embedded in a Newtonian solvent or in an effective medium - representing other particles contribution - use a divergence-free background velocity field as limiting boundary condition \citep{hinch1977averaged, kim1985modelling}. 
Since $\div \textbf{u} = 0$, it follows that the velocity field used in these theoretical study corresponds to $\textbf{u}$. 
Consequently, the disturbance stress derived in those studies is $\bm\sigma_f^{(2)} = \bm\sigma_f^0 - \bm\Sigma$.
In opposition, the stress decomposition based on $\bm\Sigma_f$, involves the fields $\textbf{u}_f$ which is not divergence free. 
Hence, this `far field' or `undisturbed' velocity boundary condition far from the test particle does not correspond to the usual boundary condition assumed in most of the theoretical problems. 
In conclusion, it is important to recognize that the vector field \(\textbf{u}\) corresponds precisely to the 'background velocity' typically employed in theoretical derivations. 
Consequently, the closure terms obtained in such analyses -such as the hydrodynamic forces, the stresslet, and higher-order moments- are formulated in terms of the difference \(\bm\sigma_f^0 - \bm\Sigma\).
Of course, it is always possible to express \(\bm\Sigma_f\) in terms of \(\bm\Sigma\) (see \ref{eq:diff_sigma1}), meaning that the choice of formulation remains free. 
Nevertheless, we must be cautious when claiming that a specific expression of the momentum exchange term is exactly equivalent to a closure term reported in the literature especially for non-dilute suspension.

Furthermore, given that various stress decomposition have already been introduced in the literature, we propose to examine their respective advantages and limitations. 
In the seminal work by \citet{zhang1997momentum}, as well as in numerous subsequent studies, the disturbance stress is defined as $\bm\sigma_f' = \bm\sigma_f^0 - \bm\sigma_f$, a formulation that may be regarded as the "intuitive" definition. 
However, it is important to note that, by employing \ref{eq:stress_closure}, one obtains
\begin{equation}
    \bm\sigma_f'
    = \bm \sigma _f ^{(1)}
+ \frac{\mu_f}{\phi_f} \avg{\delta_\Gamma( \textbf{u}_f'  \textbf{n} +  \textbf{n} \textbf{u}_f' )}.
    \label{eq:stress_closure_zhang}
\end{equation}
The last term represents a contribution from the stress induced by the dispersed phase, as illustrated below. 
To better understand this term, we consider the case of a suspension composed of rigid solid particles, for which the strain rate tensor of the dispersed phase vanishes, i.e., $\textbf{e}_d = 0$.
By subtracting \ref{eq:stress_closure} from \ref{eq:stress_closure1}, we directly obtain
\begin{equation}
    \avg{\delta_\Gamma (\textbf{n} \textbf{u}_f'+  \textbf{u}_f' \textbf{n})}
    = \textbf{E}_f - \textbf{E}
    - \phi_d \textbf{E}_f 
    =
    (\textbf{u}_f - \textbf{u}_d)\grad \phi_d + \grad \phi_d (\textbf{u}_f - \textbf{u}_d)   
    -  \phi [\grad \textbf{u}_d+ (\grad \textbf{u}_d)^\dagger ]. 
\end{equation} 
where we have used \ref{eq:u_mean}.
Therefore, at least in the case of solid particles, this term becomes non-zero whenever there are significant gradients in volume fraction and mean particle velocity. The same observations hold if one considers expression \ref{eq:stress_closure1} in the analysis above. 
Since $\bm\sigma_f$ already includes closure contributions associated with the dispersed phase, directly subtracting the full expression of $\bm\sigma_f$ from $\bm\sigma_f^0$ can lead to inconsistencies unless the closure problem is handled with sufficient care. 
Note that the last term of \ref{eq:stress_closure_zhang} is proportional to the interface area concentration, hence it is of $O(\phi_d)$.
Thus, when integrating $\bm\sigma_f'\cdot \textbf{n}$ on the interface of the droplets, the contribution from that term to the mean momentum exchange term is of $O(\phi_d^2)$. 

Another commonly adopted approach in the literature is $\bm \sigma ^{(4)} = \bm \sigma _f ^0 - p_f\bm\delta$ \citep{simonin1996,lhuillier2009rheology,morel2015mathematical,guazzelli2018rheology}.
This definition leads to the expression
\begin{equation}
    \bm\sigma_f^{(4)}  = -p_f' \bm\delta + \mu_f (\grad \textbf{u}_f^0 + ^\dagger \grad \textbf{u}_f^0),
\end{equation}
which implies that the closure is based on the absolute local fluid velocity $\textbf{u}_f^0$, rather than the fluctuating component $\textbf{u}_f'$.
Without going into the details, note that numerous closures are based on the reciprocal theorem formulation \citep{kim2013microhydrodynamics,stone2001inertial,raja2010inertial}. The closures (drag forces, stresslet etc... ) provided by the reciprocal theorem are by construction expressed in terms of the disturbance fields ($p_f'$,$\textbf{u}^*_f$), because they must decay to zero far from the test particle. 
For example, the Faxen contribution to the drag force in the case of a spherical solid particle of radius $a$ is given by $\textbf{f} = \pi a^3 \mu_f \grad^2 \textbf{u}$. 
This formulation is obtained  by considering the contribution from the disturbance stress, $\bm\sigma ^{(2)}  = -p_f' \bm\delta + \mu_f (\grad \textbf{u}_f^* + ^\dagger \grad \textbf{u}_f^*)$, which vanish far from the test particle. 
If one uses the formulation based on $\bm\sigma_f^{(4)}  = -p_f' \bm\delta + \mu_f (\grad \textbf{u}_f^0 + ^\dagger \grad \textbf{u}_f^0)$, the Faxen contribution to the drag force becomes $\textbf{f} = v_p \div \bm\Sigma + \pi a^3 \mu_f \grad^2 \textbf{u} = \pi a^3 \mu_f (1+ \frac{4}{3}) \grad^2 \textbf{u}$ where the second term is the contribution from the mean stress.
Although both formulations are equally valid, the second one appears to be less commonly used and could potentially lead to misinterpretations or inconsistencies if not carefully handled.

In light of these considerations, we adopt the formulation based on $\bm \sigma^{(2)}$-i.e., \ref{eq:two_fluid_momentum2}-for the remainder of this work, as it is less prone to errors when considering the closure problem. 
For simplicity, we will denote $\bm \sigma^{(2)}$ as $\bm \sigma^{*}$ throughout the rest of the paper.
The last two terms on the right-hand side of \ref{eq:two_fluid_momentum2} can be further expanded into a Taylor series using \ref{eq:f_exp_delta}.
Doing so leads us to the hybrid formulation of the continuous phase momentum  equation, \begin{align}
    \phi_f \rho_f(\pddt + \textbf{u}_f  \cdot \grad) \textbf{u}_f
    &= \phi_f 
    \left(\div \bm{\Sigma}
    + \rho_f \textbf{g}\right)
    + \div \bm\sigma_f^{\text{eff}}
    - \pSavg{\bm\sigma_f^{*}\cdot \textbf{n}}, 
    \label{eq:dt_uf2}
\end{align}
where,
\begin{align}
    \bm{\sigma}^{\text{eff}} 
    &= 
    - \avg{\chi_f\rho_f \textbf{u}_f'\textbf{u}_f'} 
    + \pavg{\intS{\textbf{r}\bm\sigma^{*}_f\cdot \textbf{n}} - \delta_p\intO{2\mu_f\textbf{e}_d^*}}\nonumber\\
    &- \div
        \pavg{ \frac{1}{2}\intS{\textbf{rr}\bm\sigma^{*}_f\cdot \textbf{n}}
        - \delta_p\intO{2\mu_f \textbf{r} \textbf{e}_d^*}}
        + \grad\grad (\ldots). 
    \label{eq:def_sigma_eff_f2}
\end{align}
Under this form, the left-hand side of \ref{eq:dt_uf2} represents the total derivative of $\textbf{u}_f$, while on the right-hand side we find: (1) the mean Newtonian stress contribution based on the bulk velocity $\bm\Sigma$, (2) the mean buoyancy force, (3) the Reynolds stress term, and (4) the moments of momentum exchange terms, which are computed based on $\bm\sigma_f^*$.
The $(\ldots)$ refers to the higher order moments.

The averaged mass of droplets ($m_p$), the averaged center of mass velocity ($\textbf{u}_p$), the averaged second moment of mass ($\textbf{M}_p$), and the averaged first moment of momentum ($\textbf{P}_p$), 
obey conservation laws that are given according to \ref{eq:avg_hybrid_q}, \ref{eq:avg_hybrid_q_1} and \ref{eq:avg_hybrid_q_n} (for conservation laws at the local scale, refer to the previous section).
They read, 
\begin{align}
    (\pddt + \textbf{u}_p \cdot \grad)n_p
    &=
    - n_p \div \textbf{u}_p\label{eq:mass_p}\\
    n_p (\pddt + \textbf{u}_p \cdot \grad) \textbf{M}_p
    +\div  \pavg{\textbf{u}_\alpha'\textbf{M}_\alpha}
    &=
    n_p2  (\textbf{S}_p+P_p\bm\delta)
    \label{eq:dt_hybrid_Mp}\\
    \label{eq:dt_hybrid_up}
    m_p n_p(\pddt + \textbf{u}_p \cdot \grad)\textbf{u}_p
    + \div \pavg{m_p \textbf{u}_\alpha'\textbf{u}_\alpha'}
    &=
    m_p n_p \textbf{g}
+ \pSavg{\bm\sigma_f^* \cdot \textbf{n}} + \pSavg{\bm\Sigma \cdot \textbf{n}}\\
    \label{eq:dt_hybrid_mup}
    n_p (\pddt + \textbf{u}_p \cdot \grad) \bm{\mu}_p
    +\div  \pavg{\textbf{u}_\alpha'\bm\mu_\alpha}
    &=
    \pSavg{\textbf{r}\times(\bm\sigma_f^*\cdot \textbf{n})}
    \\
n_p (\pddt + \textbf{u}_p \cdot \grad) \textbf{S}_p
    +\div  \pavg{\textbf{u}_\alpha'\textbf{S}_\alpha}
    &=
    \rho_d \pOavg{
        \textbf{w}_d^0  \textbf{w}_d^0 
        -\frac{1}{3} (\textbf{w}_d^0 \cdot  \textbf{w}_d^0) \bm\delta
    }
    - \pOavg{2 \mu_d\textbf{e}_d^*} \nonumber \\
    &+\pSavg{\frac{1}{2}(\textbf{r}\bm\sigma_f^*+^\dagger\textbf{r}\bm\sigma_f^*-\frac{2}{3}(\bm\sigma_f^* \cdot \textbf{r})\bm\delta)\cdot \textbf{n}}\nonumber\\
    &-  \pSavg{\gamma (\frac{1}{3}\bm\delta - \textbf{nn})}
     + (1-\lambda)\pOavg{2\mu_f\textbf{E}}\nonumber \\
     &+ \frac{1}{2}\pOavg{\textbf{r}(\div\bm\Sigma)+ (\div\bm\Sigma) \textbf{r}},
    \label{eq:dt_hybrid_Sp}
\end{align}
The presented system of equations extends the averaged equations developed for non-spherical solid particles \citep{curtiss1956kinetic}. 
It is worth noting that, in the case of solid particles, \ref{eq:dt_hybrid_Sp} becomes redundant, as the tensor $\textbf{S}_p$ can be expressed in terms of $\textbf{M}_p$, $\bm{\mu}_p$ or the mean angular velocity, and the correlations of their fluctuations.
In this case, \ref{eq:dt_hybrid_Sp} must be solved to determine the unknown internal stress of the solid particles.

The set of equations \ref{eq:mass_init}, \ref{eq:dt_uf2}, \ref{eq:mass_p}-\ref{eq:dt_hybrid_Sp} is completed by the following relations \begin{align}
    \phi_f + \phi_d &= 
    \phi_f + \phi  + \frac{1}{2}\grad\grad : (\textbf{M}_p n_p) + \ldots = 1,
    \label{eq:volume_conservation}\\
    \textbf{u} &= \textbf{u}_f\phi_f + 
    \phi\textbf{u}_p - \frac{1}{\rho_d} \div  (\textbf{P}_p n_p) + \ldots
    \label{eq:velocity_conservation}
\end{align}
where we have introduced the notation $\phi = n_pv_p$. 
\ref{eq:velocity_conservation} is obtained by inserting expansion \ref{eq:f_exp_chi} in \ref{eq:u_mean}.

\subsection{Symmetry of the effective stress tensor}
We remark that the second moment and higher-order moments in \ref{eq:def_sigma_eff_f2} appear under two divergence operators in \ref{eq:dt_uf2}. 
Hence, if we note $\Sigma_{ijk}$ the third rank tensor that represent these moments, then only the vector $\partial_k \partial_j\Sigma_{ijk}$ is of physical significance in the momentum balance \eqref{eq:dt_uf2}.
Thus, one can demonstrate that \citep{lhuillier1996contribution}
\begin{equation}
    \partial_j \partial_k \Sigma_{ijk}
    = \partial_j \partial_k \Sigma_{i(jk)}
    =
    \partial_j \partial_k \left[
        \Sigma_{i(jk)}
        + \Sigma_{j(ik)}
        - \Sigma_{k(ij)}
    \right],
    \label{eq:sym_proof}
\end{equation}
where $\Sigma_{i(jk)} = \frac{1}{2}[\Sigma_{ijk} + \Sigma_{ikj}]$ represents the symmetric part of $\Sigma_{ijk}$ over the index $jk$, as indicated by the parenthesis (and so on for the other tensor). 
This expression is allowed because $\partial_j \partial_k (\Sigma_{ijk} - \Sigma_{ikj}) = 0$ and $\partial_j \partial_k (\Sigma_{j(ik)} - \Sigma_{k(ij)}) = 0$. 
This manipulation highlight the fact that the effective stress due to the second order moments remains symmetric over the indices $ij$, in all circumstances.
Hence, as already demonstrated by \citet{lhuillier1996contribution} only the hydrodynamic torque can induce skew-symmetric stresses in the effective stress of the averaged momentum equation.

\section{Closure for dilute suspensions of droplets in viscous dominated flows}
\label{sec:closure}
We now consider the closures for a dilute, monodisperse suspension of spherical droplets with radius $a$ in Stokes flow. 
Our analysis recovers the results of \citet[Appendix B]{zhang1997momentum}. 
We extend these results by considering the influence of Marangoni effects on the suspension dynamics. 
Additionally, we demonstrate how higher-order moments can characterize droplet shapes and relate them to the averaged equations governing the continuous phase. 
We also address the closure of the various covariance terms arising in the averaged equations.

\begin{table}
    \centering
\begin{tabular}{|c|c|}\hline
    Velocity scale & $U$ \\
    Macroscopic length scale & $L$ \\
    Droplets radius & $a$ \\
    Reynolds number & $Re = \rho_f a U / \mu_f$   \\
    Capillary number & $Ca = \mu_f U / \gamma$ \\
    Marangoni number & $Ma =  a|\grad\gamma| / \mu_f U$ \\\hline
Viscosity ratio & $\lambda = \mu_d / \mu_f$ \\
Density ratio & $\zeta = \rho_d / \rho_f$ \\
\hline
    \end{tabular}
    \caption{Definition of physical quantities and dimensionless parameters.
    Note that the choice of velocity scales depends on the specific problem under consideration. 
    For example, in the case of sedimenting droplets in a quiescent fluid, the characteristic velocity is typically $U \sim |\textbf{u}_p|$.}
    \label{tab:dimensionless_para}
\end{table}

The closure terms in the equations above are expressed in terms of $p_f'$ and $\textbf{u}_f^*$. 
Our objective is to determine the disturbance pressure and velocity fields generated by a spherical droplet immersed in an arbitrary flow. 
Specifically, the pair $(\textbf{u}_f^*, p_f')$ represents the solution of the \textit{single-particle conditionally averaged} Navier–Stokes equations \citep{hinch1977averaged, zhang1994averaged, fintzi2025}.
When deriving averaged equations accurate to $O(\phi)$ we can neglect of droplet–droplet interactions \citep{hinch1977averaged, zhang1994averaged}. 
Furthermore, we assume negligible inertia in the closure problem by omitting all terms of $O(Re)$ in the closure equations, and correspondingly, all terms of $O(Re\phi)$ in the averaged equations (a summary of the dimensionless parameters is provided in \ref{tab:dimensionless_para}).
Although surface tension gradients are included, we assume spherical droplet shapes. 
This allows us to neglect all terms of $O(Ca)$ in the closure problem. 
As a result, we focus on the motion of an isolated spherical droplet in an arbitrary Stokes flow, subject to tangential stress discontinuities at the interface.
The analytical solution to this problem is well established and available in various sources, including \citet{
Subramanian_1985,
nadim1991motion,
pozrikidis1992boundary,
leal2007advanced,
raja2010inertial,
pozrikidis2011introduction,
kim2013microhydrodynamics}, and forms the basis for evaluating the closure terms.
For clarity, we detail the closure problem and its solution in \ref{ap:singularity_solution}.

In the following, $L$ denotes the characteristic length scale over which the averaged quantities may vary \citep{jackson1997locally}. 
Each averaged property may possess a different variation length scale.
For instance, consider $\grad \textbf{u}$, which typically varies at the length scale of the process, whereas $\grad\gamma$ may vary over a smaller length scale. 
For example, if surface tension gradient is generated by surfactant transport, $\gamma$ typically varies over a length scale of the drop size. 
If $\gamma$ varies due to a non-constant temperature field, the length scale of variation will correspond to that of the temperature field, which is related to the process or macro scale. 
To simplify the problem, we assume that $\grad\gamma$, and $\grad \textbf{u}$ typically vary on the order of $\sim L$. 
We then consider a proper separation of scales such that $a\ll L$ and therefore all contributions proportional to  $O(a^2/L^2)$ become negligible in the averaged equations \citep{jackson1997locally,zhang1997momentum}.

\subsection{Hydrodynamic stresses closures}
\label{sec:hydro_closure}

In the first place we focus on the momentum exchange terms. 
We may directly compute the following expressions from the singularity solutions and find\footnote{
    We follow the convention of \citet{happel2012low} for the transpose operator.
    For an arbitrary order tensor \textbf{A}, its transpose is denoted either as $^\dagger A_{ijkl\ldots} = A_{jikl\ldots}$ or as $ (A_{\ldots ijkl})^\dagger = A_{\ldots ijlk}$}, 
\begin{align}
    \pSavg{\bm\sigma_f^*\cdot \textbf{n}} &
    =
    \phi
    \frac{\mu_f}{a^2}
    \frac{3(2+3\lambda)}{2(1+\lambda)}\textbf{u}_r
    + \phi\mu_f  \frac{3\lambda}{4(\lambda +1)} \grad^2 \textbf{u}+ \phi \frac{1}{a}\frac{1}{\lambda +1} \grad \gamma
    + \phi a \frac{1}{10(\lambda +1)}\grad^2(\grad\gamma),
    \label{eq:drag_forces}
    \\
    \pSavg{\textbf{r}\bm\sigma_f^*\cdot \textbf{n}} &
    = \mu_f \phi 
    \frac{3(5\lambda +2)}{5(\lambda +1)}\textbf{E}
+ \phi a \frac{9}{25(\lambda +1)}(\grad\grad \gamma- \bm\delta\grad^2 \gamma/3),
\label{eq:first_mom}
    \\
    \pSavg{\textbf{rr}\bm\sigma_f^*\cdot \textbf{n}} &
    =
    \mu_f \phi \frac{3}{5(\lambda +1)} (\textbf{u}_r \bm\delta + ^\dagger\textbf{u}_r\bm\delta )
    + \mu_f \phi \frac{3(5\lambda +2)}{10(\lambda+1)}\bm\delta \textbf{u}_r
    - \phi a\frac{2}{5(\lambda+1)}(\grad \gamma \bm\delta+  ^\dagger \grad \gamma\bm\delta)
    \nonumber\\&
    + \phi a\frac{3}{5(\lambda+1)} \bm\delta \grad \gamma,
    \label{eq:second_mom1}
\end{align}
\begin{align}
    \pSavg{ 2\mu_f \textbf{e}_d^*}
    &=
    -  \mu_f \phi \frac{2(5\lambda +2)}{5(\lambda+1)}\textbf{E}
- \phi a  \frac{6}{25(\lambda+1)} (\grad\grad \gamma- \bm\delta\grad^2 \gamma§/3),\\
    \pSavg{ 2 \mu_f \textbf{re}_d^* }
    &=
    \phi \frac{3\pi}{10(\lambda+1)}
    (\bm\delta \textbf{u}_r +  \bm\delta\textbf{u}_r^\dagger)
    -\phi \frac{\pi}{5(\lambda+1)}\textbf{u}_r\bm\delta 
    -\phi a\frac{1}{5(\lambda+1)} (\bm\delta \grad \gamma+ \bm\delta\grad\gamma^\dagger)
    \nonumber \\&
    + \phi a\frac{2}{15(\lambda+1)}\grad\gamma\bm\delta,
\label{eq:secondUN}
\end{align}
where we have introduced the relative velocity $\textbf{u}_r = \textbf{u} - \textbf{u}_p$, and the viscosity ratio $\lambda = \mu_d/\mu_f$. 
Most of the terms proportional to $\textbf{E}$ and $\textbf{u}_r$ in this expression are already well known and will not be discussed here.  
In \ref{eq:drag_forces} the third term represents the force induced by $\grad \gamma$, which typically represents the thermocapillary migration of droplets if one relates $\grad\gamma$ to a temperature gradient.  
The last term of \ref{eq:drag_forces} corresponds to a ``Faxen-like'' contribution of the Marangoni forces and involves the third derivative $\gamma$.
It is worth noting that the first Marangoni moments of hydrodynamic forces scale as $\propto \phi \grad\grad \gamma$ (see \ref{eq:first_mom}), while the second moments scale as $\propto \phi \bm\delta \grad \gamma$ \eqref{eq:second_mom1}.
These tensorial form could have been anticipated based on symmetry considerations. 
Note that we have neglected the terms proportional to $\grad^2 \textbf{E}$ in the first moment expression, and to $\grad\grad \textbf{u}$ or $\grad\grad\grad \gamma$ in the second moment expression, as they are of order $O(a^2/L^2)$ or smaller.
The complete expressions for the second moments, including these higher-order contributions, are provided in \ref{ap:singularity_solution}.

The above closure provides the contribution from the disturbance fields, however in the dispersed phase equations \eqref{eq:dt_hybrid_up,eq:dt_hybrid_Sp} one need the contribution from the total momentum exchange including the contribution of the mean stress $\bm\Sigma$. 
These terms can be obtained following the procedure outlined in \citep{zhang1997momentum,morel2015mathematical}, and reads, 

\begin{align}
    \pSavg{\bm\Sigma\cdot \textbf{n}}
    &= \phi\div\bm\Sigma + O(a^2/L^2),\\
    \pOavg{\textbf{E}}
    &= \phi \textbf{E}+ O(a^2/L^2),\\
    \pSavg{\textbf{r}(\div \bm\Sigma)}
    &= O(a^2/L^2). 
    \label{eq:mean_contributions}
\end{align}
Where we have used the approximation of $\bm\Sigma_f(\textbf{x}+\textbf{r}) = \bm\Sigma_f(\textbf{x}) + \textbf{r}\cdot\grad \bm\Sigma_f|_{\textbf{r}=0} + \ldots$ and neglected the $O(a^2/L^2)$ terms.

\subsection{Velocity variance and covariance closures}

The disturbance velocity field $\textbf{u}_f'$ is proportional to $\propto \textbf{u}_r$, $\grad\textbf{E}_f$ and $\grad\grad \textbf{u}_f$ depending on the problem at hand.
Additionally, the Reynolds stress tensor $\avg{\chi_f \textbf{u}_f'\textbf{u}_f'}$ is a symmetric second-order tensor. 
We deduce that the functional form of the Reynolds stress must be 
\begin{align}
    \avg{\chi_f \rho_f \textbf{u}_f' \textbf{u}_f'}
    =&
    C_{uu}^1(\phi,\lambda) \rho_f \textbf{u}_{r} \textbf{u}_{r}
    + C_{uu}^2(\phi,\lambda) \rho_f (\textbf{u}_{r}\cdot  \textbf{u}_{r})\bm\delta\\
    &+a^2 C_{EE}^1(\phi,\lambda) \rho_f\textbf{E}\cdot \textbf{E} 
    +  a^2 C_{EE}^2(\phi,\lambda) \rho_f (\textbf{E} : \textbf{E})\bm\delta.
    + \ldots
    \label{eq:Reynolds_stress_functional_form}
\end{align}
where the remaining terms indicated by the $\ldots$ represent linear combination of terms proportional to $a^4\grad\grad \textbf{u}_f:\grad\grad \textbf{u}_f$. 
However, since these terms are proportional to $a^2$, they scale as $O(a^2/L^2)$ in the averaged equations. 
As a result, only the first two terms in \ref{eq:Reynolds_stress_functional_form} are significant in the momentum equation\footnote{
    Note that since $\textbf{u}_f' \propto \frac{a \grad \gamma}{\mu_f}$ it follows that $\textbf{u}_f'\textbf{u}_f' \propto O(a^2/L^2)$. Thus, the contribution from $\grad\gamma$ ends up being negligible under the current modeling assumptions. 
}.
The exact values of the coefficients $C_{EE}$ are provided in \citet{raja2010inertial}. 
Because the Reynolds stress is defined as an average over the continuous-phase domain (denoted by $\chi_f$), the disturbance velocity fields $\textbf{u}_f'$ cannot be directly integrated to determine the constants $C_{uu}^1$ and $C_{uu}^2$. 
Nonetheless, based on experimental measurements by \citet{cartellier2009induced}, particle-resolved simulations by \citet{fintzi2025}, and theoretical results in triply periodic domains by \citet{hill2001first}, it is reasonable to expect that these constants follow the scaling:
\begin{equation}
C_{uu}^1, C_{uu}^2 \propto \phi^{2/3} \frac{(2+3\lambda)^2}{(\lambda+1)^2}.
\end{equation}
In the present study, we have neglected droplet-droplet interactions in the closure problem, and therefore expect $\pavg{\textbf{u}_\alpha'\textbf{u}_\alpha'} = 0$. 
Nonetheless, by invoking symmetry arguments and drawing on experimental observations reported in the literature \citep{guazzelli2011fluctuations}, we conclude that:
\begin{equation}
    \pavg{m_p \textbf{u}_\alpha'\textbf{u}_\alpha'}
    =
    \rho_d C^1_{up}(\phi,\lambda)\textbf{u}_r\textbf{u}_r
    + \rho_d C^2_{up}(\phi,\lambda) \bm\delta(\textbf{u}_r\cdot \textbf{u}_r)
    \label{eq:upup}
\end{equation}
where $C_{up}^1$ and $C_{up}^2$ are unknown constants which are $\propto \phi^{2/3}$\citep{guazzelli2011fluctuations}. 
This result indicates that, due to the long-range interactions between droplets, the velocity variance of the particles remains non-zero even at order $O(\phi)$.
It is important to note that both $\pavg{m_p \textbf{u}_\alpha'\textbf{u}_\alpha'}$ and $\avg{\rho_f \textbf{u}_f'\textbf{u}_f'}$ represent inertial contributions by construction.
However, when expressed in dimensionless form, both terms scale as $O(Re  \phi^{2/3})$.
Given that $O(\phi^{2/3}Re) \gg O(Re\phi)$ in the dilute limit, we conclude that these velocity variance terms must be retained to achieve closure at order $O(Re\phi)$.

The covariance terms appearing on the left-hand side of  \ref{eq:dt_hybrid_Mp} to \ref{eq:dt_hybrid_Sp}, reflect the correlation between the shape of the droplet ($\textbf{M}_\alpha$), its angular momentum ($\bm\mu_\alpha$), and its stretching of momentum ($\textbf{S}_\alpha$), with its center of mass velocity $\textbf{u}_\alpha$. 
If any of these quantities is statistically independent of $\textbf{u}_\alpha$, the corresponding covariance term vanishes. 
In the dilute Stokes regime, a purely translating spherical droplet remains undeformed due to the balance of normal stresses at its surface \citep{leal2007advanced}. 
Similarly, translation of a spherical particles does not induce hydrodynamic torque, and thus no angular momentum is generated. 
Therefore, in the Stokes regime, the quantities  $\textbf{M}_\alpha$, $\textbf{P}_\alpha$ are uncorrelated with $\textbf{u}_\alpha$,  implying that the covariance terms $\pavg{\textbf{u}_\alpha' \textbf{M}_\alpha'},\pavg{\textbf{u}_\alpha' \bm\mu_\alpha'}$ and $\pavg{\textbf{u}_\alpha' \textbf{S}_\alpha'}$ equal zero. 
However, these conclusions no longer hold at finite inertia. 
In that case, the translational-rotational coupling \citep{rubinow1961transverse} leads to nonzero force on the particle, and the droplet can deform as a result of its motion relative to the surrounding fluid \citep{taylor1964deformation}.

\subsection{Closed form of the hybrid model}

Note that to leading order in the expansion in $Ca$ and $Re$, the linear momentum equations are decoupled from the dispersed phase moments ($\textbf{S}_p, \bm{\mu}_p$, and $\textbf{M}_p$).
As a result, \ref{eq:dt_hybrid_Sp,eq:dt_hybrid_Mp,eq:dt_hybrid_mup} are not required to compute the mean variables ($\textbf{u}_p, \textbf{u}, p_f, \phi$).
Therefore, one can directly substitute expressions \ref{eq:drag_forces}-\ref{eq:mean_contributions} into equations \ref{eq:dt_uf2} and \ref{eq:dt_hybrid_up} to obtain a closed form of the hybrid model which read,

\begin{align}
    \label{eq:firstSys}
    \phi_f + \phi &= 1\\
    \phi_f\textbf{u}_f + 
    \phi\textbf{u}_p&=\textbf{u}\\
    \div \textbf{u} &= 0\\
    (\pddt + \textbf{u} \cdot \grad)\phi
    &=
    \div (\phi \textbf{u}_r)\\
    \rho_d \phi (\pddt + \textbf{u}_p \cdot \grad)\textbf{u}_p
&=
    \phi(\div \bm\Sigma
    + \rho_d  \textbf{g})
    + \div \bm\sigma_p^\text{eff}
    + \textbf{F}
    \\
    \phi_f \rho_f(\pddt + \textbf{u}_f  \cdot \grad) \textbf{u}_f
&= \phi_f 
    \left(\div \bm\Sigma
    + \rho_f \textbf{g}\right)
    + \div \bm\sigma_f^\text{eff}
    -\textbf{F}
    \label{eq:lastSys}
\end{align}
where the closures are given by,
\begin{align}
    \textbf{F}=&
    \phi
    \frac{\mu_f}{a^2}
    \frac{3(2+3\lambda)}{2(1+\lambda)}\textbf{u}_r
    + \phi\mu_f  \frac{3\lambda}{4(\lambda +1)} \grad^2 \textbf{u}
    + \phi \frac{1}{a}\frac{1}{\lambda +1} \grad \gamma
    + \phi a \frac{1}{10(\lambda +1)}\grad^2(\grad\gamma)\\
\bm\sigma_f^\text{eff}
    =&
     \mu_f \phi \frac{5\lambda +2}{2(\lambda+1)} \textbf{E}
    - \mu_f \frac{3\lambda}{4(\lambda+1)} [
    \grad(\phi \textbf{u}_r)
    + \grad(\phi \textbf{u}_r)^\dagger]
    + \mu_f \frac{3\lambda - 2}{4(\lambda+1)} \div(\phi \textbf{u}_r)  \bm\delta\nonumber\\
    &+ a\phi \frac{3}{5(\lambda+1)}(\grad\grad\gamma-\bm\delta \grad^2\gamma/3)
    - a \frac{1}{(\lambda+1)}\grad (\phi \grad \gamma)
    - a \frac{5}{6(\lambda+1)}\div (\phi \grad \gamma)
    \nonumber \\
    &-\rho_f C^1_{uu}(\phi,\lambda)  \textbf{u}_r \textbf{u}_r
    -\rho_f C^2_{uu} (\phi,\lambda) (\textbf{u}_r \cdot \textbf{u}_r)\bm\delta
    \label{eq:sigma_feffff}\\
    \bm\sigma_p^\text{eff}
    =&
    -\rho_d C^1_{up}(\phi,\lambda) \textbf{u}_r \textbf{u}_r
    -\rho_d C^2_{up}(\phi,\lambda) (\textbf{u}_r \cdot \textbf{u}_r)\bm\delta
\end{align}
This system is constituted of 6 unknowns ($\phi_f,\phi,\textbf{u},\textbf{u}_f,p_f,\textbf{u}_p$), and 6 equations (\ref{eq:firstSys} to \ref{eq:lastSys}).  
Upon the precise knowledge of $C^1_{up}, C^2_{up}, C^1_{uu}$ and $C^2_{uu}$ the system can be considered closed and accurate to order $O(\phi)$. 

First note that most of the terms related to the phase relative motion (i.e. those $\propto \textbf{u}_r$ or $\textbf{E}$) are already presented in \citet[Appendix B]{zhang1997momentum}. 
As already noted in several studies $\grad \gamma$ contributes to the hydrodynamic forces on the droplets \citep{Subramanian_1985}, hence also contributes as a source term in the equation of $\textbf{u}_f$. 
What has been unnoticed before, however, is that terms of the form $\sim \nabla\nabla \gamma$ also appear to induce non-newtonian stresses within the suspension.
Although the problem considered in this work differs from that typically encountered in industrial processes $-$here, a fixed surface tension gradient is imposed rather than solving a scalar transport equation that governs surface tension locally$-$ it is still insightful to understand under which conditions surface tension gradients (i.e., Marangoni effects) become significant.
First, it is worth noting that Marangoni effects become negligible when the viscosity ratio is high, meaning the dispersed phase (e.g., droplets) is much more viscous than the surrounding fluid. Consequently, one expects Marangoni flows to be more prominent for bubbles than for viscous droplets. 
Consider a single bubble rising in a quiescent fluid under the influence of gravity and an imposed uniform surface tension gradient, or equivalently, a constant concentration gradient of a surfactant species at infinity, with a linear dependence of surface tension on concentration. 
Balancing the Marangoni force with the buoyant force gives the scaling,$\nabla \gamma \sim (\rho_d - \rho) g a$.
Assuming the surface tension gradient scales as $\nabla \gamma \sim \Delta \gamma / L$, where $\Delta \gamma $ is the typical variation in surface tension over a characteristic length $L$, we can estimate when Marangoni effects become significant.
For instance, in an aqueous emulsion (with $\rho \approx 1000 \text{ kg/m}^3$, $\mu \approx 10^{-3} \text{ Pa·s}$, and $\rho_d - \rho \approx 100 \text{ kg/m}^3$), and assuming $\Delta \gamma \approx 10^{-2} \text{ N/m}$ and $a \approx 10\ \mu\text{m}$, Marangoni and buoyancy forces become comparable when $L \approx 1 \text{ m}$. 
This suggests that even when surface tension varies over meter-scale distances, Marangoni forces can significantly influence the dynamics of very small bubbles or low viscosity ratio droplets.
Next, we consider a droplet subjected to a shear flow with an imposed second-order spatial variation of surface tension. 
Balancing the effective shear stress due to the flow with the stress induced by Marangoni effects yields the scaling $\nabla^2 \gamma \sim \frac{\mu \dot{\gamma}}{a},$
where $\dot{\gamma}$ is the shear rate. 
For a typical shear rate of $\dot{\gamma} \approx 1\ \text{s}^{-1}$ and droplet size $a \approx 10\ \mu\text{m}$, the characteristic length scale over which surface tension must vary for Marangoni stresses to match shear-induced stresses is approximately $L \sim 1 \text{ cm}$.
Therefore, in processes where surface tension varies over centimeter-scale distances, Marangoni effects can generate effective stresses comparable in magnitude to those from shear-induced viscosity. 
This highlights the importance of accounting for surface tension gradients in modeling interfacial flows, particularly in systems involving small bubbles or droplets.

For completness the averaged mass and second moment of mass read as, 
\begin{align}
    n_p m_p &= \rho_d \phi   
    \label{eq:volume}
    \\ 
    \label{eq:second_moment_of_mass0}
    n_p \textbf{M}_p &= \rho_d\phi \frac{a^2}{5}\bm\delta
\end{align}
Substituting these closures into  \ref{eq:dt_hybrid_mup} and \ref{eq:dt_hybrid_Sp} yields
\begin{align}
    \textbf{S}_p &= 0 \label{eq:S_eq_zerp}\\
    n_p (\pddt + \textbf{u}_p \cdot \grad)\bm\mu_p &= 0,
\end{align}
as no hydrodynamic torque is exerted on the droplet.
\subsection{Small droplet deformations}

The usual procedure to determine the droplet deformation is to use the normal stress balance \citep{nadim1991motion,nadim1996concise}. 
However, to highlight the significance of the first-moment of momentum equation, we propose an alternative method based on equations \ref{eq:dt_hybrid_Sp,eq:dt_hybrid_Mp,eq:dt_hybrid_mup}. 
As we now demonstrate, with the closure relations derived in the previous section, equation \ref{eq:dt_hybrid_Sp} can be used to compute the first-order correction to the droplet spherical shape. 
In other words, using closure terms accurate to $O(Ca)$, this approach yields the leading order contribution of $Ca$ to droplet deformation.

Let us describe points lying in the droplet $\alpha$ using the parametric equation in the local spherical reference frame ($r,\theta,\varphi$),
\begin{equation}
    \textbf{r}(r,\varphi,\theta) = r [1+ Ca f_\alpha(\varphi,\theta)] \textbf{e},
    \label{eq:parametrization}
\end{equation}
where $0<r<a$ is the radial parameter, $\theta$ the polar angle, and $\varphi$ the azimutal angle. 
$f_\alpha(\theta,\varphi)$ is the shape function of the droplet, and $\textbf{e} = \cos\varphi\sin\theta \textbf{e}_x + \sin\varphi\sin\theta\textbf{e}_y+ \cos\theta \textbf{e}_z$ the radial unit vector. 
Using the parametrization given by \ref{eq:parametrization} one can eventually compute the volume $d\Omega$ and surface $d\Gamma$ element in terms of $r,\varphi,\theta$ and the deformation function $f_\alpha(\theta,\varphi)$.
The result are, $d\Gamma =a^2 (1+2Ca f_\alpha(\theta,\varphi)) \sin\theta d\theta d\varphi$ and $d\Omega = (1+3Ca f_\alpha(\theta,\varphi)) r^2\sin\theta drd\theta d\varphi$ at the leading order in $Ca$. 
Following, \citet{nadim1996concise,nadim1991motion} we then expand $f_\alpha(\varphi,\theta)$ in a series of surface harmonics centered at the droplet center of mass, namely 
\begin{equation}
    f_\alpha(\textbf{e}) = 
    \sum_{n=2}^\infty\textbf{S}^{(n)}:\textbf{H}_\alpha^{(n)},
    \label{eq:f_definition}
\end{equation} 
with $\textbf{S}^{(n)} = \frac{(-1)^n r^{n+1}}{1\cdot3\ldots(2n-1)}\grad^{(n)}(\frac{1}{r})$ the $n^{th}$ order surface spherical harmonic, and $\textbf{H}_\alpha^{(n)}$ an $n^{th}$ order symmetric and traceless tensor to be determined. 
Inserting \ref{eq:parametrization} in the definition of the second moment of mass yields,
\begin{equation}
    n_p \textbf{M}_p = \rho_d\phi \frac{a^2}{5}(\bm\delta+2Ca \textbf{H}_p^{(2)}),
    \label{eq:second_moment_of_mass1}
    \end{equation}
where we used the definition $n_p \textbf{H}_p^{(n)} = \pavg{\textbf{H}_\alpha^{(n)}}$. 
To obtain, closed-form expression for the tensors $\textbf{H}_p^{(2)}$, we consider the first moment of the forces, as given by equation \ref{eq:dt_hybrid_Sp}.
At leading order in $Ca$,$Re$ and $a^2/L^2$, we have $\textbf{S}_p =0$ and the first and last term on the left-hand side of \ref{eq:dt_hybrid_Sp} are negligible.
Then \ref{eq:dt_hybrid_Sp} yields,
\begin{align}
\pSavg{\gamma (\frac{1}{3}\bm\delta - \textbf{nn})}
=&- \pOavg{2 \mu_d\textbf{e}_d^*} \nonumber
+\pSavg{\frac{1}{2}(\textbf{r}\bm\sigma_f^*+^\dagger\textbf{r}\bm\sigma_f^*-\frac{2}{3}(\bm\sigma_f^* \cdot \textbf{r})\bm\delta)\cdot \textbf{n}}\nonumber\\
&+ (1-\lambda)\pOavg{2\mu_f\textbf{E}}\nonumber 
\label{eq:dt_hybrid_Sp1}
\end{align}
All the terms on the right-hand side of this equation are provided by the closure introduced in \ref{sec:hydro_closure}.
However to close this equation, we must express the quantity $\pSavg{\gamma (\bm\delta/3 - \textbf{nn})}$ as a function of $\textbf{H}_p^{(n)}$.
We first Taylor expand the surface tension about the point $\textbf{x}$ as $\gamma |_{\textbf{x}_\alpha} = \gamma|_\textbf{x}+\textbf{r}\cdot \grad \gamma|_\textbf{x}+\frac{1}{2}\textbf{rr}:\grad\grad\gamma|_\textbf{x}+\ldots$ to get 
\begin{equation}
\pSavg{\gamma (\bm\delta/3 - \textbf{nn})} = \gamma \pSavg{(\bm\delta/3 - \textbf{nn})} +\grad \gamma \cdot  \pSavg{\textbf{r}(\bm\delta/3 - \textbf{nn})} + \ldots
\label{eq:closure_surface_tension}
\end{equation}
Then, by injecting the approximation $\textbf{n} = \textbf{e} - Ca (\bm\delta - \textbf{e}\textbf{e})\cdot \grad f_\alpha + O(Ca^2)$ into the expression of $f_\alpha$ given by \ref{eq:f_definition} yields after some algebra  
\begin{align}
    \pSavg{ (\bm\delta/3 - \textbf{nn})}  
    &=
    \frac{1}{a}\frac{8}{5} Ca \phi \textbf{H}_p^{(2)} \label{eq:closure_surface_tension2}\\
    \pSavg{\textbf{r} (\bm\delta/3 - \textbf{nn})}  
    &=   \frac{6}{7} Ca  \phi \textbf{H}^{(3)}_p
    \label{eq:closure_surface_tension3}\\
    \pSavg{ (\bm\delta/3 - \textbf{nn})_{ij}(\textbf{rr})_{kl}}  
    &= a  \frac{4}{15} Ca \phi
    (\textbf{H}_p^{(2)}\bm\delta)^\text{p}_{ijkl}
    +
    a Ca \phi \frac{16}{35} \textbf{H}^{(4)}_{ijkl}. 
    \label{eq:closure_surface_tension4}
\end{align}
In the last equation we have introduced the shorthand, 
\begin{equation}
    (\textbf{H}_p^{(2)}\bm\delta)^\text{p}_{ijkl}
    =  \{
        H_{ij}\delta_{kl}
        + \frac{2}{9} H_{kl}\delta_{ij}
- \frac{1}{6} (H_{ik}\delta_{jl}+ H_{jl}\delta_{ik}+H_{il}\delta_{jk}+H_{jk}\delta_{il})
\}.
\end{equation}
Injecting \ref{eq:closure_surface_tension2,eq:closure_surface_tension3,eq:closure_surface_tension4} and the closures derived in \ref{sec:hydro_closure}  into \ref{eq:dt_hybrid_Sp} gives directly,
\begin{align}    
    \textbf{H}_p^{(2)}
+
    \frac{30}{56}\frac{a}{\gamma}\grad\gamma\cdot \textbf{H}^{(3)}_p
+ \ldots
&=
    \frac{19 \lambda + 16}{8 \left(\lambda + 1\right)}
    \left(\frac{a \mu_f}{\gamma Ca}\right)
    \textbf{E}
+ 
    \frac{3}{40}\frac{(2\lambda +3)}{(\lambda+1)} 
    \frac{a^2}{Ca\gamma} 
    (\grad\grad\gamma-\grad^2\gamma\bm\delta/3)
    \label{eq:def_H2}
\end{align}
Discarding first the surface tension gradients in \ref{eq:def_H2} , we recover the classical results of \citet{taylor1932viscosity} (see also \citet{rallison1984deformation}) for droplet deformation in a pure linear flow.
As a result, the mean deformation tensor  $\textbf{H}_p^{(2)}$  is found to scale with the mean rate-of-strain tensor  \textbf{E}, non-dimensionalized by the shear rate scale $a \mu_f /(Ca \gamma)= a/U$ \footnote{While the contribution from the mean cubic flow $\grad^2\textbf{E}$, could in principle, be accounted for using Faxen relations applied to the first hydrodynamic moment, this term is $a^2/L^2$ smaller than this other and is therefore negligible in the present context.}.
Gradients in surface tension introduce considerable complexity into \ref{eq:def_H2}.
The first consequence of the surface tension gradient is the appearance of $\textbf{H}_p^{(3)}$, along with all other $\textbf{H}_p^{(n)}$, on the left-hand side of Equation \ref{eq:def_H2}, thereby coupling the equation for $\textbf{H}_p^{(2)}$ with those of the other $\textbf{H}_p^{(n)}$ terms.
The second consequence comes from the second term on the right-hand side of \ref{eq:def_H2}. 
This term represents the contribution from the hydrodynamic stress generated by $\grad\grad\gamma$. 
However, one may perform several simplifications by noting first that all the terms in $\textbf{H}_p^{(n)}$ with $n\geq 4$ are of order $a^2/L^2$ or smaller.
The same comment applies to the second term on the right-hand side of \ref{eq:def_H2}, which is of order $a^2/L^2$.
In the current modeling hypothesis one may neglect all the $O(a^2 / L^2)$  terms, hence \ref{eq:def_H2} reduces to
\begin{align}    
        \textbf{H}_p^{(2)}
        +
        \frac{30}{56}\frac{a}{\gamma}\grad\gamma\cdot \textbf{H}^{(3)}_p
        =
        \frac{19 \lambda + 16}{8 \left(\lambda + 1\right)}
        \left(\frac{a \mu_f}{\gamma Ca}\right)
        \textbf{E}. 
        \label{eq:H2_simplified}
\end{align}
The effect of the surface tension gradient is smaller than the effect of strain by a factor of $a/L$, yet it remains non-negligible within the current assumptions framework.
To compute the components of $\textbf{H}^{(3)}$, we rely on the second moment of the momentum equation (see \ref{eq:second_mom}), along with the closure relations for both the second moment of hydrodynamic forces and the first moment of internal shear rate. 
Given the complexity of the derivation, full details are provided in \ref{ap:singularity_solution}. 
Since buoyancy effects are not central to our discussion, they are neglected in this calculation. 
The resulting expression is,
\begin{multline}
    \label{eq:defH3}
    Ca \frac{8}{7} \gamma (\textbf{H}_p^{(3)})_{ijk}
    +
    Ca \; a (\grad \gamma)_l(  \frac{16}{35} 
    (\textbf{H}_p^{(2)}\bm\delta)^\text{p}_{ijkl}
    +
    \frac{64}{105} \textbf{H}^{(4)}_{ijkl})
=\\
- \mu_f a^2 \frac{5\lambda+2}{21(\lambda+1)} (\bm\delta \grad^2 \textbf{u})_{ijk}
    + \mu_f a^2 \frac{2(\lambda-1)}{21(\lambda+1)} [
        (\bm\delta \grad^2 \textbf{u})_{ikj}
        + (\bm\delta \grad^2 \textbf{u})_{jki}
        ]  \\
    + \mu_f \frac{11\lambda+10}{21(\lambda+1)} [(\grad\grad \textbf{u})_{ijk}+(\grad\grad \textbf{u})_{ikj}+(\grad\grad \textbf{u})_{kji}] 
    + \mu_f\frac{2(3\lambda+2)}{5(\lambda+1)}[
        (\bm\delta \textbf{u}_r)_{ikj}
        +(\bm\delta \textbf{u}_r)_{jki}
        - \frac{2}{3}(\bm\delta \textbf{u}_r)_{jki}
        ] \\
    +a^3\frac{4(16\lambda + 19)}{735(\lambda +1)}(\grad\grad\grad \gamma)_{ijk}
    +a^3\frac{2(66\lambda + 109)}{3675(\lambda +1)}
    [
        (\bm\delta \grad^2\grad \gamma)_{jki}
        + (\bm\delta \grad^2\grad \gamma)_{ikj}
    ] \\
    - a^3 \frac{8(73\lambda+102)}{11025(\lambda +1)} 
    (\bm\delta \grad^2\grad \gamma)_{ijk}
+a \frac{4(2\lambda +3)}{15(\lambda +1)}  
    [
        (\bm\delta\grad \gamma)_{ikj}
        + (\bm\delta\grad \gamma)_{jki}
        - \frac{2}{3}(\bm\delta\grad \gamma)_{ijk}
    ].
\end{multline}
At first glance, it may be seen that $\textbf{H}^{(3)}$ can be explicitly computed in terms of $\grad\gamma, \textbf{u}_r, \grad\grad\grad\gamma$ and $\grad \grad \textbf{u}$. 
However, it is important to note that, in Stokes flow, the relative translation ($\textbf{u}_r$) does not contribute to the deformation of the droplet \citep{taylor1964deformation,nadim1991motion}. 
To clarify this point, we consider the steady-state momentum balance on a force-free droplet \eqref{eq:drag_forces}, that is 
\begin{equation}
    (2+3\lambda)\textbf{u}_r
    = 
    -\frac{\lambda}{2} a^2\grad^2 \textbf{u}
    -\frac{2a}{3 \mu_f } \grad \gamma
    -\frac{2 a^3}{30\mu_f }\grad^2(\grad\gamma).
\end{equation} 
Inserting this expression in \ref{eq:defH3} yields, 
\begin{multline}
    (\textbf{H}_p^{(3)})_{ijk}
    +
    \frac{2a}{5\gamma}(\grad \gamma)_l[  
    (\textbf{H}_p^{(2)}\bm\delta)^\text{p}_{ijkl}
    +
    \frac{4}{3} \textbf{H}^{(4)}_{ijkl}]
    +\ldots
    =\\
    \frac{a^2 \mu_f }{Ca\gamma}\frac{11\lambda+10}{24(\lambda+1)} [
        (\grad \grad \textbf{u})_{ijk}
        +(\grad \grad \textbf{u})_{ikj}
        +(\grad \grad \textbf{u})_{kji}
        - \frac{1}{5}(
            (\bm\delta \grad^2 \textbf{u})_{ijk}
            + (\bm\delta \grad^2 \textbf{u})_{ikj}
            + (\bm\delta \grad^2 \textbf{u})_{jki}
        )
        ]
    \\
    +\frac{a^3 }{Ca\gamma}\frac{(16\lambda +19)}{210(\lambda+1)}(\grad\grad\grad\gamma)_{ijk}
    +\frac{a^3 }{Ca\gamma}\frac{(11\lambda +10)}{350(\lambda+1)}
    [(\bm\delta \grad^2\grad \gamma)_{jki}+ (\bm\delta \grad^2\grad \gamma)_{ikj}]\\
    - \frac{a^3 }{Ca\gamma} \frac{146\lambda +155}{3150(\lambda+1)}(\bm\delta \grad^2\grad \gamma)_{ijk}
    +\frac{a}{Ca \gamma} \frac{8}{15}(
        (\bm\delta \grad\gamma)_{kji}
        + (\bm\delta \grad\gamma)_{kij}
        -\frac{2}{3}(\bm\delta \grad\gamma)_{ijk}
    )
\end{multline} 
Examining the first line, one can identify the results of \citet{nadim1991motion} for the deformation of a force-free droplet embedded in a quadradic flow\footnote{Note that $- 15 \textbf{H}_p^{(3)}$  corresponds to \citet{nadim1991motion}'s $\textbf{H}^{(3)}$, thereby recovering the same result.}.  
The second line represents the contributions arising from deformation due to the first and third derivatives of the surface tension gradient.
All terms contributing to $\textbf{H}_p^{(3)}$ are of order  $O(a/L)$, implying that in \ref{eq:H2_simplified} the term involving $\textbf{H}_p^{(3)}$ scales as $O(a^2/L^2)$ and is therefore negligible. Including terms of order $O(a^2/L^2)$ significantly increases the complexity of the problem for two main reasons: (1) source terms involving gradients of $\gamma$ are present on the right-hand sides of \ref{eq:def_H2} and \ref{eq:defH3}; and (2) the equations governing each $\textbf{H}_p^{(n)}$ become coupled.

Once the averaged deformation is determined via the moment of momentum equations, it becomes possible to express the closure terms of the hybrid model (i.e., from \ref{eq:drag_forces} to \ref{eq:secondUN}) in terms of the tensors $\textbf{H}_p^{(n)}$, as illustrated, for example, in \citet{haber1971dynamics}. 
This leads to a hybrid model that is accurate to $o(Ca)$ and will be the focus of future work.

 \section{Conclusion}
\label{sec:conclusion}

In this work, we provided a comprehensive presentation of the averaged equations and the derivation of closure relations for complex dispersed phases. 
The general derivation, developed from \ref{sec:local_eq} to \ref{sec:equivalence}, offers a detailed explanation of the structure of the averaged system under the so-called hybrid formulation. 
To illustrate our methodology, \ref{sec:averaged_surface} and \ref{sec:closure} examine the case of a dilute, viscous-dominated droplet suspension, incorporating effects due to non-uniform surface tension. 
We showed that surface tension gradients give rise to a momentum source, driving the Marangoni drift of the dispersed phase, and also introduce non-Newtonian behavior in the momentum equations. 
Additionally, we derived a method to compute the averaged droplet shape using the moment of momentum equations. 
A natural extension of this work involves introducing transport equations to describe the surfactant concentration or temperature field, thereby linking surface tension gradients to the corresponding scalar fields.

There has been limited effort to extend existing results to finite Reynolds number regimes that fully address the closure problem, despite the fact that most flows of interest exhibit significant inertial effects. 
\citet{stone2001inertial} showed that in a dilute suspension of solid spherical particles immersed in a general linear flow, finite inertia gives rise to normal stresses . This prediction was extended to the case of drops by \citet{raja2010inertial}. 
However, both studies focus exclusively on neutrally buoyant particles, thereby neglecting the influence of relative motion on the suspension rheology.
To begin exploring how inertia may influence both suspension rheology and droplet deformation, it is necessary to determine at least the first-order correction in $O(Re)$ to the closure terms in \ref{eq:dt_hybrid_Sp} and \ref{eq:dt_uf2}. 
Based on symmetry arguments - similar to those applied in  \ref{eq:Reynolds_stress_functional_form}- one may state that the functional form of the Stresslet due to to pure relative motion is given by, 
\begin{equation}
    \pSavg{\textbf{r}\bm\sigma^*_f\cdot \textbf{n}}
    -\pOavg{2\mu_f\textbf{e}^*_f}
    =
    Re \phi 
    \left[
       C_s^1 \textbf{u}_{r}\textbf{u}_{r} 
    +  C_s^2 (\textbf{u}_{r}\cdot \textbf{u}_{r})\bm\delta
    \right]
    + O(Re^2,\phi^2),
    \label{eq:final}
\end{equation}
at first order in Reynolds number. 
Here, $C_s^1$ and $C_s^2$ are functions of $\lambda$ that must be determined.
We conclude that at finite Reynolds values, relative motions between droplets induce additional stresses in the suspension, scaling as $\sim \textbf{u}_r\textbf{u}_r$, in addition to the Reynolds stress contribution given in \eqref{eq:Reynolds_stress_functional_form}. 
These relative motions also lead to droplet deformation, consistent with the predictions of \eqref{eq:dt_hybrid_Sp} and in agreement with the results of \citet{taylor1964deformation}.
In an ongoing work, we explore how droplet translation at finite Reynolds numbers affects the Stresslet. 
In particular, we determine the exact values of the constants $C_s^{1}$ and $C_s^{2}$ appearing in \ref{eq:final} (see also \citet{fintzi2025}).

\section*{Acknowledgement}
It is a great pleasure to dedicate this paper to Daniel Lhuillier whose work on the "hybrid formalism" has greatly influenced the vision of both co-authors and the present work. 
As a scientist, Daniel Lhuillier serves as a mentor for young researchers. 
He constantly challenges us to question the significance and novelty of our findings and encourages us to explore papers that the authors may have overlooked.
His constant guidance have also been a source of motivation to complete this work.
The authors are also grateful for many in-depth discussions with Professor St\'ephane Popinet as well.

\appendix

\section{Topological equations for $\delta_\Gamma$}
\label{ap:delta_I}
In this appendix, we derive some useful relation related to $\delta_\Gamma$. In the sense of distribution \citep{appel2007}, 

\begin{equation}
<\grad \delta_\Gamma,\varphi> = - <\delta_\Gamma,\grad \varphi>.
\end{equation}
Hence,
\begin{equation}
<\grad \delta_\Gamma,\varphi> = - \int_{\Gamma} \grad \varphi d\Gamma.
\label{eq:1grad_delta_I}
\end{equation}
Since $\textbf{n}$ is a unit vector, we have $\grad  \textbf{n} \cdot \textbf{n} =\bm 0$ and$\grad \varphi= \grad  (\varphi \textbf{n}) \cdot \textbf{n},$
which yields 
\begin{equation}
<\grad \delta_\Gamma,\varphi> = - \int_{\Gamma} \grad  (\varphi \textbf{n}) \cdot \textbf{n} d\Gamma. 
\end{equation}
By applying the result of multiplying a distribution by a function we get \citep{appel2007} \begin{equation}
<\grad \delta_\Gamma,\varphi> = <\grad  (\delta _I \textbf{n}) \cdot \textbf{n},\varphi>.
\end{equation}
With a slight abuse of notation the preceding relation may be rewritten as
\begin{equation}
    \grad\delta_\Gamma 
    =   \grad  (\textbf{n} \delta_\Gamma) \cdot \textbf{n}
    \label{eq:grad_delta_I_app}
\end{equation}
Another relation may be obtained for the surface gradient operator of $\delta_\Gamma$. By definition,
\begin{equation}
  \gradI \delta_\Gamma  = \grad\delta_\Gamma - \textbf{nn}\cdot\grad\delta_\Gamma
\end{equation}
Since, $\textbf{nn}\cdot\grad\delta_\Gamma = \grad  (\delta_\Gamma\textbf{n})\cdot \textbf{n}-\grad \textbf{n}\cdot \textbf{n} \delta_\Gamma$ and using \ref{eq:grad_delta_I_app} we obtain
\begin{equation}
  \gradI \delta_\Gamma  = (\grad \textbf{n}\cdot \textbf{n}) \delta_\Gamma
\label{eq:gradI_deltaI}
\end{equation}

\section{Generalized form of the conservation equation on the interface}
\label{ap:interface_proof}
In this appendix we derive \ref{eq:dt_delta_I_f_I}. Multiplying \ref{eq:dt_f_I} by $\delta_\Gamma$ yields
\begin{equation}
    \delta_\Gamma
    \left[ \pddt f_\Gamma^0 
    + f_\Gamma^0 (\textbf{u}_\Gamma^0\cdot \textbf{n})  (\div \textbf{n})
    +\divI
    (f_\Gamma^0 \textbf{u}_{\Gamma||}^0
    - \mathbf{\Phi}_{\Gamma||}^0 )
    \right]
    = \delta_\Gamma s_I^0
    - \delta_\Gamma\Jump{
    f_k^0 (\textbf{u}_I^0 - \textbf{u}_k^0)
    + \mathbf{\Phi}_k^0}.
    \label{eq:delta_I_step1}
\end{equation}
We first focus on the first two terms on the left-hand side of \ref{eq:delta_I_step1}. Since, $\delta_\Gamma\pddt f_\Gamma^0 = \pddt (f_\Gamma^0\delta_\Gamma) - f_\Gamma^0\pddt\delta_\Gamma$ and using \ref{eq:dt_delta_I} yields
\begin{equation}
\delta_\Gamma
    \left[ \pddt f_\Gamma^0 
    + f_\Gamma^0 (\textbf{u}_\Gamma^0\cdot \textbf{n})  (\div \textbf{n}) \right] = \pddt (f_\Gamma^0\delta_\Gamma) + f_\Gamma^0\div [(\textbf{u}_I^0\cdot \textbf{n}) \textbf{n}\delta_\Gamma ].
\end{equation}
Since the gradient of $f_\Gamma^0$ lies in the plane parallel to the interface, we have  $\grad f_\Gamma^0 \cdot \textbf{n} = 0$. Consequently, the expression $f_\Gamma^0\div [(\textbf{u}_I^0\cdot \textbf{n}) \textbf{n}\delta_\Gamma ]$ simplifies to $\div [(\textbf{u}_I^0\cdot \textbf{n}) \textbf{n}f_\Gamma^0\delta_\Gamma ]$ which yields 
\begin{equation}
\delta_\Gamma
    \left[ \pddt f_\Gamma^0 
    + f_\Gamma^0 (\textbf{u}_\Gamma^0\cdot \textbf{n})  (\div \textbf{n}) \right] = \pddt (f_\Gamma^0\delta_\Gamma) + \div [(\textbf{u}_I^0\cdot \textbf{n}) \textbf{n}f_\Gamma^0\delta_\Gamma ].
\label{eq:delta_I_step1_2}
\end{equation}
Next, we turn our attention to the third term on the left-hand side of \ref{eq:delta_I_step1}. By applying the chain rule
\begin{multline}
    \delta_\Gamma \divI (f_\Gamma^0 \textbf{u}_{\Gamma||}^0
    - \mathbf{\Phi}_{\Gamma||}^0 ) = 
    \div [\delta_\Gamma (f_\Gamma^0 \textbf{u}_{\Gamma||}^0
    - \mathbf{\Phi}_{\Gamma||}^0 )] \\
    - \textbf{n}(\textbf{n}\cdot\grad)\cdot [\delta_\Gamma (f_\Gamma^0 \textbf{u}_{\Gamma||}^0
    - \mathbf{\Phi}_{\Gamma||}^0 )]
    - (f_\Gamma^0 \textbf{u}_{\Gamma||}^0
    - \mathbf{\Phi}_{\Gamma||}^0 )\cdot\gradI\delta_\Gamma.
\label{eq:delta_I_step2}
\end{multline}
Since $ (f_\Gamma^0 \textbf{u}_{\Gamma||}^0 - \mathbf{\Phi}_{\Gamma||}^0 )\cdot\gradI\delta_\Gamma  = (f_\Gamma^0 \textbf{u}_\Gamma^0 - \mathbf{\Phi}_{I}^0 )\cdot\gradI\delta_\Gamma$ and using \ref{eq:gradI_deltaI} we may rewrite the last term of \ref{eq:delta_I_step2} as $(f_\Gamma^0 \textbf{u}_\Gamma^0 - \mathbf{\Phi}_{I}^0 )\cdot(\grad \textbf{n}\cdot \textbf{n}) \delta_\Gamma$. This yields,
\begin{multline}
    \delta_\Gamma \divI (f_\Gamma^0 \textbf{u}_{\Gamma||}^0
    - \mathbf{\Phi}_{\Gamma||}^0 ) = 
    \div [\delta_\Gamma (f_\Gamma^0 \textbf{u}_{\Gamma||}^0
    - \mathbf{\Phi}_{\Gamma||}^0 )] \\
    - \textbf{n}(\textbf{n}\cdot\grad)\cdot [\delta_\Gamma (f_\Gamma^0 \textbf{u}_{\Gamma||}^0
    - \mathbf{\Phi}_{\Gamma||}^0 )]
    - (f_\Gamma^0 \textbf{u}_\Gamma^0 - \mathbf{\Phi}_{I}^0 )\cdot(\grad \textbf{n}\cdot \textbf{n}) \delta_\Gamma.
\label{eq:delta_I_step3}
\end{multline}
By noticing that $f_\Gamma^0 \textbf{u}_{\Gamma||}^0
    - \mathbf{\Phi}_{\Gamma||}^0 = (\bm\delta - \textbf{nn})\cdot (f_\Gamma^0 \textbf{u}_\Gamma^0
    - \mathbf{\Phi}_{I}^0 )$ we can expand the second term on the right hand side of \ref{eq:delta_I_step3} which yields
\begin{equation}
    - \textbf{n}(\textbf{n}\cdot\grad)\cdot [\delta_\Gamma (f_\Gamma^0 \textbf{u}_{\Gamma||}^0
    - \mathbf{\Phi}_{\Gamma||}^0 )]
    = (f_\Gamma^0 \textbf{u}_\Gamma^0 - \mathbf{\Phi}_{I}^0 )\cdot(\grad \textbf{n}\cdot \textbf{n})\delta_\Gamma \label{eq:delta_I_step4}
\end{equation}
To derive the last equation, we used the following properties in indicial notation: $n_in_j(\delta_{ik}-n_in_k) =0$ and $n_in_j\partial_{j}(n_in_k)=\partial_{j}(n_k)n_j$. 
Inserting \ref{eq:delta_I_step4} in \ref{eq:delta_I_step3} yields
\begin{equation}
    \delta_\Gamma \divI (f_\Gamma^0 \textbf{u}_{\Gamma||}^0
    - \mathbf{\Phi}_{\Gamma||}^0 ) = 
    \div [\delta_\Gamma (f_\Gamma^0 \textbf{u}_{\Gamma||}^0
    - \mathbf{\Phi}_{\Gamma||}^0 )]
\label{eq:delta_I_step5}
\end{equation}
Inserting \ref{eq:delta_I_step1_2} and \ref{eq:delta_I_step5} in \ref{eq:delta_I_step1} gives

\begin{equation}
    \pddt (\delta_\Gamma f_\Gamma^0)  
    + \div (
        \delta_\Gamma f_\Gamma^0 \textbf{u}_I^0
        - \delta_\Gamma \mathbf{\Phi}_{\Gamma||}^0 
        )
    = 
    \delta_\Gamma s_I^0
    - \delta_\Gamma\Jump{
    f_k^0 (\textbf{u}_I^0 - \textbf{u}_k^0)
    + \mathbf{\Phi}_k^0},
\end{equation}
which is the final form of the conservation equation on the interface.

\section{Arbitrary order moments equations}
\label{ap:Moments_equations}
In this appendix, we extend the Lagrangian conservation laws to an arbitrary order moment equation. 
Let us first define the arbitrary moment of the Eulerian field $f_d^0$, in index notation it reads, 
\begin{equation*}
    (\textbf{q}_\alpha^{(n)})_{i_1\ldots i_n}
    = \intO{
    \pri{1}{n} f_d^0 
    }, 
\end{equation*}
where we recall that $r_{i_m} = x_{i_m} - (\textbf{x}_\alpha)_{i_m}$. 
Applying the Reynolds transport theorem \eqref{eq:reynolds_transport} we show that :
\begin{multline}
    \ddt {(\textbf{q}_\alpha^{(n)})_{i_1\ldots i_n}}
    =\intO{
        \left[ \partial_t \left(f_d^0 \pri{1}{n}\right) 
    + \div \left(\textbf{u}_d^0 f_d^0 \pri{1}{n}\right) \right]
    }\\
    +\intS{ \pri{1}{n} f_d^0 \left(\textbf{u}_\Gamma^0 - \textbf{u}_d^0\right)\cdot \textbf{n}_d }. 
\end{multline}
Applying the product rule to the terms inside the first integral results in the following expression: 
\begin{multline}
    \ddt {(\textbf{q}_\alpha^{(n)})_{i_1\ldots i_n}}
    =\intO{ 
        f_d^0 \left[ \partial_t \left(\pri{1}{n}\right) 
        + (\textbf{u}_d^0\cdot \grad) \left( \pri{1}{n}\right) \right]
    }\\
    +\intO{ 
        \pri{1}{n} 
        \left[ \partial_t f_d^0
    +  \div \left(\textbf{u}_d^0 f_d^0 \right) \right]
    }
    +\intS{ \pri{1}{n} f_d^0 \left(\textbf{u}_\Gamma^0 - \textbf{u}_d^0\right)\cdot \textbf{n}_d }. 
\end{multline}
By applying the product rule, once again, to the terms within the first integral on, noting that $\pddt \textbf{r}_{i_e} = (\textbf{w}_d^0)_{i_e}$, and utilizing the conservation equation \eqref{eq:dt_f_k} on the second integral, leads us to the relation: 
\begin{multline}
    \ddt {(\textbf{q}_\alpha^{(n)})_{i_1\ldots i_n}}
    = \sum_{e=1}^{n} \intO{ 
        f_d^0 \prod^{n}_{\substack{ m=1 \\   m \neq e}} r_{i_m} (\textbf{w}_d^0)_{i_e}
        }
    +\intO{\pri{1}{n} (\div\bm{\Phi}_d^0)}\\
    + \intO{ \pri{1}{n} s_d^0}
    +\intS{ \pri{1}{n} f_d^0 \left(\textbf{u}_\Gamma^0 - \textbf{u}_d^0\right)\cdot \textbf{n}_d }.
\end{multline}
The second term on the right-hand side of this equation can be reformulated as,
\begin{align*}
    \intO{ \pri{1}{n} (\div\bm\Phi_d^0) }
    &= \intO{ \div \left(\pri{1}{n} \bm\Phi_d^0 \right)}
    - \intO{ \bm\Phi_d^0 \cdot \grad \left(\pri{1}{n} \right)}\\
    &= \intS{ \pri{1}{n} (\bm\Phi_d^0 \cdot \textbf{n}_d)}
    -\sum_{e=1}^{n} 
    \intO{ (\bm\Phi_d^0)_{i_e}  \prod^{n}_{\substack{ m=1 \\m \neq e}} r_{i_m}  }
\end{align*}
Including this relation into the former equation yields, 
\begin{multline}
    \ddt {(\textbf{q}_\alpha^{(n)})_{i_1\ldots i_n}}
    = \sum_{e=1}^{n} 
    \intO{
        \prod^{n}_{\substack{ m=1 \\m \neq e}} r_{i_m} (f_d^0 \textbf{w}_d^0  - \bm\Phi_d^0)_{i_e}
    }
+ \intO{ \pri{1}{n} s_d^0 }\\
    +\intS{ \pri{1}{n} [\bm\Phi_d^0 + f_d^0 \left(\textbf{u}_\Gamma^0 - \textbf{u}_d^0\right)]\cdot \textbf{n}_d }.
    \label{eq:dt_q_n}
\end{multline}
This represents the final form of the Lagrangian conservation equation for the $n^{th}$ order moment of the quantity $f_d^0$ within the particle. 
When $f_d^0$ is a scalar quantity, this expression indicates that $\ddt \textbf{q}_\alpha^{(n)}$ is entirely symmetric since it involves only sums of products of the $r_{i_n}$ times a scalar quantity. 
Regarding the seemingly non-symmetric terms involving $\textbf{w}_d^0$ and $\bm\Phi_d^0$, it is important to note that, due to the summation operator's presence, these terms also become symmetric.

Regarding the surface property conservation equations, the derivation is similar and will not be displayed here. 
The result yields, 
\begin{multline}
    \ddt {(\textbf{q}_{\alpha\Gamma}^{(n)})_{i_1\ldots i_n}}
    = \sum_{e=1}^{n} 
    \intS{
        \prod^{n}_{\substack{ m=1 \\m \neq e}} r_{i_m} (f_\Gamma^0\textbf{w}_\Gamma^0 - \bm\Phi_{||\Gamma}^0)_{i_e}
    }
    + \intS{ \pri{1}{n} (\textbf{s}_\Gamma^0)_k }
    \\
    +\intS{ \pri{1}{n} \Jump{\bm\Phi_k^0 + f_k^0 \left(\textbf{u}_\Gamma^0 - \textbf{u}_k^0\right)\cdot \textbf{n}_d}}.
    \label{eq:dt_Qgamma_n}
\end{multline}
where we have defined, 
\begin{equation*}
    (\textbf{q}_{\alpha\Gamma}^{(n)})_{i_1\ldots i_n}
    = \intS{
    \pri{1}{n} f_I^0 
    }. 
\end{equation*}

Summing \ref{eq:dt_q_n} and \ref{eq:dt_Qgamma_n} one obtains the equation for the total $n^{th}$ order moment, $\textbf{Q}_{\alpha}^{(n)} = \textbf{q}_{\alpha\Gamma}^{(n)}+\textbf{q}_{\alpha}^{(n)}$, namely, 
\begin{multline}
    \ddt {(\textbf{Q}_{\alpha}^{(n)})_{i_1\ldots i_n}}
    = 
    \sum_{e=1}^{n} 
    \intO{
        \prod^{n}_{\substack{ m=1 \\m \neq e}} r_{i_m} (f_d^0\textbf{w}_d^0  - \bm\Phi_d^0)_{i_e}
    }
    + \intO{ \pri{1}{n} (\textbf{s}_d^0)_k }\\
    +     
    \sum_{e=1}^{n} 
    \intS{
        \prod^{n}_{\substack{ m=1 \\m \neq e}} r_{i_m} (f_\Gamma^0\textbf{w}_\Gamma^0 - \bm\Phi_{||\Gamma}^0)_{i_e}
    }
    + \intS{ \pri{1}{n} (\textbf{s}_\Gamma^0)_k }
    \\
    +\intS{ \pri{1}{n} ([\bm\Phi_f^0 + \textbf{f}_f^0 \left(\textbf{u}_\Gamma^0 - \textbf{u}_f^0\right)]\cdot \textbf{n}_d)_k }. 
    \label{eq:dt_Q_n}
\end{multline}

By averaging \ref{eq:dt_Q_n}, we obtain the particle averaged equation for $\pavg{\textbf{Q}_\alpha^{(n)}} = n_p\textbf{Q}_p^{(n)} $, it yields,
\begin{multline}
    \pddt \pavg{(\textbf{Q}_\alpha^{(n)})_{i_1\ldots i_n}^\alpha}
    + \div  \pavg{\textbf{u}_\alpha (\textbf{Q}_\alpha^{(n)})_{i_1\ldots i_n}^\alpha}
    = \sum_{e=1}^{n} 
    \pOavg{
        \prod^{n}_{\substack{ m=1 \\m \neq e}} r_{i_m} (f_d^0\textbf{w}_d^0  - \bm\Phi_d^0)_{i_e}
    }\\
    + \pOavg{ \pri{1}{n} (\textbf{s}_d^0)_k }
    +     
    \sum_{e=1}^{n} 
    \pSavg{
        \prod^{n}_{\substack{ m=1 \\m \neq e}} r_{i_m} (f_\Gamma^0\textbf{w}_\Gamma^0 - \bm\Phi_{||\Gamma}^0)_{i_e}
    }\\
    + \pSavg{ \pri{1}{n} (\textbf{s}_\Gamma^0)_k }
    +\pSavg{ \pri{1}{n} ([\bm\Phi_f^0 + \textbf{f}_f^0 \left(\textbf{u}_\Gamma^0 - \textbf{u}_f^0\right)]\cdot \textbf{n}_d)_k }. 
    \label{eq:avg_hybrid_q_n}
\end{multline}

 \section{Expansion of phase and surface quantities as series of moments}\label{app:expansion}

We first provide a concise proof of the following relation
\begin{equation}
    \delta(\textbf{x} - \textbf{x}_\alpha - \textbf{r})
    = \delta(\textbf{x} - \textbf{x}_\alpha)
    - \textbf{r}\cdot\grad \delta(\textbf{x} - \textbf{x}_\alpha)
    + \frac{1}{2}\textbf{rr}:\grad\grad\delta(\textbf{x} - \textbf{x}_\alpha) 
    - \ldots
\label{eq:exp_delta}
\end{equation}
which is widely used in the literature (see for example \citet{zhang2023evolution}). 
We begin by applying the Dirac delta function $\delta(\textbf{x} - \textbf{x}_\alpha - \textbf{r})$, to a test function $\varphi(\textbf{x})$.
This results in the following expression,
\begin{align*}
    < \delta(\textbf{x} - \textbf{x}_\alpha - \textbf{r}), \varphi(\textbf{x})>
    =
    \varphi(\textbf{x}_\alpha + \textbf{r}).  
\end{align*}
By applying the Taylor expansion to the test function $\varphi(\textbf{x}_\alpha + \textbf{r})$ around the point $\textbf{r} = 0$ , we obtain the following expression
\begin{equation}
    <\delta(\textbf{x} - \textbf{x}_\alpha - \textbf{r}), \varphi(\textbf{x})>
    =
    \varphi(\textbf{x}_\alpha) 
    + \textbf{r} \cdot \grad \varphi|_{\textbf{x}_\alpha}
    + \frac{1}{2}\textbf{r}\textbf{r} : \grad\grad \varphi|_{\textbf{x}_\alpha}
    + \ldots
    \label{eq:first_step}
\end{equation}
By the definition of the Dirac delta function and its derivative \citep{appel2007}, we have  $  (\delta_\alpha, \varphi) = \varphi(\textbf{x}_\alpha)$, $(\grad\delta_\alpha, \varphi) = -\grad\varphi|_{\textbf{x}_\alpha} $,
$...$  
Substituting these definitions into \ref{eq:first_step} , we obtain in the distributional sense \ref{eq:exp_delta}.

Let us now turn our attention to the demonstration of \ref{eq:fd_asympt0}. 
We can show that, 
\begin{align}
    <(f_d^0\chi_\alpha)(\textbf{x}),\varphi(\textbf{x})>
    &= <(f_d^0\chi_\alpha)(\textbf{x}_\alpha+\textbf{r}),\varphi(\textbf{x}_\alpha+\textbf{r})> \label{eq:first_equality}\\
    &= 
    <(f_d^0\chi_\alpha)(\textbf{x}_\alpha + \textbf{r}) ,<\delta(\textbf{x} - \textbf{x}_\alpha - \textbf{r}), \varphi(\textbf{x})>>.
    \label{eq:second_equality}
\end{align}
The first equality follows directly from the change of variables $\textbf{x}=\textbf{x}_\alpha + \textbf{r}$. 
The second equality holds by the definition of the Dirac delta functions and of the convolution product in the distributional sense \citep{appel2007}.
Substituting \ref{eq:exp_delta}  into the previous equation gives, 
\begin{equation}
    f^0_d \chi_\alpha
    = 
    \delta_\alpha
    \intO{
        f^0_d
    }
    - \div\left(    
    \delta_\alpha
    \intO{
    \textbf{r}
    f^0_d}
    \right)
    + \frac{1}{2}\grad\grad :\left(\delta_\alpha\intO{\textbf{rr} f^0_d}\right)
    - \ldots
\end{equation}
The previous proof can be easily extended to surface quantities by following the same steps as in \ref{eq:first_equality} and \ref{eq:second_equality},  leading to
\begin{equation} 
    <f_\Gamma^0 \delta_{\Gamma\alpha},\varphi(\textbf{x})> = <(f_\Gamma^0 \delta_{\Gamma\alpha})(\textbf{x}_\alpha + \textbf{r}) ,<\delta(\textbf{x} - \textbf{x}_\alpha - \textbf{r}), \varphi(\textbf{x})>>,
\end{equation}
from which we obtain
\begin{equation} 
f_\Gamma^0 \delta_{\Gamma\alpha} 
=\delta_\alpha\intS{f^0_\Gamma}
- \div\left(\delta_\alpha\intS{\textbf{r} f^0_\Gamma}\right)
+ \frac{1}{2}\grad\grad :\left(\delta_\alpha\intS{\textbf{rr} f^0_\Gamma}\right)
-\ldots 
\end{equation}

 \section{Arbitrary order equivalence}
\label{ap:equivalence}
In this appendix, we provide a general proof of \ref{eq:scheme_equivalence}. 
Let's begin by re-writing the phase averaged equation:
\begin{equation}
        \pddt \avg{\chi_d f_d^0}
        = \div \avg{\chi_d \bm\Phi_d^0 - \chi_d f_d^0 \textbf{u}_d^0}
        + \avg{\chi_d s_d^0}
        + \avg{\delta_\Gamma\left[
            \bm\Phi_d^0
            + f_d^0
            \left(
                \textbf{u}_\Gamma^0
                - \textbf{u}_d^0
            \right)
        \right]
        \cdot \textbf{n}_d}.
        \label{eq:dt_f_d_O}
\end{equation}
Our objective here is to demonstrate how \ref{eq:dt_f_d_O} is related to the Lagrangian moments' equation, given by \ref{eq:dt_q_n}. 
The first step is to expand each term of \ref{eq:dt_f_d_O} using the relation \ref{eq:f_exp} which directly gives,
\begin{align*}
        0 &=
        - \pddt \expo{f_d^0} \\
        &+\div \expo{(\bm\Phi_d^0  - f_d^0 \textbf{u}_d^0)}\\
        &+ \expo{ s_d^0}\\
        &+ \expoS{\left[
            \bm\Phi_d^0
            + f_d^0
            \left(
                \textbf{u}_\Gamma^0
                - \textbf{u}_d^0
            \right)
        \right]
        \cdot \textbf{n}_d}. \\
\end{align*}
The third term can be reformulated using the decomposition : $\textbf{u}_d^0 = \textbf{u}_\alpha + \textbf{w}_d^0$, which gives,
\begin{multline}
    \expo{f_d^0 \textbf{u}_d^0}\\
    =     \expoU{f_d^0 }\\
    +     \expo{f_d^0 \textbf{w}_d^0}.
\end{multline}
Injecting this formulation in the former equation yields,
\begin{align}
    & \pddt \expo{f_d^0} \\
    &+ \div \expoU{f_d^0}\\
    &= \div \expo{(\bm\Phi_d^0 - f_d^0 \textbf{w}_d^0)}\\
    &+ \expo{ s_d^0}\\
    &+ \expoS{\left[
        \bm\Phi_d^0
        + f_d^0 
        \left(
            \textbf{u}_I^0
            - \textbf{u}_d^0
        \right)
    \right]
    \cdot \textbf{n}_d}. \\
    \label{eq:nearly_done}
\end{align}
Then, note that the first term on the right-hand side can be re-written when evaluated at the order $n-1$, as follows,
\begin{multline*}
    \div \expo[{(n-1)}]{(\bm\Phi_d^0 - f_d^0 \textbf{w}_d^0)}
    = \\
    \frac{(-1)^{n}}{{(n)}!} \partialp{1}{n}  n \pOavg{ \pri{1}{n-1}(f_d^0 \textbf{w}_d^0 - \bm\Phi_d^0)_{i_{n}} }
\end{multline*} 
Upon using that expression in \ref{eq:nearly_done} we can factor out the gradient operators, i.e. the $\frac{(-1)^n}{n!} \partialp{1}{n}$, which gives, 
\begin{multline}
    0 = \frac{(-1)^n}{n!}
    \partialp{1}{n}
    \left[
        - \partial_t
        \pavg{(\textbf{q}_\alpha^{(n)})_{i_1\ldots i_n}}
        - \div \pavg{\textbf{u}_\alpha (\textbf{q}_\alpha^{(n)})_{i_1\ldots i_n}}
    \right.\\\left.
        +n\pavg{\int_{\Omega_\alpha} \pri{1}{n-1} (f_d^0 \textbf{w}_d^0-\bm\Phi_d^0) d\Omega}
        +\pavg{\int_{\Omega_\alpha} \pri{1}{n} s_d^0 d\Omega}
        \right.\\\left.
        +\pavg{\int_{\Omega_\alpha} \pri{1}{n} \left[
            \bm\Phi_d^0
            + f_d^0
            \left(
                \textbf{u}_I^0
                - \textbf{u}_d^0
            \right)
        \right]
        \cdot \textbf{n}_d d\Omega}
    \right].
    \label{eq:exp_f_d_O}
\end{multline}
At this stage, one might immediately recognize \ref{eq:dt_q_n} in the square bracket. 
However, note that the third term of \ref{eq:exp_f_d_O} differs from the second term of \ref{eq:dt_q_n}. 
Indeed, 
\begin{equation}
    n\pavg{\int_{\Omega_\alpha} \pri{1}{n-1} ( f_d^0 \textbf{w}_d^0-\bm\Phi_d^0)_{i_n} d\Omega}
    \neq
    \sum_{e=1}^{n} 
    \avg{
        \intO{
        \prod^{n}_{\substack{ m=1 \\m \neq e}} r_{i_m} (f_d^0 \textbf{w}_d^0  - \bm\Phi_d^0)_{i_e}
        }
    }. 
    \label{eq:ineq_which_does_not_make_sens}
\end{equation}
Even if the above inequality holds as it is, it must be understood that what matter in \ref{eq:exp_f_d_O} is the gradient of that term. 
Additionally, the contraction of one of these terms with the gradient operator $\partialp{1}{n}$ makes the skew-symmetric part of these tensors (in the indices $i_1\ldots i_n$) having a vanishing contribution to the expression.
Thus, one might apply the operator $\partialp{1}{n}$ on each side of \ref{eq:ineq_which_does_not_make_sens} and by noticing that the skew-symmetric part of the left-hand side of \ref{eq:ineq_which_does_not_make_sens} vanish, this leads to, 
\begin{multline}
    \partialp{1}{n}\left[
        n\pavg{\int_{\Omega_\alpha} \pri{1}{n-1} ( f_d^0 \textbf{w}_d^0-\bm\Phi_d^0)_{i_n} d\Omega}
        \right]\\
    =
    \partialp{1}{n}\left[
    \sum_{e=1}^{n} 
    \avg{
        \intO{
        \prod^{n}_{\substack{ m=1 \\m \neq e}} r_{i_m} (f_d^0 \textbf{w}_d^0  - \bm\Phi_d^0)_{i_e}
        }
    }
    \right]. 
    \label{eq:it_make_sens_again}
\end{multline}
Injecting this last equality in \ref{eq:exp_f_d_O} and using \ref{eq:dt_Qgamma_n} to reformulate the interfacial term directly gives, 
\begin{multline}
    0 = \frac{(-1)^n}{n!}
    \partialp{1}{n}
    \left[
        - \pddt \pavg{(\textbf{Q}_\alpha^{(n)})_{i_1\ldots i_n}^\alpha}
        - \div  \pavg{\textbf{u}_\alpha (\textbf{Q}_\alpha^{(n)})_{i_1\ldots i_n}^\alpha}
        = \phantom{\pOavg{\pri{}{n}}}\right.\\\left.
            +\sum_{e=1}^{n} 
        \pOavg{
            \prod^{n}_{\substack{ m=1 \\m \neq e}} r_{i_m} (f_d^0\textbf{w}_d^0  - \bm\Phi_d^0)_{i_e}
        }
        + \pOavg{ \pri{1}{n} (\textbf{s}_d^0)_k }\right.\\\left.
        +     
        \sum_{e=1}^{n} 
        \pSavg{
            \prod^{n}_{\substack{ m=1 \\m \neq e}} r_{i_m} (f_\Gamma^0\textbf{w}_\Gamma^0 - \bm\Phi_{||\Gamma}^0)_{i_e}
        }
        + \pSavg{ \pri{1}{n} (\textbf{s}_\Gamma^0)_k } \right.\\\left.
        +\pSavg{ \pri{1}{n} ([\bm\Phi_f^0 + \textbf{f}_f^0 \left(\textbf{u}_\Gamma^0 - \textbf{u}_f^0\right)]\cdot \textbf{n}_d)_k }
    \right],
\end{multline}
which proves \ref{eq:scheme_equivalence}.
 \section{Singularity solution for droplets with surface tension gradient}
\label{ap:singularity_solution}
At the leading order in droplet volume fraction, the closure problem is equivalent to that of an isolated droplet in an infinite medium \citet{hinch1977averaged}. 
Hence, we consider the problem of an isolated droplet, immersed in an arbitrary quadratic flow without the presence of fluid inertia. 
The disturbances pressure and velocity field are noted $\textbf{u}_{out}$, $\textbf{u}_{in}$, $p_{out}$ and $p_{in}$, for the velocity outside the droplet, the velocity inside the droplet, the pressure outside the droplet and the pressure inside the droplet, respectively. 
The corresponding ``undisturbed field'' are noted $\textbf{u}$ and $p_f$, corresponding to the ensemble averaged mixture velocity and averaged pressure of the continuous phase in the main text. 

Under these hypotheses, in dimensionless form, $\textbf{u}_{out}$, $\textbf{u}_{in}$, $p_i$ and $p_o$ are governed by the Stokes equations, namely, 
\begin{align}
    \div \textbf{u}_{in} &= 0 
    & \div \textbf{u}_{out} &= 0 \\
     \grad^2 \textbf{u}_{in}  &= \grad p_{in} 
    & \grad^2 \textbf{u}_{out} &= \grad p_{out} 
\end{align}
At the surface of the droplet ($r=1$) the condition of continuity of velocity and continuity of the normal stresses read, 
\begin{align}
    \label{eq:vel_jump}
    \textbf{u}_{out} - \textbf{u}_{in} &= 0\\
\label{eq:vel_jump2}
    (\textbf{u}_{in}  + \textbf{u}_r)\cdot \textbf{n}
    &= 0\\
    \mathbf{n}\cdot 2(\textbf{e}_{out} - \lambda \textbf{e}_{in}+\textbf{E} -\lambda\textbf{E} )\cdot (\bm\delta - \textbf{nn})
    &= 
    \textbf{b}\cdot (\bm\delta - \textbf{nn})
    \label{eq:stress_jump}
\end{align}
where $\textbf{u}_r = \textbf{u} - \textbf{u}_p$ corresponds to the velocity of the mixture with respect to the droplet mean velocity. 
We recall that $\textbf{e}_{in/out} = (\grad \textbf{u}_{in/out} + ^\dagger \grad \textbf{u}_{in/out})/2 $ and $\textbf{E} = (\grad \textbf{u} + ^\dagger \grad \textbf{u})/2$, which correspond to the rate of strain inside or around or droplet, and to the ensemble averaged rate of strain, respectively. 
In \ref{eq:stress_jump} the vector \textbf{b} correspond to the dimensionless tangential stress jump across the interface and reads, 
\begin{equation}
    \textbf{b}
    =
    \frac{a \grad \gamma}{\mu_f U}
\end{equation}
Far from the droplet (centered at $\textbf{r}=\textbf{0}$) the velocity and pressure fields satisfy, 
\begin{equation}
    \lim_{r\to \infty}(\textbf{u}_{out},p_{out}) = 0. 
\end{equation}

In general $\textbf{b}$ as well as $\textbf{u}$ are not constant vectors and vary across space as a function of \textbf{r}. 
However, it can be assumed that these quantities are slowly varying functions of space at the scale of a droplet radius, hence we may introduce the simplifications 
\begin{align*}
    \textbf{u}(\textbf{n}) 
    &=  \textbf{u}_r|_{\textbf{r}=0}
    +  \textbf{r} \cdot  \grad\textbf{u}|_{\textbf{r}=0}
    +  \frac{1}{2}\textbf{rr} :  \grad\grad\textbf{u}|_{\textbf{r}=0}
    + \ldots\\
     \textbf{E}(\textbf{n}) 
    &=   \textbf{E}|_{\textbf{r}=0}
    + \textbf{r} \cdot  \grad \textbf{E}|_{\textbf{r}=0}
    + \frac{1}{2}\textbf{rr} :  \grad\grad \textbf{E}|_{\textbf{r}=0}
    + \ldots\\
     \textbf{b}(\textbf{n}) 
    &=   \textbf{b}|_{\textbf{r}=0}
    + \textbf{r} \cdot  \grad \textbf{b}|_{\textbf{r}=0}
    + \frac{1}{2}\textbf{rr} :  \grad\grad \textbf{b}|_{\textbf{r}=0}
    + \ldots
\end{align*}
Note that because $\textbf{b} \sim \grad \gamma$ the tensor: $\grad \textbf{b}$ and $\grad\grad \textbf{b}$ are symmetric tensors. 
Injecting, these expressions into \ref{eq:vel_jump,eq:vel_jump2,eq:stress_jump} one deduce that the velocity and pressure fields are linearly related to $\textbf{u}|_{\textbf{r}=0}$, $\textbf{b}|_{\textbf{r}=0}$, and to the gradients of these vectors. 

According to the linearity of the Stokes equations we deduce that $\textbf{u}_{i/o}$ and $p_{i/o}$  must be linear combination of spherical harmonics proportional to $\textbf{u}|_{\textbf{x}=0}$, $\grad \gamma|_{\textbf{x}=0}$, and their derivatives.
Hence, we can write that the disturbance velocity and pressure fields are given by, 
\begin{align*}
    \begin{pmatrix}
        \textbf{u}_{o}\\
        p_{o}\\
        \textbf{u}_{i}\\
        p_{i}
    \end{pmatrix}
    =
    \begin{pmatrix}
        \textbf{U}_{o}^{(1)} + \textbf{U}_{o}^{(2)}\cdot \grad + \textbf{U}_{o}^{(3)} :\grad\grad &
        \textbf{U}_{o}^\text{(b-1)} + \textbf{U}_{o}^\text{(b-2)}\cdot \grad + \textbf{U}_{o}^\text{(b-3)} :\grad\grad \\
        \textbf{P}_{o}^{(1)} + \textbf{P}_{o}^{(2)}\cdot \grad + \textbf{P}_{o}^{(3)} :\grad\grad &
        \textbf{P}_{o}^\text{(b-1)} + \textbf{P}_{o}^\text{(b-2)}\cdot \grad + \textbf{P}_{o}^\text{(b-3)} :\grad\grad \\
        \textbf{U}_{i}^{(1)} + \textbf{U}_{i}^{(2)}\cdot \grad + \textbf{U}_{i}^{(3)} :\grad\grad &
        \textbf{U}_{i}^\text{(b-1)} + \textbf{U}_{i}^\text{(b-2)}\cdot \grad + \textbf{U}_{i}^\text{(b-3)} :\grad\grad \\
        \textbf{P}_{i}^{(1)} + \textbf{P}_{i}^{(2)}\cdot \grad + \textbf{P}_{i}^{(3)} :\grad\grad &
        \textbf{P}_{i}^\text{(b-1)} + \textbf{P}_{i}^\text{(b-2)}\cdot \grad + \textbf{P}_{i}^\text{(b-3)} :\grad\grad \\
    \end{pmatrix}
    \cdot 
    \begin{pmatrix}
        \textbf{u}_r\\
        \textbf{b}
    \end{pmatrix}
\end{align*}
The tensors $\textbf{U}^{(n)}_{o/i}$ (resp. $\textbf{P}^{(n)}_{o/i}$) are $(n+1)^{th}$ (resp. $n^{th}$) order tensors, solely function of \textbf{r} and the viscosity ratio $\lambda$. 
The tensors $\textbf{U}_{o/i}^{(1)}$ and $\textbf{P}_{o/i}^{(1)}$ represent the functional form of the famous Hadamard-Ribczynski solution \citep{pozrikidis1992boundary,kim2013microhydrodynamics}. 
The dependency of $\textbf{u}_{i/o}$ and $p_{i/o}$ with the mean gradient velocity $\grad \textbf{u}$ is given by the tensor $\textbf{U}_{o/i}^{(2)}$ and $\textbf{P}_{o}^{(2)}$.
The disturbance velocity and pressure fields of a drop immersed in pure linear flow can be found in \citet{rallison1978note,leal2007advanced,raja2010inertial}. 
Finally, the disturbance field of a droplet immersed in a pure quadratic flow is defined through the tensor $\textbf{U}_{o/i}^{(3)}$ and $\textbf{P}_{o/i}^{(3)}$.
The definition of the former tensors may be found in \citet{nadim1991motion}.

The tensors $\textbf{U}^{(b-n)}_{o/i}$ (resp. $\textbf{P}^{(b-n)}_{o/i}$) are $(n+1)^{th}$ (resp. $n^{th}$) order tensors, solely function of \textbf{r} and the viscosity ratio $\lambda$ as well. 
The disturbance fields induced due to a constant, linear, or quadratic stress jump at the interface of the droplet is less common in the literature. 
For instance one might refer to \citet{Subramanian_1985,leal2007advanced} to compute the disturbance fields due to a constant surface tension gradient, hence defining the expression of  $\textbf{U}_{o/i}^{(b-1)}$ and $\textbf{P}_{o/i}^{(b-1)}$. 
Then to express the disturbance fields in terms of $\grad \textbf{b}$ or higher order derivatives one might refer to the solution given in \citet[Appendix C]{raja2010inertial}.
Without loss of generality we consider only the symmetric traceless part of $(\grad \textbf{b}_{jk})$ in the following. 
Based on the classic method of spherical harmonic \citep[chapter 8]{leal2007advanced} we could derive the expression for the disturbance fields due to an imposed $\grad\grad \textbf{b}$, it reads,
\begin{align*}
    (\textbf{U}_o^{(b-3)})_{ijkl}
    &=
    -\frac{7r^4+10r^2-9}{1260(\lambda+1)r^5}(
        \delta_{ij}\delta_{kl}
        + \delta_{ik}\delta_{jl}
        + \delta_{il}\delta_{kj}
        )
    -\frac{7r^4-60r^2+45}{1260(\lambda+1)r^7}(
        x_ix_j\delta_{kl}
        + x_ix_k\delta_{jl}
        + x_ix_l\delta_{kj}
        )\\
    &+ \frac{r^2 -3}{84(\lambda +1)r^7}(
        \delta_{ij}x_kx_l
        + \delta_{ik}x_jx_l
        + \delta_{il}x_kx_j
    )
    - \frac{5r^2 - 7}{28(\lambda+1)r^9}x_ix_jx_kx_l
    \\
    (\textbf{P}_o^{(b-3)})_{jkl}
    &=
    -\frac{7r^2 - 45}{630(\lambda+1)r^5}(
        \delta_{jk}x_l
        + \delta_{jl}x_k
        + \delta_{lk}x_j
    )
    - \frac{5}{14(\lambda+1)r^7}x_jx_kx_l
    \\
    (\textbf{U}_i^{(b-3)})_{ijk}
    &=
    \frac{9r^4-20r^2+7}{630(\lambda+1)}(
        \delta_{ij}\delta_{kl}
        + \delta_{ik}\delta_{jl}
        + \delta_{il}\delta_{kj}
    )
    +\frac{9r^2-5}{630(\lambda+1)}(
        x_ix_j\delta_{kl}
        + x_ix_k\delta_{jl}
        + x_ix_l\delta_{kj}
        )\\
    &- \frac{3r^2-2}{42(\lambda +1)}(
        \delta_{ij}x_kx_l
        + \delta_{ik}x_jx_l
        + \delta_{il}x_kx_j
    )
    + \frac{1}{14(\lambda+1)}x_ix_jx_kx_l
    \\
    (\textbf{P}_i^{(b-3)})_{jkl}
    &=
    \frac{54r^2 - 35}{315(\lambda+1)r}(
        \delta_{jk}x_l
        + \delta_{jl}x_k
        + \delta_{lk}x_j
    )
    - \frac{6}{7(\lambda+1)}x_jx_kx_l. 
\end{align*}
It is implied that each of these tensors must be contracted with the full symmetric tensor $(\grad\grad\textbf{b})_{ijk} \sim (\grad\grad\grad\gamma)_{ijk}$ to obtain the velocity or pressure fields. 

Hence, based on these expressions and the solutions  provided in the literature, one can compute the force traction at the surface of the droplet ($\bm\sigma_{out}\cdot \textbf{n}$), as well as the internal shear rate ($\textbf{e}_{in}$), and obtains the following expressions, 
\begin{align}
    \intS{\bm\sigma_{out}\cdot \textbf{n}} &
    =
    2\pi\frac{2+3\lambda}{1+\lambda}\textbf{u}_r
    + \pi \frac{\lambda}{\lambda +1} \grad^2 \textbf{u}
    +
    \frac{4\pi}{3}\frac{1}{\lambda +1}\textbf{b}
    + \frac{2\pi}{15(\lambda +1)}\grad^2\textbf{b}
    \\
    \intS{ \textbf{x}\bm\sigma_{out}\cdot \textbf{n}} &
    =
    \frac{2\pi(5\lambda +2)}{5(\lambda +1)}[\grad \textbf{u}+ (\grad \textbf{u})^\dagger]
    + \frac{12\pi}{25(\lambda +1)} (\grad \textbf{b} -\frac{1}{3}\bm\delta\div \textbf{b}) 
    \\
    \intS{\textbf{xx}\bm\sigma_{out}\cdot \textbf{n}} &
    =
    \frac{4\pi}{5(\lambda +1)} (\textbf{u}_r \bm\delta + ^\dagger\textbf{u}_r \bm\delta)
    + \frac{2\pi(5\lambda +2)}{5(\lambda+1)}\bm\delta \textbf{u}_r
    - \frac{8\pi}{15(\lambda +1)} ( \textbf{b}\bm\delta + ^\dagger\textbf{b}\bm\delta )
    + \frac{4\pi}{5(\lambda+1)}\bm\delta \textbf{b}
    \label{eq:nd_mom}
\end{align}
and, 
\begin{align}
    \intO{2\textbf{e}_{in}}
    &=
    -\frac{4\pi}{15}\frac{5\lambda +2}{\lambda+1}
    [\grad \textbf{u}+ (\grad \textbf{u})^\dagger]
    - \frac{8\pi}{25}\frac{1}{\lambda+1}
    (\grad \textbf{b} -\frac{1}{3}\bm\delta\div \textbf{b})
\\
    \intO{2\textbf{r}\textbf{e}_{in}}
    &=
    \frac{2\pi}{5(\lambda+1)}
    (\bm\delta \textbf{u}_r +  \bm\delta\textbf{u}_r^\dagger)
    -\frac{4\pi}{15(\lambda+1)}\textbf{u}_r\bm\delta 
    -\frac{4\pi}{15}\frac{1}{\lambda+1}
    (\bm\delta \textbf{b} +  \bm\delta\textbf{b}^\dagger)
    +\frac{8\pi}{45}\frac{1}{\lambda+1}
    \textbf{b}\bm\delta. 
    \label{eq:st_mom}
\end{align}
In these expressions we did not include the dependence on $\propto \grad\grad \textbf{u}$ and $\propto \grad\grad \textbf{b}$ in \ref{eq:nd_mom} and \ref{eq:st_mom} because they turn out to be of $O(a^2/L^2)$, hence negligible in the averaged equation. 
Nevertheless, they will be needed to compute the deformation of the droplets, hence we display here the higher order terms, namely 
\begin{multline}
    \intS{(\textbf{xx})_{jk}(\bm\sigma_{out}\cdot \textbf{n})_i} 
    =
    \frac{7\lambda^2+190\lambda+88}{105(\lambda^2+5\lambda+4)}[
        (\grad\grad \textbf{u})_{ijk}
        + (\grad\grad \textbf{u})_{ikj}
    ]
    + \frac{119\lambda^2+190\lambda-24}{105(\lambda^2+5\lambda+4)}(\grad\grad \textbf{u})_{jki}\\
    - \frac{7\lambda^2+80\lambda-72}{105(\lambda^2+5\lambda+4)}[
        (\bm\delta \grad^2 \textbf{u})_{ijk}
        + (\bm\delta \grad^2 \textbf{u})_{ikj}
    ]
    + \frac{2(13\lambda-4)}{21(\lambda^2+5\lambda+4)}(\bm\delta\grad^2 \textbf{u})_{jki}\\
    \frac{76}{735(\lambda+1)}(\grad\grad \textbf{b})_{ijk}
    + \frac{218}{3675(\lambda+1)} (\bm\delta\grad^2\textbf{b})_{jki}
    - \frac{272}{3675(\lambda+1)} [(\bm\delta\grad^2\textbf{b})_{ijk}+(\bm\delta\grad^2\textbf{b})_{ikj}],
    \label{eq:second_order_closuresS}
\end{multline}
\begin{multline}
    \intS{(\textbf{x})_{k}(\textbf{e}_{in})_{ij}} 
    =
    - \frac{7\lambda^2+20\lambda+3}{105(\lambda^2+5\lambda+4)}[
        (\grad\grad \textbf{u})_{kij}
        + (\grad\grad \textbf{u})_{kji}
    ]
    - \frac{8(2\lambda+1)}{21(\lambda^2+5\lambda+4)}(\grad\grad \textbf{u})_{ijk}\\
    - \frac{\lambda-10}{21(\lambda^2+5\lambda+4)}[
        (\bm\delta \grad^2 \textbf{u})_{jki}
        + (\bm\delta \grad^2 \textbf{u})_{ikj}
    ]
    + \frac{2(3\lambda-2)}{21(\lambda^2+5\lambda+4)}(\bm\delta\grad^2 \textbf{u})_{ijk}\\
    - \frac{32}{735(\lambda+1)}(\grad\grad \textbf{b})_{ijk}
    + \frac{292}{11025(\lambda+1)} (\bm\delta\grad^2\textbf{b})_{ijk}
    - \frac{22}{1225(\lambda+1)} [(\bm\delta\grad^2\textbf{b})_{jki}+(\bm\delta\grad^2\textbf{b})_{ikj}].
    \label{eq:second_order_closuresE}
\end{multline}

\subsubsection*{Deformation of the droplets}

In the body of the text, we use the second moment of the momentum equation to determine the third mode of deformation $\textbf{H}_p^{(3)}$ and show that it is negligible. 
Here, we provide additional information on the derivation of the second moment of the momentum equation. 
Based on the general formula \ref{eq:dt_Q_n} one obtain by neglecting the inertial effects  as well as the body forces,  
\begin{equation}
    \intO{ r_{j}(\bm{\sigma}^0_d)_{ik}+r_{k}(\bm{\sigma}^0_d)_{ji}}
    +\intS{ r_{j}(\bm{\sigma}^0_\Gamma)_{ik}+r_{k}(\bm{\sigma}_\Gamma^0)_{ji}}
    = 
    \intS{  r_{k}r_{j} (\bm{\sigma}_f^0\cdot\textbf{n})_i }
\label{eq:step_1}
\end{equation}
Using the approach outlined by \citet{lhuillier1996contribution}, we can rewrite this equation by summing and subtracting the permutations:  $B_{ijk} + B_{jik} - B_{kij}$, where $B_{ijk}$ corresponds to \ref{eq:step_1}.
This yields:
\begin{align}
    \intO{2 r_{k}(\bm{\sigma}^0_d)_{ij}}
    +\intS{2\gamma r_{k}(\bm\delta - \textbf{nn})_{ij}}
    &= 
    +\intS{
        r_{k}r_{j} (\bm{\sigma}_f^0\cdot\textbf{n})_i 
        + r_{k}r_{i} (\bm{\sigma}_f^0\cdot\textbf{n})_j 
        - r_{j}r_{i} (\bm{\sigma}_f^0\cdot\textbf{n})_k 
    }
\end{align}
Taking the traceless part of this equation on the index $ij$, and substituting $\bm\sigma_f^0$ by $\bm\sigma_f^* + \bm\Sigma$ and $\textbf{e}_d^0$ by $\textbf{e}_d^* + \textbf{E}$ one obtains, 
\begin{align}
    \intS{2\gamma r_{k}(\bm\delta/3 - \textbf{nn})_{ij}}
    &= 
    +\intS{
        r_{k}r_{j} (\bm{\sigma}^*_f\cdot\textbf{n})_i 
        + r_{k}r_{i} (\bm{\sigma}^*_f\cdot\textbf{n})_j 
        - r_{j}r_{i} (\bm{\sigma}^*_f\cdot\textbf{n})_k 
    }\nonumber \\
    &-\frac{1}{3}\delta_{ij}\intS{
        r_{k}r_{l} (\bm{\sigma}^*_f\cdot\textbf{n})_l
        + r_{k}r_{l} (\bm{\sigma}^*_f\cdot\textbf{n})_l 
        - r_{l}r_{l} (\bm{\sigma}^*_f\cdot\textbf{n})_k 
    }\nonumber \\
    &+\mu_f (1-\lambda)\intO{ 4\textbf{E}_{ij} r_{k}  }
    - \mu_f\lambda \intO{2 r_{k}(\textbf{e}_d^*)_{ij}},
\label{eq:second_mom}
\end{align}
where we have noticed that $\div\bm\Sigma = -  \rho_f \textbf{g}$ at $O(1)$ in $\phi$, and because we neglected the body forces in this example  $\div\bm\Sigma =0$. 
Injecting \ref{eq:second_order_closuresS,eq:second_order_closuresE} in \ref{eq:second_mom} one arrive at \ref{eq:defH3}.

\bibliography{Bib/bib_bulles.bib}

\end{document}